\documentclass[fleqn,usenatbib]{mnras}

\usepackage{newtxtext,newtxmath}

\usepackage[T1]{fontenc}

\DeclareRobustCommand{\VAN}[3]{#2}
\let\VANthebibliography\thebibliography
\def\thebibliography{\DeclareRobustCommand{\VAN}[3]{##3}\VANthebibliography}

\usepackage{amsfonts}
\usepackage[nopatch]{microtype}
\usepackage[table]{xcolor}
\usepackage{booktabs}
\usepackage{physics}
\usepackage{caption}
\usepackage{subcaption}
\usepackage{hyperref}
\usepackage{xcolor}
\usepackage{soul}
\usepackage{graphics}
\usepackage{float}
\usepackage{amsmath}
\usepackage{pdfpages}
\usepackage{aas-macros}

\title[Jet feedback modelling]{On the consistency of jet feedback modelling across different astrophysics hydrodynamical codes}

\author[Maragkakis et al.]{N. Maragkakis $^{1,2}$\thanks{E-mail: nick.maragkakis@research.uwa.edu.au},
M. A. Bourne$^{3,4,5}$,
C. Power $^{1,2}$,
F. Huško$^6$,
A. Ludlow$^1$, \&
S. Shabala$^7$
\\
$^{1}$International Centre for Radio Astronomy Research, The University of Western Australia, 35 Stirling Highway, Crawley, Western Australia 6009, Australia\\
$^{2}$ARC Centre of Excellence for All Sky Astrophysics in 3 Dimensions (ASTRO 3D)\\
$^{3}$Centre for Astrophysics Research, Department of Physics, Astronomy and Mathematics, University of Hertfordshire, College Lane, Hatfield, AL10 9AB, UK\\
$^{4}$Institute of Astronomy, University of Cambridge, Madingley Road, Cambridge CB3 0HA, UK\\
$^{5}$Kavli Institute for Cosmology (KICC), University of Cambridge, Madingley Road, Cambridge CB3 0HA, UK\\
$^6$Leiden Observatory, Leiden University, PO Box 9513, 2300 RA Leiden, Netherlands\\
$^7$School of Natural Sciences, University of Tasmania, Private Bag 37, Hobart, Tasmania 7001, Australia\\
}

\date{Accepted XXX. Received YYY; in original form ZZZ}

\pubyear{\the\year{}}

\begin{document}
\label{firstpage}
\pagerange{\pageref{firstpage}--\pageref{lastpage}}
\maketitle

\begin{abstract}
Active Galactic Nuclei (AGN) feedback is essential in cosmological simulations of galaxy formation, yet its implementation has to rely on subgrid models due to limited resolution. We present a novel subgrid jet-launching method for galaxy formation simulations and implement it in three hydrodynamical codes: the smoothed particle hydrodynamics (SPH) code {\sc swift}, the moving-mesh code {\sc arepo}, and the Eulerian grid code {\sc pluto}. To isolate the impact of hydrodynamical solvers on jet evolution, we compare idealised jets and their remnants in uniform and stratified media across resolutions and jet parameters. In uniform media, all jets drive bow shocks, inflate hot lobes, exhibit backflows, and evolve self-similarly. For the parameters explored, {\sc swift} lobes are shorter, wider, and hotter; {\sc arepo} lobes are longer, thinner, and cooler; while {\sc pluto} lobes display complex flows with intermediate characteristics. In stratified media, jets deviate from self-similar evolution, inflating longer and thinner lobes due to lower external ram pressure. After switch-off, {\sc swift} jets evolve into smooth cylindrical bubbles, {\sc arepo} jets produce long filamentary remnants, and {\sc pluto} jets yield intermediate-length remnants with varying degrees of mixing. Despite such differences, all jets and remnants have a similar impact on the ambient medium. We conclude that variations in lobe properties between codes emerge even for identical subgrid prescriptions, since the coupling of jet feedback to resolvable scales and the effective resolution depend on the hydrodynamical method. In structure formation simulations, these solver differences are likely subdominant to uncertainties in subgrid modelling and calibration, while averaging over galaxy populations may lessen their impact.
\end{abstract}

\begin{keywords}
galaxies: formation -- galaxies:jets -- methods: numerical
\end{keywords}



\section{Introduction}
\label{sec:introduction}

\par Numerical simulations are a powerful tool for the theoretical astrophysicist, allowing them to study highly complex and non-linear problems. This is particularly true in modelling galaxy formation, where feedback---the coupling of mass, momentum, and energy injected by stars and black holes into their ambient medium---can drive behaviour that is highly sensitive to physical conditions on a range of spatial and temporal scales \citep{Vogelsberger_2020}. Cosmological hydrodynamical simulations and semi-analytic models have revealed that feedback plays a key role in shaping the properties of the galaxy population, such as the galaxy luminosity function \citep[e.g.][]{Cole_2000, Benson_2003, Croton_2006, Bower_2006} and (equivalently) the galaxy stellar mass function (GSMF) and the stellar-to-halo mass relation \citep[e.g.][]{Schaye_2014, Dubois_2016, McCarthy_2016, Pillepich_2017, Henden_2018, Dave_2019}. Additionally, feedback contributes to establishing scaling relations such as those between black hole mass and stellar bulge mass \citep[e.g.][]{Booth_2010, Schaye_2014, Sijacki_2015, Volonteri_2016}, galaxy size and mass \citep[e.g.][]{Schaye_2014, Furlong_2016, Genel_2017, Arjona_2025}, and \textsc{h\,i} mass and halo mass \citep[e.g.][]{Chauhan_2020}. Feedback from active galactic nuclei (AGN) is now generally accepted as a key mechanism for ensuring that star formation in massive galaxies is quenched (and remains so), and that galaxies attain colours and morphologies consistent with observations \citep[e.g.][]{Dubois_2013, Sijacki_2015, Dubois_2016, Donnari_2020, Goubert_2024, Byrne_2024}.

\par AGN feedback is driven by the accretion of matter onto the central supermassive black holes of galaxies, and its effects are evident in galaxy groups and clusters \citep[e.g.][]{Fabian_2012, Eckert_2021}. It acts in a variety of ways: driving accretion-disc winds \citep{Pounds_2003, King_2003, King_2015, Costa_2020}; exerting direct radiation pressure \citep{Ishibashi_2015, Bieri_2017, Costa_2018}; and/or launching fast bipolar jets \citep{Scheuer_1974, Hardcastle_2020, Bourne_2023, Mukherjee_2025}, which heat the ambient medium \citep{Sazonov_2005, Arrigoni_2018} and eject massive outflows \citep{Rupke_2011, Spilker_2025}. AGN jets can contribute to the quenching of star formation in galaxies and the regulation of heating and cooling in galaxy groups and clusters \citep[e.g.][]{McNamara_2007, Fabian_2012, Hlavacek-Larrondo_2022, Bourne_2023}. Evidence for such AGN jet feedback is provided by X-ray cavities that are filled with radio-emitting plasma, observed in galaxy clusters, such as Perseus \citep{Böehringer_1993, Fabian_2000, Fabian_2003, Fabian_2006, Hitomi_2016, van_Weeren_2024} and Virgo \citep{Forman_2005, Forman_2007}. 

\par The theoretical basis for AGN feedback as a key regulator of galaxy growth was originally outlined by \citet{Silk_1998}. AGN feedback in galaxy formation modelling was then subsequently included in semi-analytic models as radio-mode feedback. This injected energy into the hot gaseous atmospheres surrounding massive galaxies in high-mass halos, offsetting gas cooling and preventing them from becoming overly massive and luminous \citep{Croton_2006, Bower_2006}. It differed from hydrodynamical simulations, which modelled AGN feedback as quasar-mode feedback via thermal energy injections \citep[e.g.][]{Di_Matteo_2005, Di_Matteo_2008}. While such models are still in use, many modern hydrodynamical galaxy formation simulations adopt a two-mode AGN feedback scheme, which combines the quasar and radio modes---at high (low) black hole accretion rates, the radiatively efficient (inefficient) quasar (radio) mode is active \citep{Sijacki_2007, Dubois_2012, Sijacki_2015, Weinberger_2016, Dave_2019}. 

\par Despite this progress, implementing AGN feedback in hydrodynamical simulations remains challenging because the relevant physical processes span a large spatial range, from sub-parsec to megaparsec scales \citep[see e.g.][]{Gaspari_2020, Oei_2024}. The limited resolution of cosmological galaxy formation simulations means that AGN feedback, by necessity, is included as a subgrid model whose primary aim is to capture its impact on scales resolvable in the computational domain. This demands simplifying prescriptions for how feedback is deposited into the gas surrounding black holes, and it is commonplace to calibrate models to reproduce present-day ($z=0$) large-scale observables such as the GSMF \citep{Schaye_2014, Pillepich_2017, Dave_2019}, the black hole mass--stellar bulge mass relation \citep{Schaye_2014, Dubois_2016, Dave_2019}, and, more recently, the baryon content of groups and clusters \citep{McCarthy_2016, Henden_2018, Kugel_2023, Bigwood_2025}. Moreover, assumptions made in galaxy formation models (e.g. the choice of cooling function or star formation model) as well as the underlying hydrodynamical solver, will affect the implementation and physics of feedback \citep{Bourne_2015}. Together, these factors contribute to substantial uncertainties in the first-principles modelling of AGN feedback in cosmological hydrodynamical simulations.

\par At the same time, variations in feedback models and parameter choices can lead to differences in the resulting galaxy populations. While global galaxy statistics across large-volume cosmological simulations---such as EAGLE \citep{Schaye_2014}, Horizon-AGN \citep{Dubois_2016},  IllustrisTNG \citep{Pillepich_2017}, and SIMBA \citep{Dave_2019}---are in broad agreement in terms of stellar mass functions and star formation rate densities, discrepancies emerge at the level of individual systems. For example, galaxy morphologies \citep{Genel_2019}, cold gas fractions \citep{Dave_2020}, black hole growth histories \citep{Habouzit_2021}, and gas inflows and outflows around central galaxies in halos \citep{Wright_2024} can differ significantly between simulations. Moreover, additional differences arise in the predicted population statistics of quasars and AGN jets, as quantified by the quasar and AGN radio luminosity functions \citep{Raouf_2017, Fanidakis_2010, Thomas_2021}. 

\par This frames the twofold challenge that galaxy formation simulations face. First, AGN feedback cannot be modelled from first principles on galactic scales and requires subgrid models. Second, the effects of these models are highly sensitive to numerical choices and parameters, making interpretation and comparison of results difficult. In this context, comparison projects involving different hydrodynamical codes and different models of cooling, star formation, and feedback are a powerful numerical tool to help us address these challenges. 

\par Accurate modelling of AGN jets requires a careful treatment of strong shocks and fluid instabilities, both of which are sensitive to the underlying hydrodynamical method. In general, the choice of hydrodynamical solver can impact the outcome in several key aspects of astrophysical problems. Because of this, there have been a number of studies exploring the differences between SPH, moving-mesh/meshless, and grid-based approaches in hydrodynamical simulations \citep[e.g.][]{Frenk_1999, Voit_2005, Agertz_2007, Wadsley_2008, Sijacki_2012, Power_2014, Hopkins_2015, Sembolini_2016}. Some studies have examined standard hydrodynamical tests (such as a dense gas cloud impinging on a diffuse wind, the so-called blob test) and showed that grid codes \citep{Agertz_2007} and moving-mesh/meshless codes \citep{Sijacki_2012, Hopkins_2015} perform better than traditional SPH codes. However, recent improvements in the SPH methods \citep[e.g.][]{Rosswog_2007, Read_2012, Saitoh_2013, Beck_2015, Wadsley_2017, Price_2018, Borrow_2021, Sandnes_2025} make them increasingly competitive in their performance against standard benchmarks \citep[e.g.][]{Braspenning_2023}.

\par Traditionally, grid codes have handled interacting fluids, mixing, and instabilities better \citep{Agertz_2007} while being able to sufficiently resolve high-density contrasts, making them the preferred computational method for high-resolution AGN jet feedback simulations \citep{Bourne_2023}. Building upon this, moving-mesh and meshless codes have the potential to improve upon the fixed-mesh approaches and, when combined with refinement schemes, have proven effective at resolving high-resolution jets in idealised \citep{Weinberger_2017, Bourne_2017, Ehlert_2022} as well as more realistic cluster environments \citep{Bourne_2019, Su_2021, Bourne_2021}. More recently, an SPH code has been used to model high-resolution idealised jets in \cite{Husko_2023}, and jets and remnants in \cite{Husko_2023b}, highlighting the ease and efficiency of particle-based methods.

\par In this paper, we examine the AGN jets produced by three independent, state-of-the-art, astrophysical codes:
\begin{itemize}
    \item {\sc swift}~\citep{SWIFT}, which represents the Lagrangian or smoothed particle hydrodynamics approach,
    \item {\sc arepo}~\citep{Arepo}, which represents the moving-mesh approach, and
    \item {\sc pluto}~\citep{PLUTO,PLUTO_AMR}, which represents the Eulerian or adaptive mesh-refinement approach.
\end{itemize}
To do this, we employ a novel feedback model that is designed for resolutions achievable in modern cosmological and galaxy formation simulations. 

\par We note that pure Eulerian codes are still the most widely used approach in AGN jet simulations and are well suited to capturing the complex structure of jet lobes, including jet recollimation, backflows, and turbulence \citep[e.g.][]{Yates-Jones_2021, Yates-Jones_2023}. In contrast, moving-mesh codes require specialised refinement schemes to properly resolve the low-density lobes \citep[e.g.][]{Weinberger_2017, Bourne_2017}, while SPH codes are expected to be less effective for jet simulations without improvements in mixing and particle-splitting techniques. Nonetheless, it is useful to quantify any differences between codes and test different parameters and resolutions to assess their consistency. This is especially relevant for cosmological hydrodynamical simulations where different numerical approaches may offer unique advantages and where AGN jet modelling is only one aspect of more complicated systems.

\par In Section~\ref{sec:astrophysical hydrodynamical codes}, we briefly review the key features of the three codes we consider: {\sc swift}, {\sc arepo}, and {\sc pluto}. In Section~\ref{sec:jet feedback across codes}, we model our idealised AGN jets using the aforementioned codes with our novel launching module. We present our results in Section~\ref{sec:results}, including uniform-medium runs, stratified-medium runs and remnant runs. Finally, we discuss our results in Section~\ref{sec:discussion}, and summarise our conclusions in Section~\ref{sec:conclusions}.

\section{Astrophysical hydrodynamical codes}
\label{sec:astrophysical hydrodynamical codes}

\par We use the public versions of three astrophysical codes, considering only pure hydrodynamics coupled to a minimal jet model (see Section~\ref{ssec:jet model}).

\subsection{{\sc swift}}
\label{ssec:swift}
\par {\sc swift} \citep{SWIFT} is a Lagrangian, SPH code that solves the equations of motion for the fluid using particle-carried and kernel-smoothed quantities. We adopt the default SPHENIX \citep{SPHENIX} scheme that incorporates artificial viscosity to capture shocks, as well as artificial conductivity to promote mixing between different fluid layers. Both are supplemented with limiters that adapt their strength accordingly, to avoid excessive viscosity in shear flows and unwanted energy dissipation in feedback injections. Additionally, we make use of the particle-splitting module in {\sc swift} to keep all SPH particle masses roughly the same throughout the simulations.

\subsection{{\sc arepo}}
\label{ssec:arepo}
\par {\sc arepo} \citep{Arepo, Arepo2, ArepoCode} is a quasi-Lagrangian, moving-mesh code that solves hydrodynamics on an unstructured Voronoi mesh using a second-order, finite-volume discretisation with an exact Riemann solver. The mesh can move with the flow, and in its standard operating state, a roughly constant mass resolution is maintained via a cell refinement and de-refinement scheme. Previous approaches implemented within {\sc arepo} to study AGN-driven jet evolution \citep[e.g.,][]{Weinberger_2017, Bourne_2017, Talbot_2021} have employed additional super-Lagrangian refinement criteria to enhance resolution within the jets and lobes. However, we use the standard quasi-Lagrangian approach (i.e., fixed-mass resolution) to ensure our model remains applicable to a broad range of galaxy formation simulations.

\subsection{{\sc pluto}}
\label{ssec:pluto}
\par {\sc pluto} \citep{PLUTO,PLUTO_AMR} is an Eulerian, grid-based simulation code developed for high-Mach-number astrophysical fluid flows. The fluid is evolved on a static three-dimensional Cartesian grid by solving the conservation laws using the HLLC (Harten-Lax-van Leer-Contact) approximate Riemann solver with linear reconstruction, and a second-order Runge-Kutta time-integration scheme. To increase the simulation robustness, we use the shock-flattening feature to switch to the HLL solver and the MINMOD limiter in the presence of strong shocks. The code supports several different physics modules, including special relativistic hydrodynamics and magnetohydrodynamics. However, we only use non-relativistic hydrodynamics in the absence of magnetic fields to ensure broad applicability.

\section{Jet feedback theory and modelling}
\label{sec:jet feedback across codes}

\par We focus on simulating idealised hydrodynamical jets in simplified environments to facilitate robust code comparisons. To maintain control over the setup, we do not couple jet injection to accretion onto a supermassive black hole but, instead, introduce jets with fixed powers and durations, using parameters set by hand. These include jet power, injection velocity, opening angle, and ambient-medium properties, chosen to produce jets that resemble Fanaroff–Riley type II (FR-II) sources and inflate lobes of hot gas. This idealised approach enables a clean and controlled comparison between codes, avoiding complications arising from differences in accretion physics. While the jets may exhibit features of real AGN sources, our setup is deliberately simplified to isolate the effects of jet propagation and lobe inflation. This allows us to determine their impact on the thermodynamic profiles of stratified media, and to explore implications for cosmological simulations.

\par We adopt an adiabatic equation of state and neglect gas cooling, self-gravity, and magnetic fields. Additionally, our jets are injected with sub-relativistic speeds ($< 0.5 \, c$). Although launching jets at relativistic speeds and including special relativistic effects can influence jet stability and even lobe properties \citep{Komissarov_1998, English_2016, Perucho_2017, Yates-Jones_2023}, classical jets capture important aspects of the evolution, especially the self-similar scaling of lobes \citep[see e.g.][]{Turner_2023a}. This is relevant for cosmological galaxy formation simulations that typically do not implement special relativistic effects.

\subsection{Self-similar theory}
\label{ssec:self-similar theory}

\par Our simulations consist of jets propagating in ambient media with power-law density profiles of the form
\begin{equation}
  \rho = \rho_0 \left(\frac{r}{r_\mathrm{c}}\right)^{-\alpha},
\end{equation}
where $\rho_0$, $r_\mathrm{c}$, and $\alpha$ are constants. We consider two cases with $\alpha < 2$ to study jets, lobes, and remnants.  We include runs in a uniform medium with $\alpha = 0$, and in a stratified medium with $\alpha \sim 1$. We compare our jets with each other and with an analytic self-similar solution that is derived by making a number of simplifying assumptions. In the self-similar regime, the jet length $L_\mathrm{jet}$ depends only on the jet power, $P_\mathrm{j}$, and the ambient-medium density, as the product $\rho_0 r_\mathrm{c}^{\alpha}$. Using dimensional analysis, we can derive its evolution with time as \citep{Falle_1991}
\begin{equation}
    L_\mathrm{jet} \sim l \equiv \left(\frac{P_\mathrm{j}}{\rho_0 r_\mathrm{c}^\alpha}\right)^{1/(5-\alpha)} t^{3/(5-\alpha)}.
\end{equation}
For a uniform medium, $\alpha = 0$, and hence 
\begin{equation}
    L_\mathrm{jet} \sim l = \left(\frac{P_\mathrm{j}}{\rho_0}\right)^{1/5} t^{3/5}.
    \label{eq:3}
\end{equation}
The exact formula, valid for cylindrical jets, can be found by including a dimensionless constant of order unity, $C$, in the above formula as  \citep{Kaiser_1997, Komissarov_1998, Kaiser_2007}
\begin{equation}
    L_\mathrm{jet} = C \times l = C \, \left(\frac{P_\mathrm{j}}{\rho_0}\right)^{1/5} t^{3/5},
    \label{eq:4}
\end{equation}
where 
\begin{equation}
    C = \left\{\frac{A^4}{18\pi}\frac{(\gamma + 1)(\gamma - 1)(5-\alpha)^3}{9[\gamma + (\gamma - 1)A^2/2] - 4 - \alpha}\right\}^{1/(5-\alpha)}.
    \label{eq:5}
\end{equation}
Here, $\gamma = 5/3$ is the ideal-gas adiabatic index and $A$ is the lobe aspect ratio (i.e. the ratio of the jet-lobe length to the jet-lobe width). In the self-similar solution, the lobe aspect ratio remains constant throughout the jet's evolution. We estimate $A \sim 1/\theta$, where $\theta$ is the half-opening angle of the jet \citep{Komissarov_1998}. Then the width of the jet can be expressed as 
\begin{equation}
    W_\mathrm{jet} = \frac{L_\mathrm{jet}}{A}.
    \label{eq:6}
\end{equation}
While this analytic solution ignores several key features of AGN jets, like the breakout phase, backflows, instabilities, and non-thermal pressure (e.g. from cosmic rays), it is nevertheless a useful, approximate guide for the large-scale evolution of the jet lobes. There are other analytic and simulation based models of jet propagation that might better approximate specific aspects of the evolution (e.g. the early phase) \citep{Begelman_1989, Hardcastle_2018, Su_2021, Turner_2023b},  but comparing these to our results is beyond the scope of this paper.

\subsection{Jet injection model}
\label{ssec:jet model}

\par In designing a model for jet injection, it is necessary to account for the inherent differences between grid and SPH codes. There is a plethora of grid-based models in the literature that launch jets by injecting mass, momentum, and energy \citep[e.g.][]{KarenYang_2016, Bourne_2017, Martizzi_2018} or by setting a hydrodynamic state within the injection region \citep[e.g.][]{Weinberger_2017, Weinberger_2023, Yates-Jones_2023}. On the other hand, SPH jets can be realised via injection or kicking of particles \citep{Husko_2022, Husko_2023, Husko_2023b}. To ensure a clean comparison between different codes, we developed a hybrid virtual-particle model that is largely code agnostic. Virtual particles act as carriers of mass, momentum, and energy, but are hydrodynamically decoupled; they are launched from the origin and deposit the mass, momentum, and energy into the ambient medium once certain criteria are satisfied. 

\par
\begin{figure}
    \centering
    \includegraphics[width=\linewidth]{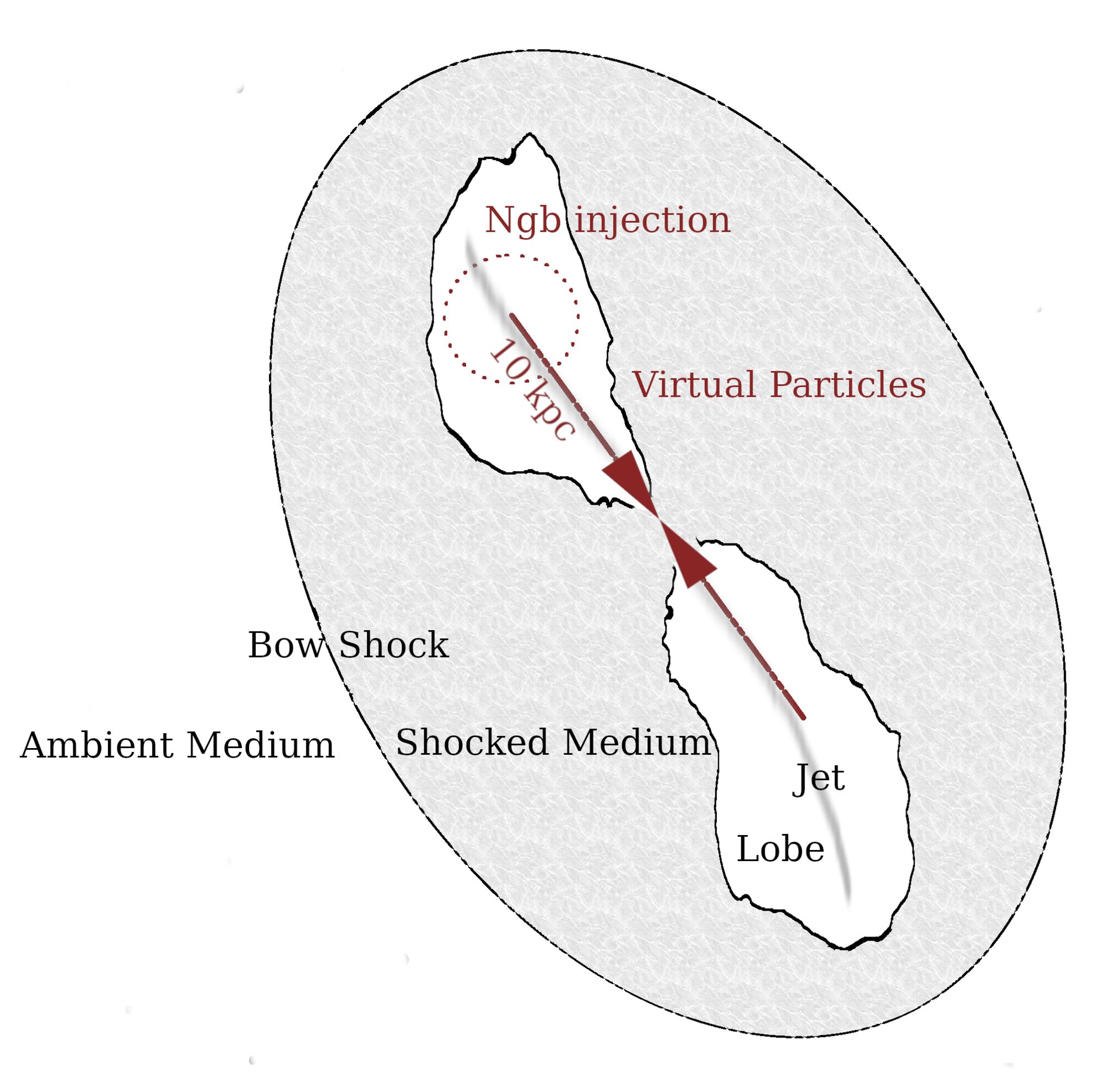}
    \caption{Sketch of our virtual-particle model for a single injection event. Two virtual particles are launched in opposite directions, inside two cones, and travel for $10 \, \mathrm{kpc}$ without interacting with the medium, before they inject mass, momentum, and energy into their gas neighbours. Denoted are the different components of the resulting jets and lobes they inflate.}
    \label{fig:1.Sketch}
\end{figure}

 We simulate a single jet episode with a series of discrete injection events. A single injection event consists of launching two virtual particles from the centre of our computational domain along opposite directions (with the first chosen at random), inside two coaxial cones of half-opening angle $\theta = 15\degr$ to produce two jets and ensure conservation of linear momentum. The virtual particles travel with constant velocity, without interacting with the gaseous medium, until they reach a pre-defined radius, $r_0 = 10 \, \mathrm{kpc}$. They then deposit mass, momentum, and energy into the $10$ nearest gas elements, in a mass-weighted fashion, and are removed from the computational domain (see Fig.~\ref{fig:1.Sketch}). 

\par A pair of virtual particles is launched whenever
\begin{equation}
    P_\mathrm{j} \Delta t \geq 2\times \frac{1}{2} m_\mathrm{j} v_\mathrm{j}^2,
    \label{eq:f}
\end{equation}
where $P_\mathrm{j}$ is the jet power, $\Delta t$ is the time elapsed since the last injection event, and $m_\mathrm{j}$ and $v_\mathrm{j}$ are the virtual-particle (jet) mass and velocity, respectively. The term $1/2 \, m_\mathrm{j} v_\mathrm{j}^2$ represents the energy carried by each particle, noting that the factor of two in Eq.~\ref{eq:f} accounts for two particles being generated and kicked in opposite directions. This energy is not added into the system until the virtual particles have propagated to the prescribed distance $r_0$. 

\par Upon reaching this radius, each particle's $10$ nearest gas elements receive mass and energy as
\begin{equation}
    \Delta m = m_\mathrm{j},
\end{equation}
and
\begin{equation}
    \Delta E = \frac{1}{2} m_\mathrm{j} v_\mathrm{j}^2,
    \label{energy eq}
\end{equation}
so that the total mass and total energy injected (from both particles) are $2\times\Delta m$ and $2\times\Delta E$. Each neighbouring gas element $i$ receives a fraction of this feedback, updating its mass and energy as
\begin{equation}
    m_i = m_{i0} + \frac{m_{i0}}{m_\mathrm{ngb}}\times m_\mathrm{j},
\end{equation}
\begin{equation}
    E_i = E_{i0} + \frac{m_{i0}}{m_\mathrm{ngb}}\times \frac{1}{2} m_\mathrm{j} v_\mathrm{j}^2,
\end{equation}
where $m_{i0}$ and $E_{i0}$ are its initial mass and energy, and $m_\mathrm{ngb}$ represents the total mass of the  neighbours in a loop. 

\par For this energy to be interpreted as kinetic, we add momentum accordingly to achieve the proper final state. Following \citet{Bourne_2017}, the momentum of each gas element is updated as
\begin{equation}
    \boldsymbol{p_i} = \boldsymbol{p_{i0}} + \left(\sqrt{2m_i K_i} - \lvert{\boldsymbol{p_{i0}}}\rvert\right) \frac{\boldsymbol{v_\mathrm{j}}}{v_\mathrm{j}}.
\end{equation}
where $\boldsymbol{p_{i0}}$ is its initial momentum, $K_i$ is the target kinetic energy ($E_i$ minus any initial thermal energy), and the resulting momentum kick acts in the direction of travel of the virtual particle. As a small caveat, this formula does not strictly conserve the injected energy due to possible momentum cancellation. In this case, an additional thermal component is added to account for the difference \citep[see][for details]{Bourne_2017}. 

\par The final velocities of the gas elements will then depend on their initial masses and kinetic energies, as well as on the parameter $v_\mathrm{j}$, and will thus vary between the codes. Nonetheless, for simplicity, we will refer to the parameter $v_\mathrm{j}$ as the (jet) injection velocity.

\par To faithfully follow the jet’s propagation in the surrounding medium, a proper numerical integration scheme must also be chosen with care. For AGN feedback to be accurately captured in simulations, time-stepping must be fine enough to resolve its effects within the feedback region; for instance, this has been shown to be an important source of error in simulations of supernova feedback \citep[e.g.][]{Saitoh_2009, Durier_2012}. To control hydrodynamical time steps, we use the standard Courant-Friedrichs-Lewy (CFL) condition, which relies on the sound speed and size of a resolution element to set a time-step limit. However, the CFL condition does not account for the extra injection of energy and supersonic propagation of jets when resolution elements first receive feedback. 

\par Thus, for codes that employ hierarchical, local time-stepping ({\sc swift} and {\sc arepo}), we enforce small and fixed time steps for all resolution elements that reside in the central regions ($r<40 \, \mathrm{kpc}$) of our domain and will be directly or indirectly affected by feedback injections. Moreover, we include the built-in time-step limiter in {\sc swift}, of the kind advocated by \citet{Saitoh_2009}, to employ a neighbour loop that wakes up inactive particles if their time steps are too large compared to their neighbours \citep{SWIFT}. Similarly, we include the non-local tree-based time-step limiter in {\sc arepo}, which aims to estimate the arrival time of waves from other locations within the simulation domain \citep{Arepo}. On the other hand, for the global time-stepping code {\sc pluto} we adopt a fixed time step for all cells, and set it small enough to accommodate jet feedback. In this setup, all codes have a fixed, uniform, and small enough time step for the resolution elements that are affected by feedback, as well as their neighbours. We note that the purpose of this study is the careful modelling and comparison of jets in idealised environments, and we reserve performance and scalability practicalities, including more specialised time-stepping schemes for realistic astrophysical or cosmological environments, to future works.

\par Finally, an important consideration when developing an AGN feedback model is refinement. While moving grid ({\sc arepo}) and fixed grid ({\sc pluto}) codes can use refinement and adaptive grids to achieve higher resolution in specific areas of interest, SPH codes ({\sc swift}) do not employ these techniques. This is relevant for high-resolution jet simulations and isolated jet studies, which are traditionally carried out with grid codes that employ a variety of refinement criteria or adaptive grid techniques to better resolve the propagation of jets and their interaction with the environment \citep[e.g.][]{Cattaneo_2007, Li_2014, KarenYang_2016, Bourne_2017, Bourne_2023}. However, in cosmological structure formation simulations, excessive refinement inside the jet region can become computationally expensive and/or present unintended negative consequences for other aspects of the simulation. For our results to be relevant for such cosmological simulations, we choose not to include any additional specialised refinement techniques in {\sc arepo}, nor an adaptive grid in {\sc pluto}. Nonetheless, to ensure particles in {\sc swift} retain similar masses after mass injections (which is important for SPH calculations), we include particle-splitting and split particles that reach twice their initial mass. Similarly, cells in {\sc arepo} are refined/de-refined to ensure that they have a consistent mass within a factor of two. These choices facilitate a fair comparison of codes under comparable effective resolutions, without introducing bias from extra refinement methods.

\subsection{Initial conditions and jet parameters}
\label{ssec:initial conditions}

\par For all simulations, we use a periodic box with a volume of $300 \times 300 \times 600 \, \mathrm{kpc^3}$ with the jet propagating along the long $z$ axis. In our uniform-medium runs, resolution elements are arranged on a regular Cartesian grid with initial density $\rho_0 = 2.5 \times 10^{-26} \, \mathrm{g/cm}^3$ and initial temperature $T = 10^7 \, \mathrm{K}$; this environment is representative of the central regions of galaxy clusters. We choose these parameters so as to produce jets that achieve self-similar behaviour (with various degrees of deviation from the analytic solution). In addition to the uniform medium, in Sections \ref{ssec:stratified medium jets} and \ref{ssec:remnants}, we perform simulations of jets and remnants in a stratified medium following an initially isothermal beta profile
\begin{equation}
    \rho = \rho_0 \left[1 + \left(\frac{r}{r_c}\right)^2\right]^{-3\beta/2}
\end{equation}
with $\rho_0 =  10^{-25} \, \mathrm{g/cm}^3$, $\beta = 0.38$ and $r_\mathrm{c} = 10 \, \mathrm{kpc}$. At large distances from the origin $\rho\sim r^{-3\beta} = r^{-1.14}$, so that the analytic self-similar solution remains valid. The beta density profile is motivated by observations of X-ray emission from galaxy groups and clusters and is frequently used in simulations of AGN jets \citep[e.g.][]{Vernaleo_2006, English_2016, Yates-Jones_2021}. In this case, we include an external gravitational field with acceleration
\begin{equation}
    \mathbf{g} = -3\beta u \frac{\gamma - 1 }{r^2+r_\mathrm{c}^2} \mathbf{r}
\end{equation}
where $ u = \frac{1}{\gamma-1} \frac{k_b T}{\mu m_p}$ is the initial internal energy, $T = 10^7 \, \mathrm{K}$ is the initial temperature, constants have their usual meaning, and we set $\mu = 0.6$. This potential is computed analytically so as to ensure that the initial configuration is in hydrostatic equilibrium.  

\par Throughout this work, we set the jet power to $P_\mathrm{j} = 5\times 10^{45} \, \mathrm{erg/s}$, typical of moderate to high-power radio-galaxy jets. In our standard, uniform-medium runs (Section~\ref{ssec:jets in three codes}), we set the virtual-particle velocity to $v_\mathrm{j} = 4 \times 10^4 \, \mathrm{km/s}$ to ensure the jets enter the self-similar regime. We set the virtual-particle mass to $m_\mathrm{j} = 4\times 10^4 \, \mathrm{M}_\odot$ to fix the number of injection events. This virtual-particle mass will be safely smaller than the masses of individual gas elements in {\sc swift} and {\sc arepo} to help keep the mass resolution relatively constant. The jets remain active for $98 \, \mathrm{Myr}$, resulting in approximately $12,000$ injection events. 

\par In Section~\ref{ssec:resolution} we compare different resolution runs in the uniform medium, in Section~\ref{ssec:jet injection velocity} we vary the injection velocity $v_\mathrm{j}$, and in Section~\ref{ssec:same neighbour mass injection} we compare with simulations that inject into a fixed neighbour mass. In the stratified-medium runs (Section~\ref{ssec:stratified medium jets}, Section~\ref{ssec:remnants}), we halve the virtual-particle mass to $m_\mathrm{j} = 2\times 10^4 \, \mathrm{M}_\odot$, keep the same velocity and shorten the jet duration to $44 \, \mathrm{Myr}$ to simulate a similar number of injection events and produce similar-length jets. In the uniform medium we use resolutions that vary from $N = 10^7$ to $N = 10^8$ resolution elements whereas in the stratified medium we instead employ $N = 2.9\times10^7 $ and $1.8\times10^7$ resolution elements. We summarise the complete set of our runs in Table~\ref{tab:Jets Table}. 

\begin{table*}
	\centering
	\caption{Summary of all jet simulations performed using \sc{swift}, \sc{arepo}, and \sc{pluto}.}
	\label{tab:Jets Table}
	\begin{tabular}{llcccl}
		\hline
		& Medium & Box volume $(\mathrm{kpc}^3)$ & Resolution elements & Injection velocity $(\mathrm{km/s})$ & Neighbour loop scheme \\
		\hline
		Fiducial & Uniform & $300^2 \times 600$ &  $10^8$ & $4\times10^4$ & Neighbour number \\   
		Resolution study & Uniform & $300^2 \times 600$ &  $10^7,5\times10^7,10^8$ & $4\times10^4$ & Neighbour number \\    
		Velocity study & Uniform & $300^2 \times 600$ &  $10^8$ & $2\times10^4,4\times10^4,8\times10^4$ & Neighbour number \\   
	    Neighbour mass study & Uniform & $300^2 \times 600$ &  $10^8$ & $2\times10^4,4\times10^4,8\times10^4$ & Neighbour mass \\     
		Stratified medium jets & Stratified & $300^2 \times 600$ &  $2.9\times10^7$ & $4\times10^4$ & Neighbour number \\   
		Remnants & Stratified & $900^2 \times 1200$ &  $1.8\times10^7$ & $4\times10^4$ & Neighbour number \\  
		\hline
	\end{tabular}
\end{table*}

\section{Results}
\label{sec:results}

\subsection{Comparison of jets across codes}
\label{ssec:jets in three codes}

\par To compare the large-scale evolution and morphology of jets across different hydrodynamical solvers, we now examine their properties in our fiducial simulation set. 

\par
\begin{figure*}
    \centering
    \includegraphics[width=\textwidth]{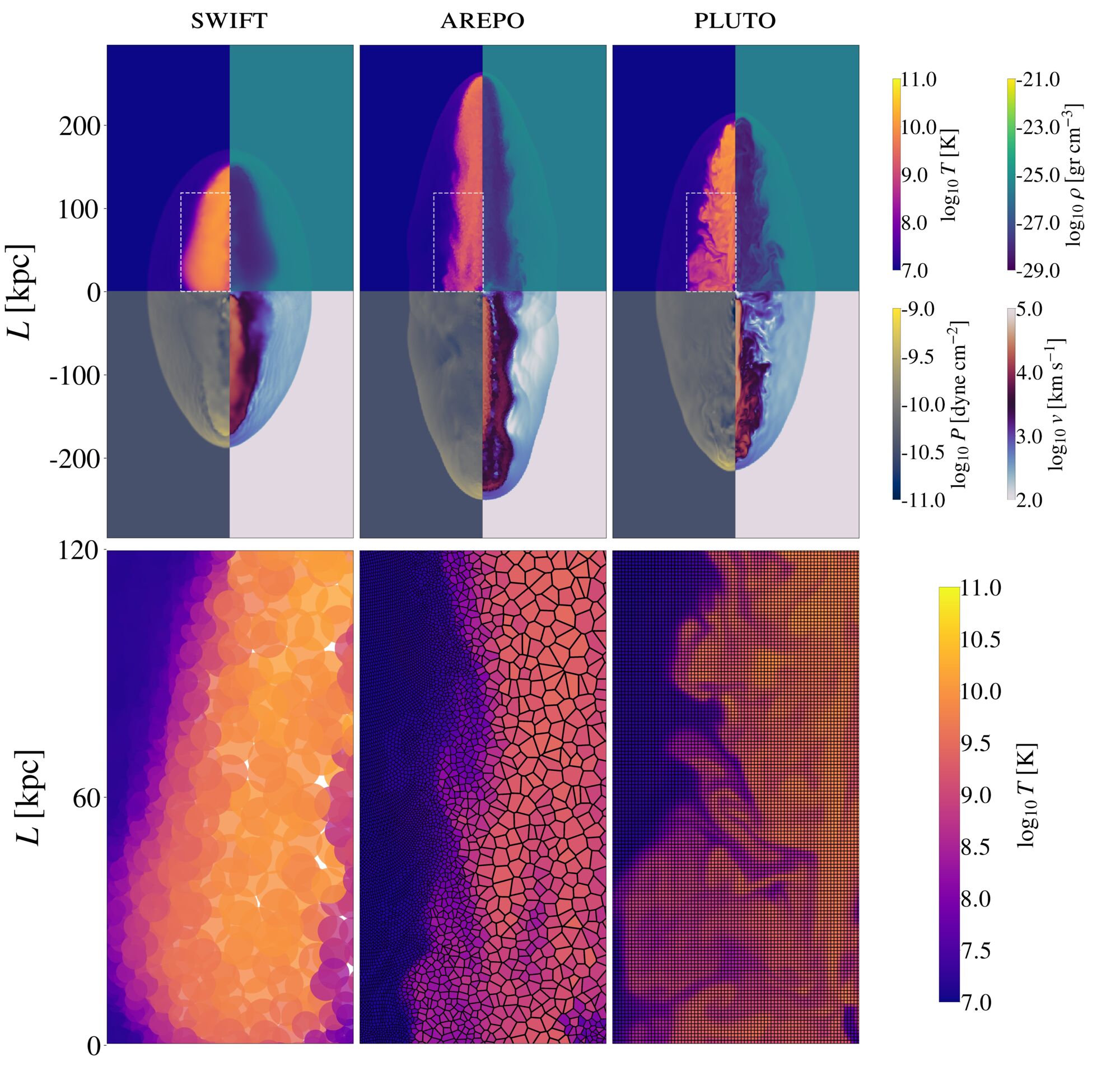}
     \caption{Comparison of {\sc swift}, {\sc arepo} and {\sc pluto} jets. \textbf{Top:} Overview slices in the $y=0$ plane in the uniform medium after $98 \, \mathrm{Myr}$ at our highest resolution ($N = 10^8$,  $M_\mathrm{gas} = 2\times 10^5 \, \mathrm{M}_\odot$) simulations. Mapped quantities are temperature (top left), density (top right), pressure (bottom left), and velocity magnitude (bottom right). All jets drive a bow shock in the ambient medium, inflate lobes, and develop backflows. {\sc swift} jets inflate shorter, wider, and hotter lobes, {\sc arepo} jets produce longer, thinner, and cooler lobes, and {\sc pluto} jets yield lobes that are of intermediate length and width and relatively hot. \textbf{Bottom:} Zoomed-in (white box in the top panel) temperature map displaying the underlying discretisation method for each code. We can distinguish differences between codes in the spatial resolution of the lobes and the medium, as well as in the development of Kelvin-Helmholtz instabilities.}
     \label{fig:2.Jets Overview}
\end{figure*}

 \noindent\textit{Visual impression:} In the top panel of Fig.~\ref{fig:2.Jets Overview} we show slices\footnote{The slices for the {\sc swift} jets are taken using the {\sc swiftsimio} python module \citep{Borrow_2020} which accounts for SPH smoothing. The slices for the {\sc arepo} and {\sc pluto} jets are generated directly from local cell properties within the slice plane.} in the $y=0$ plane, for the jets simulated with the three codes after $98 \, \mathrm{Myr}$ in the uniform medium. These simulations represent our highest resolution runs, consisting of $10^8$ initial resolution elements that have an initial mass of $M_\mathrm{gas} = 2\times 10^5 \, \mathrm{M}_\odot $. Each slice is split into four parts depicting gas temperature (top left), density (top right), pressure (bottom left) and velocity magnitude (bottom right). 

\par All of the jets drive a bow shock into the ambient medium and inflate lobes of hot gas. We can readily distinguish the different components of the jets (as labeled in Fig.~\ref{fig:1.Sketch}): a sheath of high-velocity jet material; high-temperature, low-density lobes consisting of shocked jet material; the shocked ambient medium behind the bow shock; and the yet unshocked ambient medium. All jets and lobes appear coherent, with any instabilities not significantly disrupting their structure. Inspecting the pressure slices, we infer that the jet lobes and the shocked external medium behind the bow shock achieve rough pressure equilibrium (especially in {\sc arepo} and {\sc pluto}), even though the jet heads remain over-pressured. In the velocity slices, the jet material in the central sheath reaches velocities very close to the jet injection velocity (i.e. the virtual-particle velocity), whereas the shocked material populating the jet lobes exhibits lower velocities.

\par There are some distinctions between jets simulated with different codes. {\sc swift} jets inflate short and hot lobes that are wide at the base. In contrast, {\sc arepo} jets produce longer, cooler, and thinner lobes. Finally, {\sc pluto} lobes are of intermediate length and width and relatively hot. Both {\sc arepo} and {\sc pluto} lobes show Kelvin-Helmholtz (KH) instabilities between the lobe and the shocked ambient medium that are not resolved at this resolution in {\sc swift}. 

\par Each code solves the fluid equations with a specific discretisation method that may influence the dynamics and propagation of jets. In the bottom panel of Fig.~\ref{fig:2.Jets Overview}, we zoom in on the white box region in the top panel in the temperature slice and depict the underlying structure of each of the codes. {\sc swift} lobes are resolved with particles with smoothing kernels while {\sc arepo} uses a voronoi mesh and {\sc pluto} a static grid. We observe how {\sc swift} and {\sc arepo} achieve variable spatial resolution based on the local density, with lower spatial resolution inside the lobes and higher spatial resolution at the interface of the lobe and ambient medium. In contrast, {\sc pluto} has the same spatial resolution everywhere. These different setups will also influence the development of fluid instabilities, with higher lobe resolution generally allowing for more complex flows and smaller-scale KH instabilities \citep{Husko_2023, Weinberger_2023}.

\par
\begin{figure*}
    \centering
    \includegraphics[width=\textwidth]{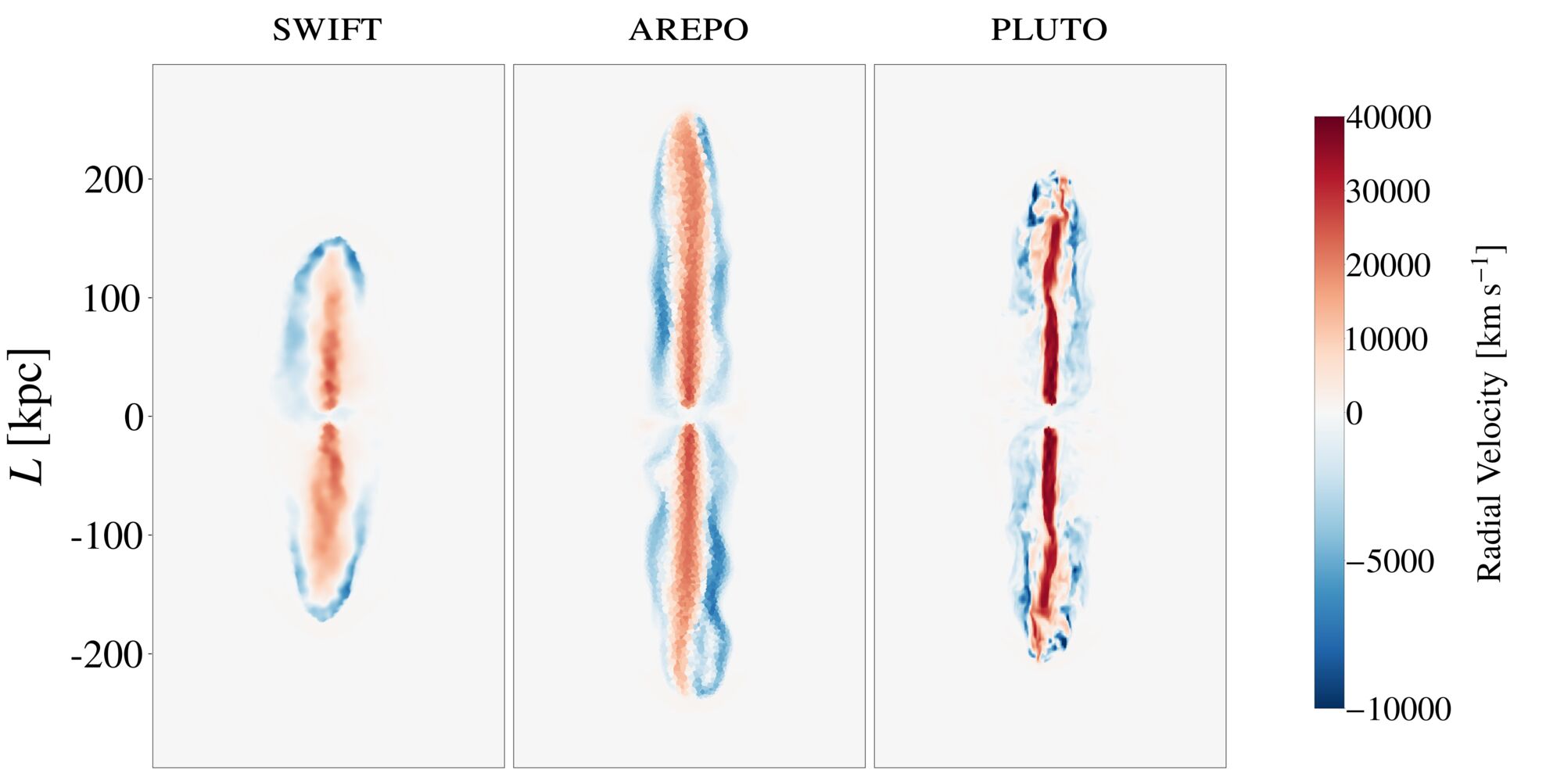}
     \caption{Radial velocity slices of jets simulated with the three codes in the uniform medium after $98 \, \mathrm{Myr}$. All jets display a central sheath of high-velocity material moving outwards along the jet axis, surrounded by backflows. {\sc swift} jets show the least coherent backflows, {\sc arepo} jets display stronger and more streamlined backflows, and {\sc pluto} jets present fast but irregular backflows.}
     \label{fig:3.Jets Radial Velocity}
\end{figure*}

 A key feature of our simulations is the presence of backflows. Backflows consist of jet material that, after being shocked at the jet head, flows backwards along the jet axis, toward the injection site. In the radial velocity maps in Fig.~\ref{fig:3.Jets Radial Velocity}, we can see that all jets exhibit high-velocity backflows around the central, outward-flowing sheath. {\sc swift} jets show the least coherent backflows, while {\sc arepo} jets display strong, streamlined backflows, and {\sc pluto} jets produce fast but irregular backflows. The observed backflows can also influence the development of KH instabilities, whose growth rate scales with the velocity shear.

\par In the case of {\sc pluto}, the central sheath of jet material reaches the highest velocities, up to the jet injection velocity ($v_\mathrm{j} = 40,000$ km/s). This is a direct consequence of our jet injection model; our injection scheme distributes kinetic energy and mass to a fixed number of gas elements (see Section~\ref{ssec:jet model}). In {\sc swift} and {\sc arepo}, these elements maintain roughly constant mass throughout the simulation, due to particle-splitting or mesh refinement, but in {\sc pluto}, cells can reach arbitrarily low densities, and hence masses, especially in the centre of the domain as the jet evacuates material. Consequently, injecting the same kinetic energy and mass into lower-mass cells leads to higher effective jet velocities in {\sc pluto}, compared to the other codes. A different, less code-dependent behaviour is observed when, instead of injecting the jet into a fixed number of neighbouring gas elements, we inject it into a fixed mass of gas elements. This results in more consistent effective jet velocities, with the {\sc arepo} and {\sc pluto} jets appearing more similar (see Section~\ref{ssec:same neighbour mass injection}). 

\par The higher-velocity flows in {\sc pluto}, in combination with the higher resolution in the lobes, may be causing the jets to disrupt before reaching the termination shock, as the jet head is more susceptible to KH instabilities \citep{Yates_2018}. These instabilities could develop between the jet and the shocked lobe material due to the high-velocity shears at the jet head. In real jets, however, magnetic fields may act to stabilise the fluid against KH instabilities \citep[e.g.][]{Hardee_2007}. 

\par
\begin{figure*}
    \centering
    \includegraphics[width=0.7\textheight]{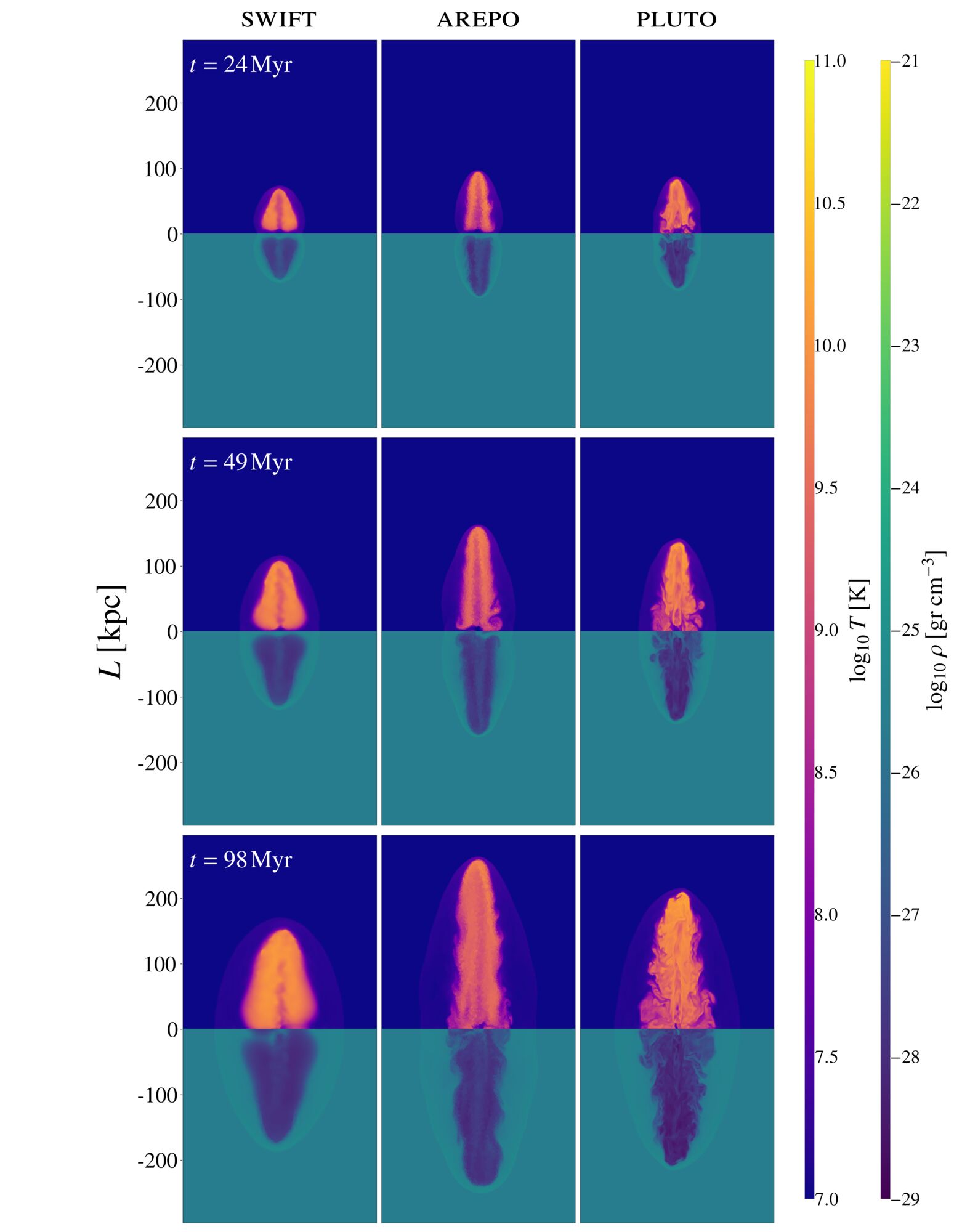}
    \caption{Temperature (top) - density (bottom) slices of the time evolution of the jets simulated with the three codes in the uniform medium. Different rows showcase the different jets at progressively later times. The {\sc swift} lobes retain their shape and evolve self-similarly from early on, while {\sc arepo} and {\sc pluto} lobes are becoming more elongated, for at least the first $50 \, \mathrm{Myr}$. The top and bottom lobes progressively fuse in all codes due to backflows and the growth of KH instabilities.}
    \label{fig:4.Jets Evolution}
\end{figure*}

 While the final state of the jets and their lobes provides useful information, examining their time evolution across the three codes can give us some insight into the dynamics driving their expansion. To this end, in Fig.~\ref{fig:4.Jets Evolution} we display temperature (top) and density (bottom) slices for the three jets in the uniform medium, for progressively later times in each row. {\sc swift} lobes appear to retain their shape from quite early in the simulation while progressively growing in size, indicating self-similar expansion. In contrast, the lobes simulated with {\sc arepo} and {\sc pluto} become noticeably elongated in the first two rows (for the first $\sim 50 \, \mathrm{Myr}$) but mostly retain their shape during their later evolution. We also note, to varying degrees, the progressive fusing of the top and bottom lobes at their base in all the codes. This could be attributed to the development of backflows but also to the growth of some modes of KH instabilities at the base of the {\sc arepo} and {\sc pluto} lobes. 
Finally, we observe the development of multiple recollimation shocks in the {\sc pluto} jet by $50 \, \mathrm{Myr}$. These shocks seem to disrupt by the end of the simulation, as the flow becomes more susceptible to instabilities.

\par \noindent\textit{Jet lobe evolution:} To quantify the differences between jets simulated with different codes, we calculate the lengths and widths of the jet lobes. 
\begin{itemize}
\item \textbf{Jet lobe} material is identified as elements (particles/cells) that have temperatures $T > 10^8 \, \mathrm{K}$ or velocity magnitudes $v > 0.5 \times 10^4 \, \mathrm{km/s}$. 
\item \textbf{Jet length} is defined as the average distance to the furthest $1\%$ of lobe resolution elements from the origin along the jet axis $(z)$. 
\item \textbf{Jet width} is defined as the average perpendicular distance of the $1\%$ of elements that lie furthest in the transverse direction $\left(\sqrt{x^2+y^2}\right)$. Note that this yields the maximum width of the lobe. 
\end{itemize}
Using the $1\%$ of the number of elements in each lobe, rather than the $1\%$ of mass or volume, is less physically motivated but conceptually simpler. In any case, we do not find any significant differences between the methods. The final values are the averages between the two lobes for each simulation. We checked that our estimates are in rough agreement with those obtained using a jet tracer in {\sc arepo} and {\sc pluto}, where lobe material is defined by tracer values $tr>0.01$. 

\par
\begin{figure*}
    \centering
    \includegraphics[width=\textwidth]{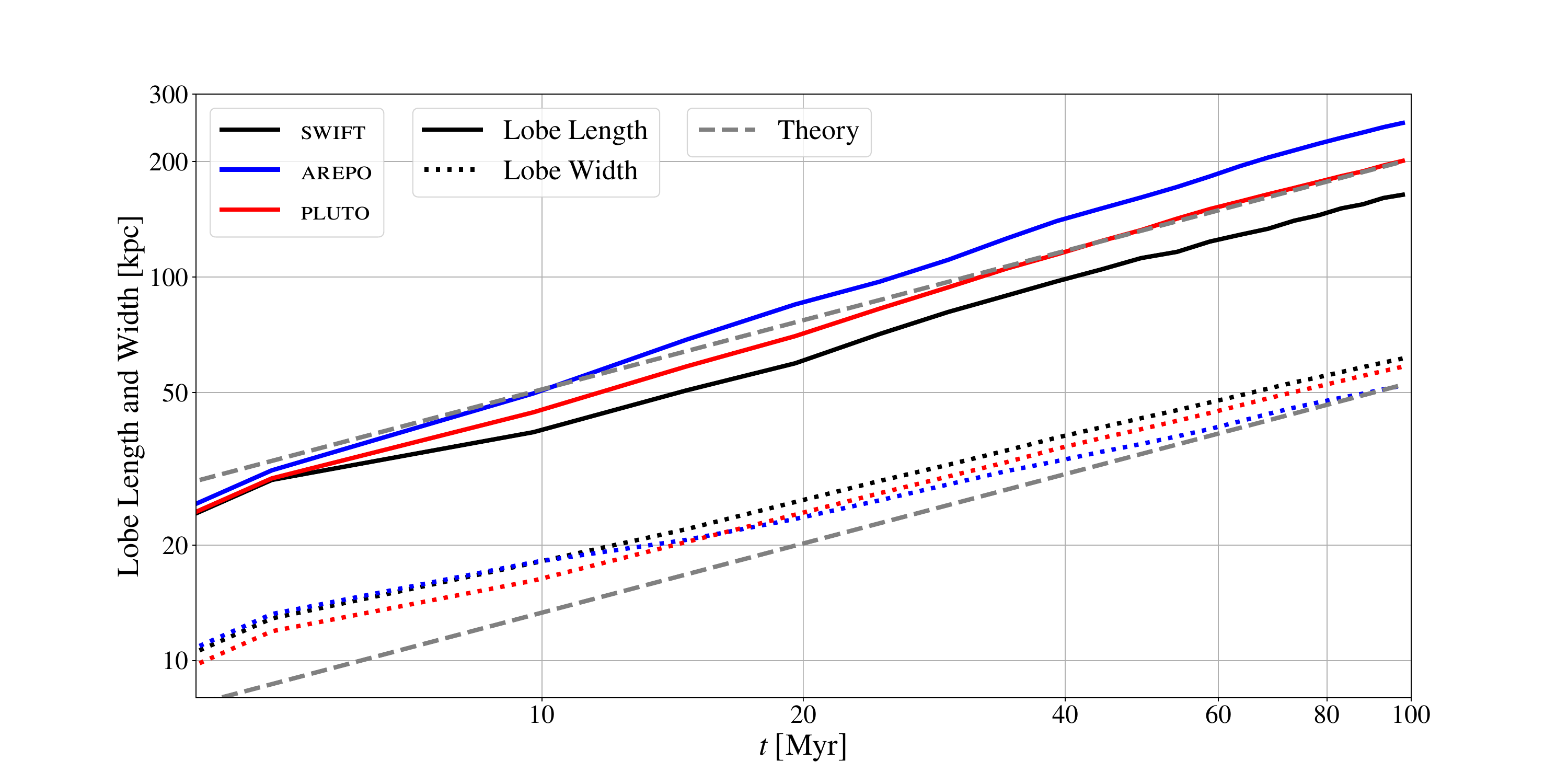}
     \caption{Plots of lobe length and width with respect to time for jets simulated with the three codes in the uniform medium. Black lines represent the lobes simulated with {\sc swift}, blue lines the {\sc arepo} lobes, and red lines the {\sc pluto} lobes. Solid lines represent estimates of length, and dotted lines represent estimates of width. The dashed grey lines represent the analytic, self-similar solution. All jets produce lobes that expand self-similarly by the end of the simulations. {\sc swift} under-predicts, {\sc arepo} over-predicts, and {\sc pluto} agrees with the normalised analytic solution for lobe length.}
    \label{fig:5.Length-Width Plot}
\end{figure*}

 In Fig.~\ref{fig:5.Length-Width Plot}, we plot the lobe lengths (solid lines) and widths (dotted lines) of the jets simulated with different codes, in different colours, and include the self-similar analytic solution, Eqs.~(\ref{eq:4})--(\ref{eq:6}).  Black lines represent the lobes simulated with {\sc swift}, blue lines the {\sc arepo} lobes, and red lines the {\sc pluto} lobes and the dashed grey line the analytic solution. All of the codes produce lobes that expand self-similarly by the end of the simulations. Specifically, for lobe length, {\sc swift} under-predicts, {\sc arepo} over-predicts, and {\sc pluto} agrees with the normalised analytic solution. At the same time, all codes over-predict the lobe width compared to the normalised solution for most of the duration of the jet episode.

\par Although agreement with the slope of the self-similar solution is a robust indicator of self-similar behaviour, we note that agreement with the normalised solution, in practice, depends on the particular combination of jet parameters \citep{Husko_2023}, especially jet injection velocity (see Section~\ref{ssec:jet injection velocity}). 

\par
\begin{figure}
    \centering
    \includegraphics[width=\columnwidth]{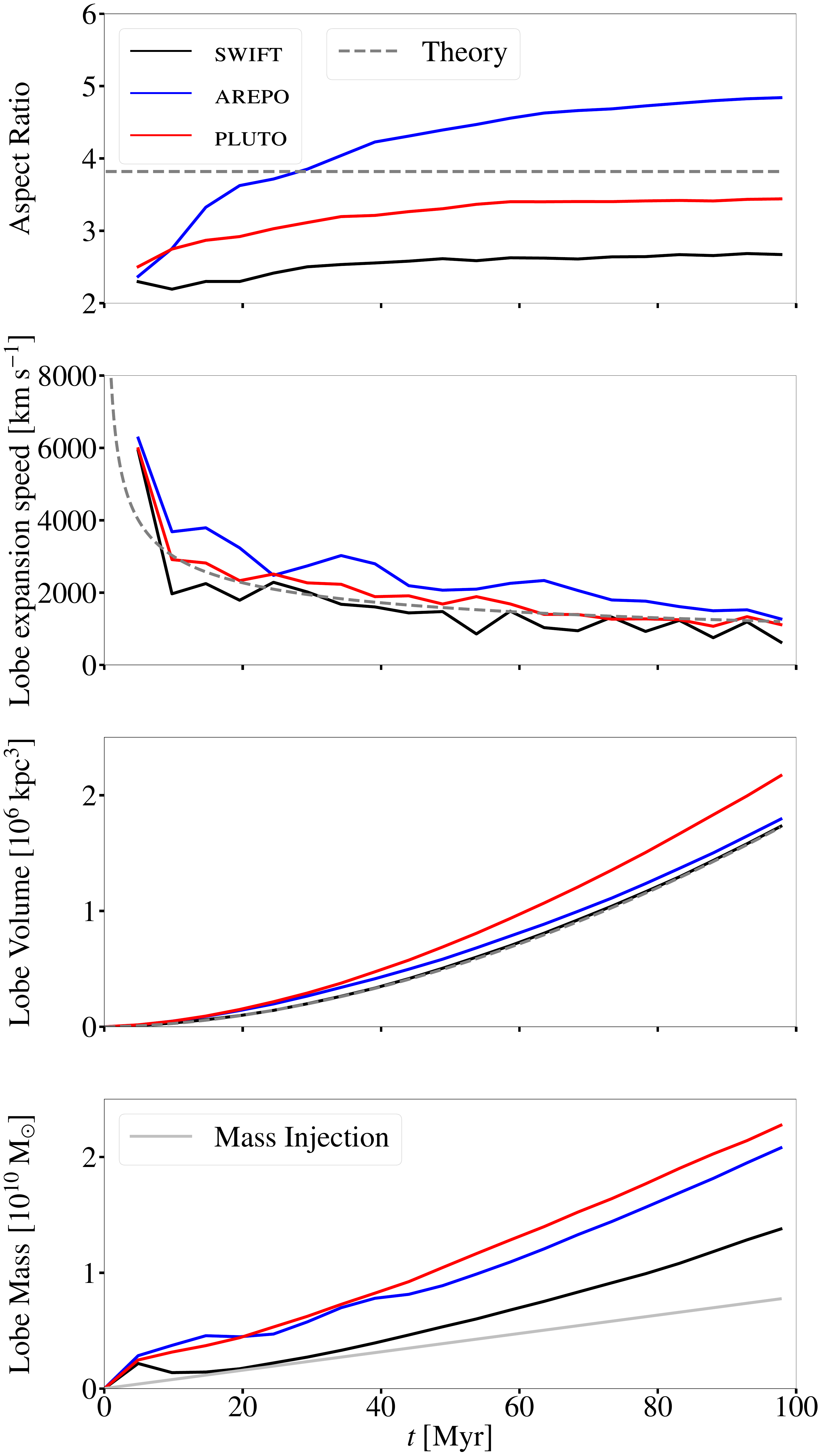}
     \caption{Lobe properties of jets produced by the three codes with respect to time. Solid coloured lines represent the three codes and dashed grey lines the analytic self-similar model. \textbf{First row:} Aspect ratio of jet lobes (lobe length divided by width) with respect to time. {\sc swift} lobes show minimal aspect ratio evolution with time, while {\sc arepo} and {\sc pluto} lobes are increasing their aspect ratio for the first $\sim 40-60 \, \mathrm{Myr}$. \textbf{Second row:} Speeds of lobes with respect to time. All lobes are decelerating to around $1500 \, \mathrm{km/s}$ by the end of the simulation. \textbf{Third row:} Lobe volume as a function of time. {\sc swift} lobes match the analytic prediction closely. {\sc arepo} lobes reach slightly larger volumes, while {\sc pluto} lobes consistently attain the largest volumes. \textbf{Fourth row:} Lobe mass as a function of time. The solid silver line represents the mass injected by the jet. {\sc arepo} and {\sc pluto} lobes achieve comparable masses, whereas {\sc swift} lobes are slightly less massive, indicating less efficient mixing and entrainment.}
    \label{fig:6.Properties Plot}
\end{figure}

 To more clearly demonstrate the self-similar evolution of the jets, we estimate how their lobe aspect ratio (the lobe length divided by the width) changes as the simulation progresses. In the first row of Fig.~\ref{fig:6.Properties Plot}, we plot the lobe aspect ratio with respect to time for the jets simulated with the three codes (solid coloured lines) and the analytic solution (dashed grey line). {\sc swift} lobes appear to expand self-similarly from very early on since their aspect ratio does not change significantly throughout the simulation, in line with the analytic solution being just a constant. In contrast, {\sc arepo} and {\sc pluto} jets showcase a more obvious, transitional, breakout phase \citep{Komissarov_1998, Hardcastle_2013, Husko_2023} where the jets inflate long and thin lobes that increase their aspect ratios---this breakout phase is not captured by the self-similar solution. Eventually, the lobes start exhibiting self-similar behaviour after around $40-60 \, \mathrm{Myr}$. 

\par In the second row of Fig.~\ref{fig:6.Properties Plot}, we plot the lobe expansion speeds of the different jets and include the analytic prediction. This speed is calculated by evaluating the change in lobe length between consecutive snapshots and dividing by the time difference. Naturally, this bulk lobe speed will always be lower than the jet velocity as the jet material shocks against the ambient medium at the jet head. The three codes show good agreement with each other and the normalised analytic result, with {\sc swift} lobes being marginally slower, {\sc pluto} lobes somewhat faster, and {\sc arepo} lobes the fastest. All of the lobes decelerate over the duration of the simulation and end up having a similar speed of $\sim 1500 \, \mathrm{km/s}$ by the end.

\par The last two rows of Fig.~\ref{fig:6.Properties Plot} show the time evolution of the lobe volume (top) and lobe mass (bottom) for jets simulated with the three codes and the self-similar prediction for lobe volume. All simulation codes show monotonic growth in both volume and mass, consistent with ongoing jet activity. {\sc pluto} produces lobes with the largest volumes, while {\sc swift} and {\sc arepo} yield smaller lobe volumes that are in better agreement with the analytic solution. For the last simulation snapshot, {\sc pluto} over-predicts lobe volume by $20\%$ compared to the analytic solution, while {\sc swift} and {\sc arepo} agree to within $1\%$ and $4\%$ respectively. In terms of lobe mass, we also show the mass injected by the jet (solid silver line), which represents the minimum possible lobe mass across all codes. {\sc swift} consistently predicts lower values than the other codes, indicating less efficient mixing and entrainment with the ambient medium.

\par
\begin{figure}
    \centering
    \includegraphics[width=\columnwidth]{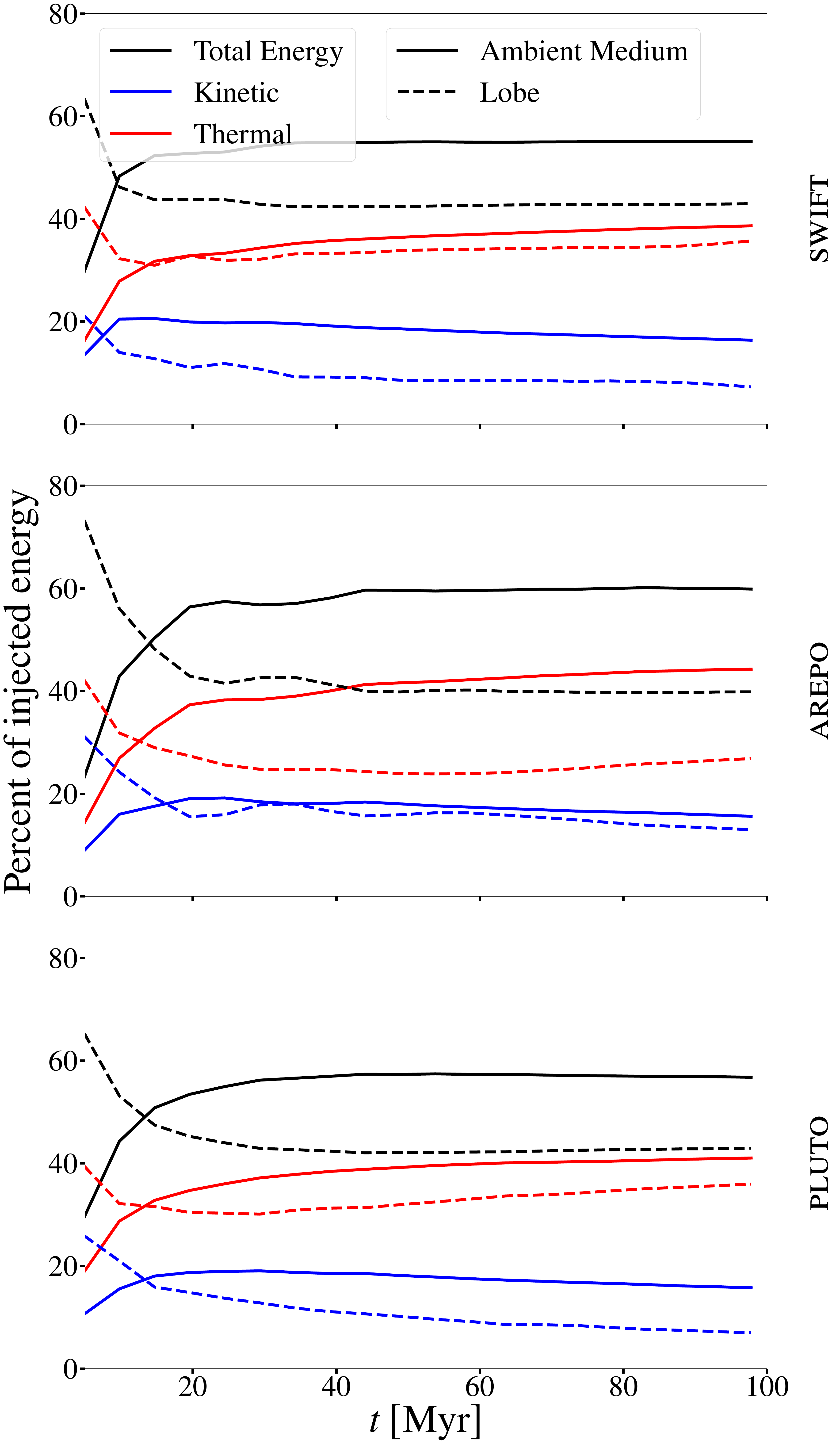}
     \caption{Evolution of the different energy components (thermal energy in red, kinetic energy in blue and total energy, as the sum of the two, in black) in the lobes (dashed line) or ambient medium (solid line) of the jets simulated with the three codes. All jets transfer around $60\%$ of their injected energy to the ambient medium, and all lobes contain more thermal energy than kinetic energy.}
    \label{fig:7.Energies Plot}
\end{figure}

 Besides the dimensions and appearance of the lobes, a critical quantity to consider in the context of AGN feedback and its impact on galaxy, group and cluster evolution is the lobe energetics. In Fig.~\ref{fig:7.Energies Plot}, we plot the different energy components inside the lobes (dashed line) and in the ambient medium (solid line) with respect to time. We denote thermal energy in red, kinetic energy in blue, and total energy, which is the sum of the two, in black. We subtract any initial thermal energy to ensure that we capture only the injected energy. Overall, we find good agreement between the three codes; all of the jets transfer around $60 \%$ of the injected energy to their ambient medium, with most of it being in thermal form. The energy in the lobes is predominantly thermal in all the codes, but with slightly different ratios; {\sc swift} and {\sc pluto} lobes contain roughly four times more thermal energy than kinetic energy, whereas {\sc arepo} lobes have about twice as much. While almost all of the energy injection in our jets is in kinetic form, the high injection velocities efficiently thermalise the gas, inflating lobes that quickly become more thermal-energy-dominated (as opposed to kinetic-energy-dominated) and expand due to both their thermal pressure laterally and to their thrust in the jet direction. We expect these trends to still hold qualitatively if cooling is included, although there could be quantitative differences.

\subsection{Sensitivity to numerical resolution}
\label{ssec:resolution}

When comparing different hydrodynamical codes, it is important to consider the inherent differences in how they discretise the fluid, as this impacts the effective mass and spatial resolutions achievable within a simulation.
\begin{itemize}
\item \textbf{Mass resolution:} {\sc swift} and {\sc arepo} maintain roughly \emph{constant mass resolution} (to within a factor of two) using particle-splitting and refinement, respectively. In contrast, {\sc pluto} cells, which maintain a fixed volume, can achieve arbitrarily low (high) masses in low- (high-)density regions of a simulation, and hence present \emph{adaptive mass resolution}. This is especially relevant for the inner parts of our simulation domain ($r < 10 \, \mathrm{kpc}$) where, due to the action of the jet, {\sc pluto} cells achieve very low densities and hence masses. This, in turn, influences the injection and evolution of the jet (see Section~\ref{ssec:same neighbour mass injection}). 

\item \textbf{Spatial resolution:} {\sc swift} and {\sc arepo} feature \emph{adaptive spatial resolution}. As an SPH code, {\sc swift} interpolates local fluid quantities via kernel smoothing over neighbouring particles, with spatial resolution set by the smoothing length; overlapping regions require independent smoothing lengths for a more conservative estimate \citep{Agertz_2007}. {\sc arepo}, instead, employs a moving-mesh whose cells adapt to the fluid flow, with a mesh-regularisation scheme that prevents excessive distortions and keeps them approximately round \citep{Arepo}; spatial resolution is then set by an effective cell radius $R_{\mathrm{cell}} = \sqrt[3]{3V_{\mathrm{cell}}/4\pi}$. In both codes, spatial resolution scales with local gas density: high-density regions are resolved more finely, while low-density regions are coarsely resolved. Consequently, in jet simulations, lobes are modelled at lower spatial resolution than the shocked ambient medium and bow shock (see bottom panel of Fig.~\ref{fig:2.Jets Overview}). In contrast, in {\sc pluto}, the use of a regular Cartesian grid with fixed cell size ensures \emph{constant spatial resolution} in both lobes and ambient medium.
\end{itemize}
These differences imply that different hydrodynamical methods will be better suited to capturing different aspects of jet evolution, such as shock structure, lobe expansion, or instability growth, and should be considered carefully when interpreting simulation results.

\par To access differences in effective spatial resolution, we focus on the simulated jet lobes, whose low densities make them challenging to resolve with Lagrangian and quasi-Lagrangian codes. In the final snapshots of our fiducial simulations, within the lobes,
\begin{itemize}
\item {\sc swift} has smoothing lengths ranging from $1 \, \mathrm{kpc}$ to $9 \, \mathrm{kpc}$ (with mean $3 \, \mathrm{kpc})$
\item {\sc arepo} has cell radii ranging from $0.5 \, \mathrm{kpc}$ to $3.5 \, \mathrm{kpc}$ (with mean $1.5 \, \mathrm{kpc}$)
\item {\sc pluto} has cells with half-side $0.4 \, \mathrm{kpc}$. 
\end{itemize}
Thus, {\sc pluto} achieves the highest spatial resolution in the lobes, followed by {\sc arepo}, and finally {\sc swift}. Even in the unperturbed ambient medium, however:
\begin{itemize}
\item {\sc swift} has smoothing lengths of $1 \, \mathrm{kpc}$
\item {\sc arepo} has cell radii of $0.5 \, \mathrm{kpc}$ 
\item {\sc pluto} has cells with half-side $0.4 \, \mathrm{kpc}$
\end{itemize}
This means that {\sc swift} starts with lower spatial resolution in the initial conditions of our simulations, due to SPH smoothing. A setup of an equal number of resolution elements across codes, therefore inherently disadvantages SPH methods, and this must be considered when assessing their applicability. Nevertheless, such comparisons remain useful in the context of cosmological simulations, where the number of elements serves as a convenient proxy for the achieved mass and spatial resolution.

\par The specific resolution element count in the lobes varies between the codes: {\sc swift} lobes are resolved with $66,550$ particles, {\sc arepo} lobes with $110,748$ cells and {\sc pluto} lobes with $4,103,820$ cells. As shown in Fig.~\ref{fig:6.Properties Plot}, the total lobe mass and volume are very similar between the {\sc arepo} and {\sc pluto} runs, while the difference in the number of lobe resolution elements across the codes reflects the interplay between mass and spatial resolution. In contrast, {\sc swift} lobes contain less mass, indicating reduced mixing and entrainment of ambient material---consistent with the apparent absence of KH instabilities along the {\sc swift} lobes. By construction, the lower levels of entrainment, and hence mass, in {\sc swift} lobes naturally result in fewer resolution elements within the lobes.

\par To illustrate the impact of resolution for each code, we present three sets of runs in the uniform medium, with different initial numbers of resolution elements (and initial gas element masses): $N = 10^7$ ($M_\mathrm{gas} = 2\times 10^6 \, \mathrm{M}_\odot $), $N = 5\times10^7$ ($M_\mathrm{gas} = 4 \times 10^5 \, \mathrm{M}_\odot $), and $N = 10^8$ ($M_\mathrm{gas} = 2\times 10^5 \, \mathrm{M}_\odot $) resolution elements. We fix all other parameters, as well as the number of injection events, to be the same as in our fiducial runs. Each code begins with uniform spatial resolution and identical initial gas element masses, ensuring a consistent starting point. However, as the simulations evolve, differences in fluid discretisation lead to the code-specific behaviour we highlighted above. Specifically, {\sc swift} and {\sc arepo} progress from uniform spatial resolution to adaptive spatial resolution based on the local density. For {\sc pluto}, the gas mass resolution can change significantly throughout the simulations depending on the local gas density.

\par Fig.~\ref{fig:8.Jets Resolution} shows slices of temperature and density for {\sc swift}, {\sc arepo} and {\sc pluto} jets in the uniform medium, at $t = 98 \, \mathrm{Myr}$, with each row representing a different resolution. The simulations with $N = 10^8$, in the bottom row, represent our fiducial runs. Comparing the lobes at this final snapshot of the simulations:

\begin{figure*}
    \centering
    \includegraphics[width=0.7\textheight]{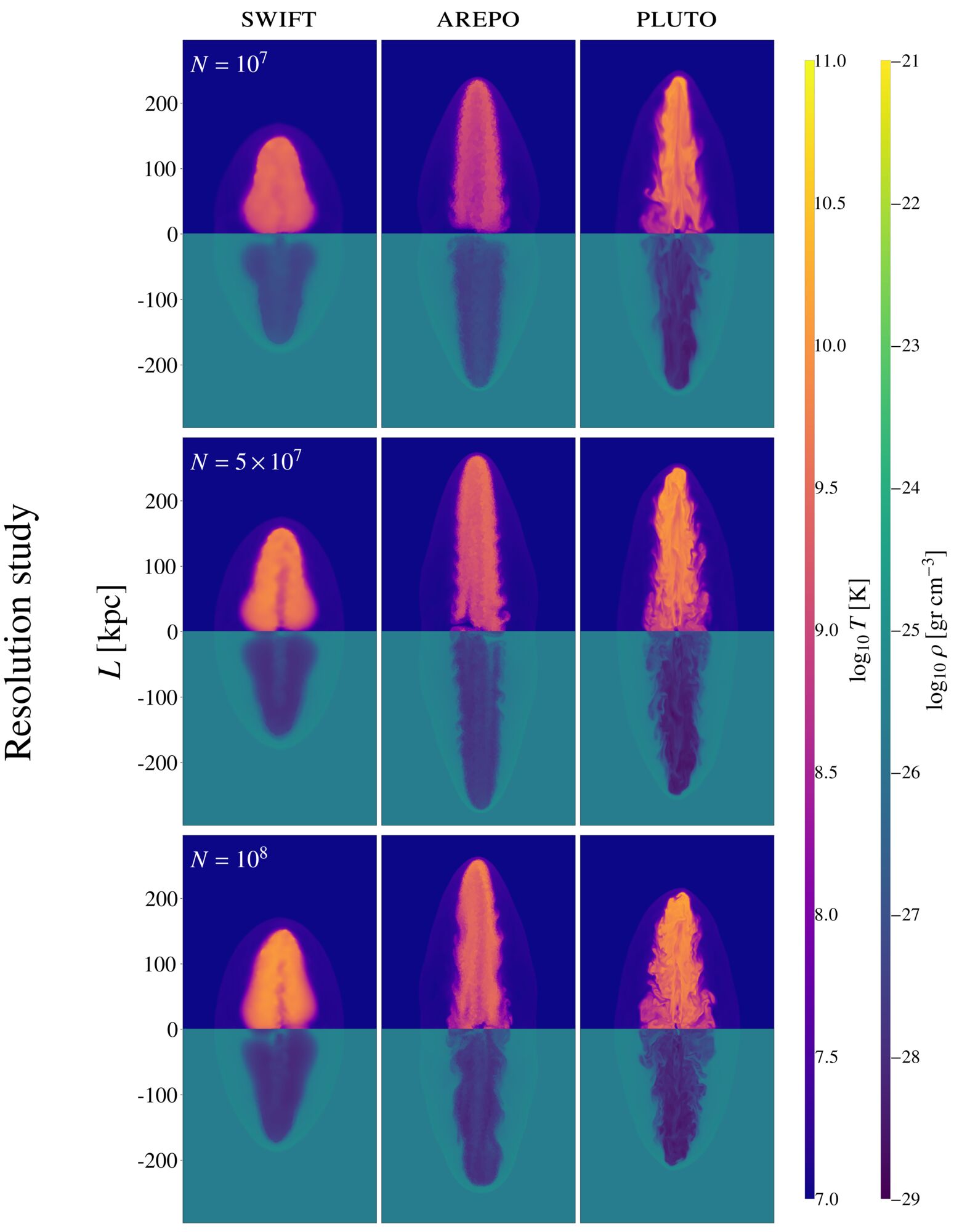}
    \caption{Temperature–density slices of jets simulated with the three codes in the uniform medium after $98 \, \mathrm{Myr}$, at three different resolutions ($N = 10^7$, $N = 5 \times 10^7$, $N = 10^8$), shown in separate rows. {\sc swift} lobes present only minor changes at different resolutions, while {\sc arepo} and {\sc pluto} lobes significantly increase and then decrease their length with increasing resolution.}
    \label{fig:8.Jets Resolution}
\end{figure*}

\par {\sc swift} jets present only minor changes across different resolutions. Their lobes become slightly longer ($+4\%$) and slightly wider ($+5\%$) from lowest to highest resolution. This mild dependence on resolution agrees with \citet{Husko_2023}, where {\sc swift} lobes showed noticeable shortening only at very low or high resolutions not probed here. 

\par By contrast, {\sc arepo} jets show a stronger resolution dependence. Their lobes get notably longer ($+14\%$) from $N=10^7$ to $N=5\times10^7$, before becoming shorter again ($-8\%$) at $N=10^8$. Their widths steadily increase ($+15\%$) from lowest to highest resolution. Increasing lobe length with resolution in {\sc arepo} jets has been noted in \citet{Weinberger_2017, Bourne_2017, Weinberger_2023}, while an eventual turnaround was observed in \citet{Weinberger_2023}. 

\par Similarly, the lobes simulated with {\sc pluto} become modestly longer ($+4\%$) from $N=10^7$ to $N=5\times10^7$, and then get significantly shorter ($-18\%$) at $N=10^8$. Their width slightly increases ($+5\%$) at the highest resolution. This reduction of lobe length with resolution in {\sc pluto} jets has been observed in \citet{Yates_2018}. 

\par
\begin{figure}
    \centering
    \includegraphics[width=\columnwidth]{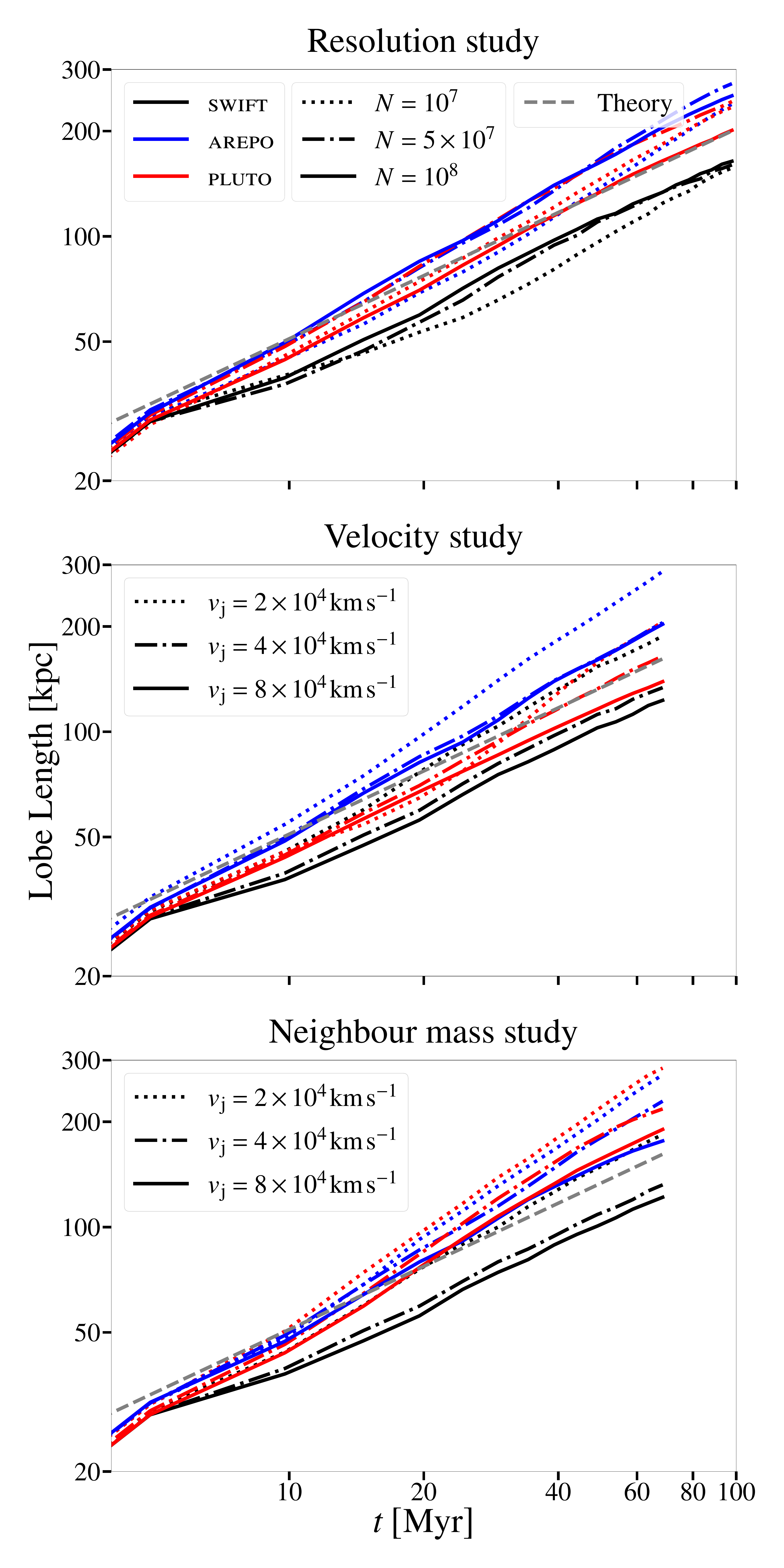}
    \caption{Lobe length as a function of time for all our runs. Coloured lines represent the three codes, line style shows the parameter variation, and the dashed grey line is the analytic prediction. \textbf{Top row:} Runs at three different resolutions. {\sc swift} lobes show only minor variations, while {\sc arepo} and {\sc pluto} lobes become longer at intermediate resolution and shorter again at the highest resolution. \textbf{Middle row:} Runs with three different injection velocities. Fast-slow jets inflate shorter lobes that expand in closer agreement with the slope of the analytic solution, whereas slow-heavy jets produce longer lobes that tend to over-predict the analytic slope. \textbf{Bottom row:} Runs with three different injection velocities using the fixed-neighbour-mass scheme. Only {\sc pluto} exhibits a systemic change relative to the fiducial scheme, now behaving very similarly to {\sc arepo}.}
    \label{fig:9.Lobe Length General Plot}
\end{figure}

 The variation of lobe length with respect to time for all our simulations (as summarized in Table~\ref{tab:Jets Table}) can be seen in Fig.~\ref{fig:9.Lobe Length General Plot}. In the top row, we plot the lobe lengths of the jets at different resolutions. The coloured lines represent the three codes, with line style showing the specific resolution and the dashed grey line the analytic solution, Eqs.~(\ref{eq:4}) and (\ref{eq:5}). Across resolutions, {\sc swift} lobes are the shortest, consistently under-predicting the normalised analytic solution. {\sc arepo} and {\sc pluto} lobes show better agreement with each other at lower resolutions while over-predicting the normalised solution. At the highest resolution, however, {\sc arepo} lobes remain quite long while {\sc pluto} lobes shorten to match the normalised result.

\par It is worth noting that our use of a fixed-neighbour-number injection scheme leads to the following effect: higher resolutions have lower masses for individual gas elements and hence lower neighbour masses for injections of the same kinetic energy, which results in higher velocities. These higher velocities could then contribute to the decrease in lobe length observed in {\sc arepo} and {\sc pluto} at the highest resolution ($N=10^8$) as explained in Section~\ref{ssec:jet injection velocity}---although the effect seems to be subdominant at lower resolutions.

\par Besides the dimensions of the jets and their lobes, we can see how their morphologies change with increased resolution. One obvious effect is noted in the {\sc arepo} and {\sc pluto} lobes, which more readily develop complex flows and KH instabilities at higher resolutions.\footnote{Even though our simulated {\sc swift} lobes are mostly smooth, KH and RT instabilities are evident in higher-resolution SPH jet simulations \citep{Husko_2023, Husko_2023b}.} We speculate that this could be another factor contributing to the decrease in lobe length observed at the highest resolution ($N=10^8$) in these codes \citep{Yates_2018, Weinberger_2023}. An additional important feature, mainly observed in the {\sc swift} and {\sc arepo} lobes, is the development of stronger backflows with increased resolution. These backflows contribute to the fusing of the top and bottom lobes at their base, and could lead to the widening of lobes \citep{Husko_2023}.

\par Interestingly, we find very good consistency between resolutions regarding lobe masses and energetics (not shown here). The only noteworthy trend is a slight increase in the lobe thermal energy, accompanied by a decrease in the lobe kinetic energy with increasing resolution. This is also evident in the slightly higher lobe temperatures in the slices. A likely explanation is that the faster flows at higher resolutions can more effectively thermalise through shocks and adiabatic expansion work, raising the thermal energy at the expense of kinetic energy.

\par Encouragingly, for code-to-code consistency, the jets in the different codes appear more similar to each other in the lower-resolution runs. These jets have initial mass resolutions close to the resolutions used in many cosmological hydrodynamical simulations (although we achieve higher spatial resolutions and simulate more injection events). This suggests that AGN feedback, as implemented in cosmological simulations, may not be very different across different codes, or at least not until higher resolution can be achieved.

\subsection{Effect of injection velocity}
\label{ssec:jet injection velocity}

\par Jet velocity is a key parameter governing the propagation of jets. Observationally, FR-II lobes are powered by relativistic jets, while there is evidence that FR-I jets are slowed down on $\sim \mathrm{kpc}$ scales \citep{Laing_2013}, indicating a direct connection between jet speed and lobe morphology. In simulations, the behaviour of AGN jets can change radically with variation of the jet injection velocity \citep{English_2016, Weinberger_2017, Husko_2023, Huško_2023}, suggesting that differences between codes can be highly dependent on the parameters used to model the AGN jets. 

\par To demonstrate this, we compare the three codes in the uniform medium and at our highest resolution, with three different jet injection/virtual-particle velocities $v_\mathrm{j}$. We keep all other parameters fixed, barring a change to the virtual-particle mass, so as to maintain the same number of injection events. Keeping the jet power, and thus the injected energy per time the same means that higher velocity injections lead to faster, lighter jets, while lower velocity injections lead to slower, heavier jets (see Eq.~\ref{energy eq}). We evolve the jets for $68 \, \mathrm{Myr}$, ensuring that their lobes stay within the simulation boxes, and include runs with $v_\mathrm{j}=2\times10^4 \, \mathrm{km/s}$ ($m_\mathrm{j} = 1.6\times 10^5 \, \mathrm{M}_\odot$), $v_\mathrm{j}=4\times10^4 \, \mathrm{km/s}$ ($m_\mathrm{j} = 4\times 10^4 \, \mathrm{M}_\odot$), and $v_\mathrm{j}=8\times10^4 \, \mathrm{km/s}$ ($m_\mathrm{j} = 1\times 10^4 \, \mathrm{M}_\odot$). 

\par
\begin{figure*}
    \centering
    \includegraphics[width=0.7\textheight]{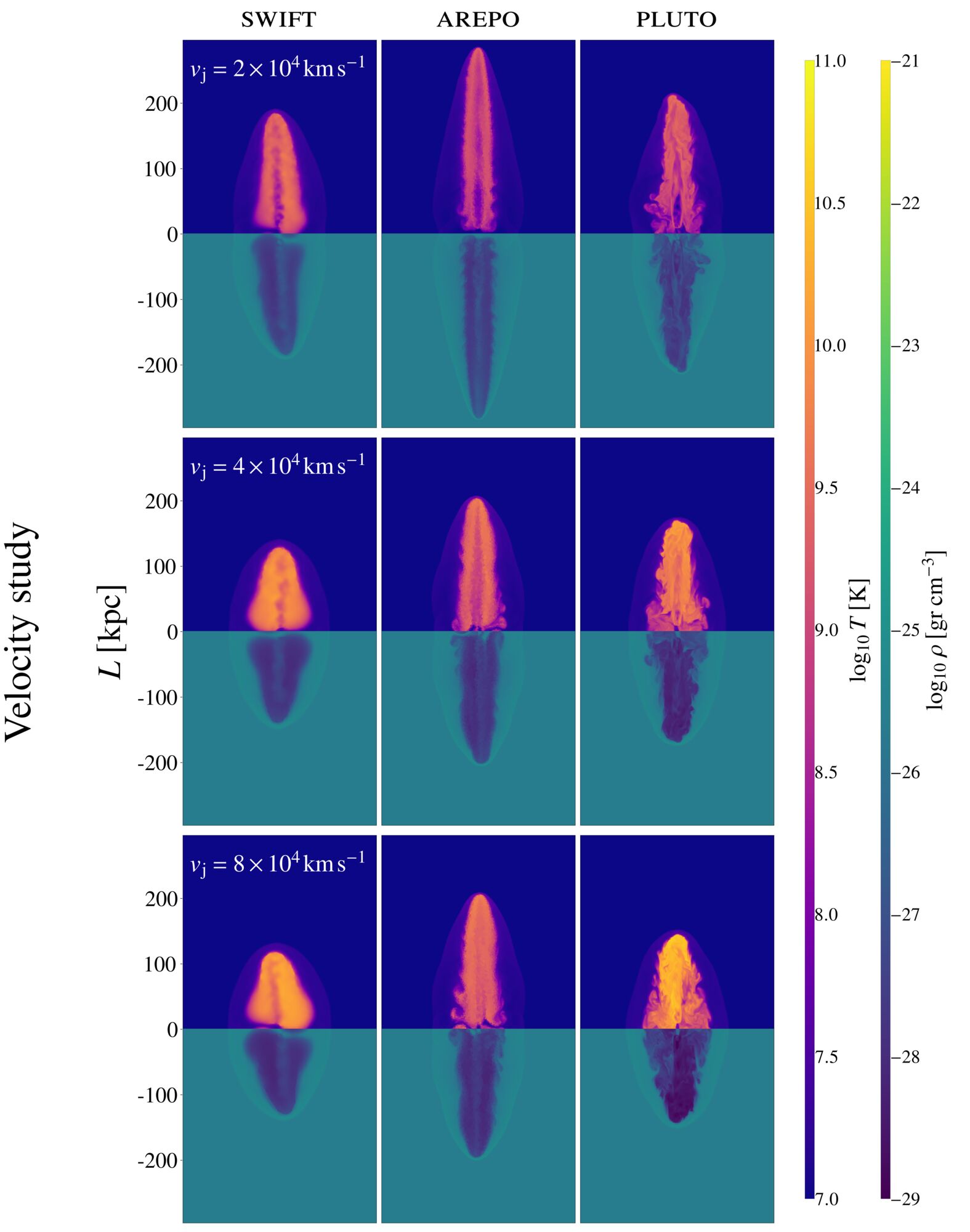}
    \caption{Temperature-density slices of jets simulated with the three codes in the uniform medium after $68 \, \mathrm{Myr}$, with three different jet injection velocities ($v_\mathrm{j}=2\times10^4 \, \mathrm{km/s}$, $v_\mathrm{j}=4\times10^4 \, \mathrm{km/s}$, $v_\mathrm{j}=8\times10^4 \, \mathrm{km/s}$), shown in separate rows. Increasing jet injection velocity results in shorter, wider and hotter lobes with more obvious backflows.}
    \label{fig:10.Jets Injection Velocity}
\end{figure*}

 Fig.~\ref{fig:10.Jets Injection Velocity} shows temperature-density slices of jets in the uniform medium at $t = 68 \, \mathrm{Myr}$, produced by the three codes, with different injection velocities shown in separate rows. The simulations with $v_\mathrm{j}=4\times10^4\mathrm{km/s}$, in the middle row, represent our fiducial runs. Fast and light jets produce lobes that are shorter, wider, and hotter, with more obvious backflows. This occurs because higher injection velocities lead to a larger fraction of the injected energy being thermalised through stronger shocking and adiabatic expansion of the lobes. Simultaneously, the higher pressures at the jet head redirect the shocked jet material in a strong backflow. In contrast, slow and heavy jets have more inertia and higher momentum flux, and are thus able to drill through the ambient medium more effectively, inflating longer and thinner lobes.

\par These trends are evident in the middle row of Fig.~\ref{fig:9.Lobe Length General Plot}, where we compare the lobe evolution with respect to time for simulations with the three injection velocities and the analytic self-similar solution. Slow-heavy jets produce lobes that deviate from the analytic self-similar solution and mostly overshoot the normalised prediction for lobe length. On the other hand, fast-light jets inflate lobes that expand self-similarly for longer and even under-predict the normalised solution. We also note a slight increase in lobe volume with increasing velocity, consistent with adiabatic expansion combined with strong backflows. In terms of energetics, the fraction of kinetic energy in the lobes decreases while the fraction of thermal energy increases with injection velocity, reflecting the greater thermalisation of the injected energy and the corresponding rise in lobe temperature. 

\par Comparing the three codes, we find that some trends persist regardless of jet injection velocity; for example, {\sc arepo} consistently produces the longest lobes and {\sc swift} the shortest. Other trends change with jet injection velocity, as some lobe properties display code-specific dependence; for instance, the lobe mass in {\sc arepo} slightly increases while in {\sc pluto} it slightly decreases with higher jet velocities. As a result, at the fastest injection velocity ($v_\mathrm{j}=8\times10^4 \, \mathrm{km/s}$), {\sc arepo} achieves the highest lobe mass, in contrast to our fiducial parameter runs ($v_\mathrm{j}=4\times10^4 \, \mathrm{km/s}$). This suggests that code differences can be dependent on the specific choices and parameters in a jet feedback scheme. 

\subsection{Neighbour mass injection}
\label{ssec:same neighbour mass injection}

\par The results of our comparison study depend critically on the implementation of the subgrid jet model---specifically, on how the injected mass, momentum, and energy couple to the computational domain. In our fiducial model, this occurs via a virtual particle initiating a neighbour-finding loop, through which the nearest $10$ gas elements receive feedback (see Section~\ref{ssec:jet model}). In this setup, the final state of the gas elements depends on their initial mass and, consequently, on the mass resolution of each code. {\sc swift} and {\sc arepo} use particle-splitting and mass refinement, respectively, so as to maintain an approximately constant mass resolution. In contrast, {\sc pluto} cells can reach arbitrarily low densities, especially in regions directly affected by jet activity. As a result, when injecting over a fixed number of neighbours, the total neighbour mass remains roughly consistent in {\sc swift} and {\sc arepo}, but becomes significantly lower in {\sc pluto} due to the lower density in the injection region. For a fixed injected energy, this leads to variations in the effective jet velocity and influences the subsequent propagation of the jet. 

\par To illustrate this, we include here a controlled comparison of the three codes in the uniform medium and at our highest resolution, where the injection happens for a fixed neighbour mass: $M_\mathrm{ngb} = 20 \times 10^5 \, \mathrm{M}_\odot$). We perform tests with three different jet injection velocities and evolve the jets for $68 \, \mathrm{Myr}$, following the same strategy as in Section~\ref{ssec:jet injection velocity}.
\par
\begin{figure*}
    \centering
    \includegraphics[width=0.7\textheight]{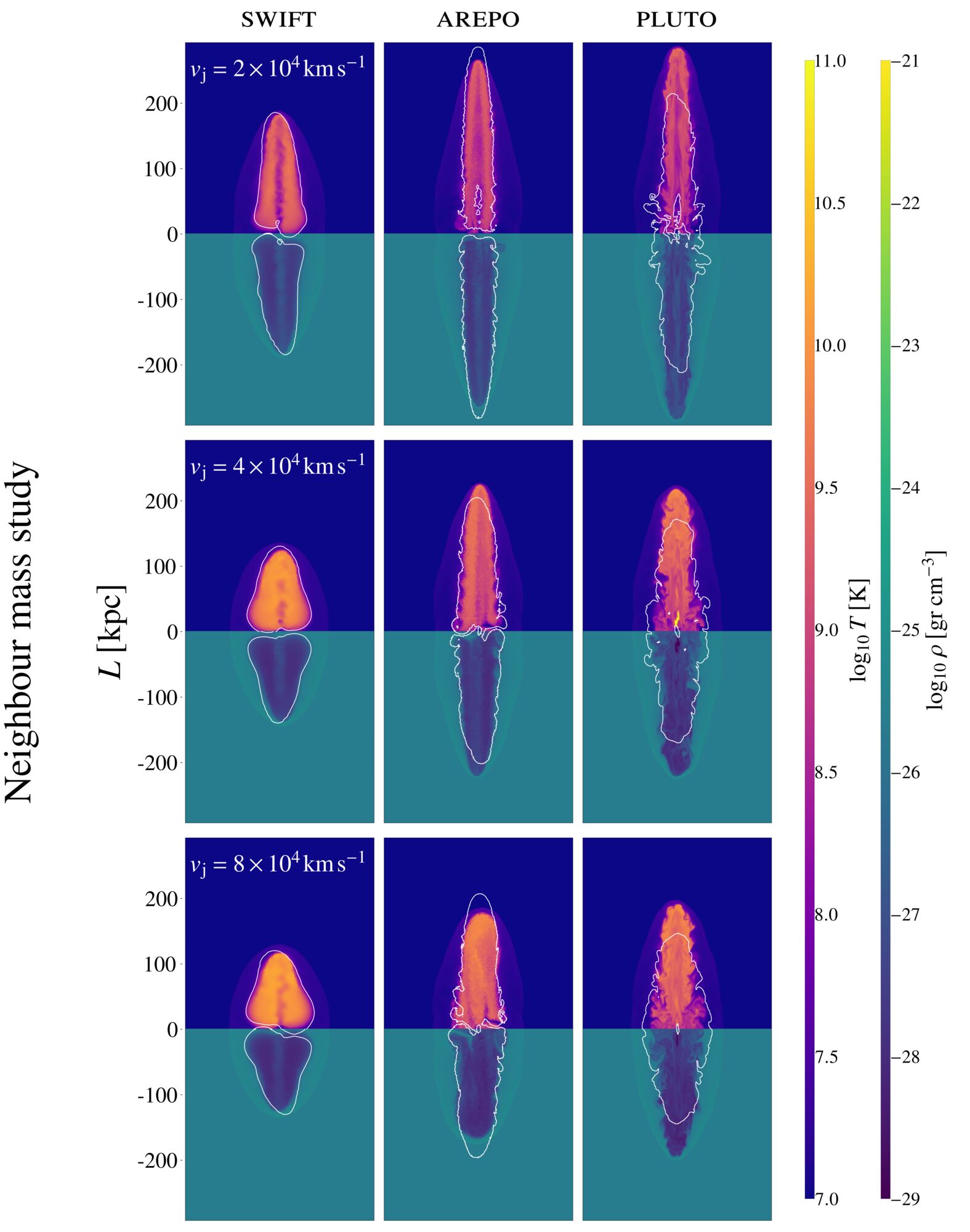}
    \caption{Temperature-density slices of jets simulated with the three codes in the uniform medium after $68 \, \mathrm{Myr}$, with injection into the same neighbour mass. We test three different jet velocities, shown in separate rows, and include the lobe contours of the fiducial fixed-number-of-neighbours scheme. {\sc swift} lobes are almost identical between the two schemes. {\sc arepo} lobes present some deviations but no systemic differences. {\sc pluto} lobes are consistently longer and cooler.}
    \label{fig:11.Jets Mass-Velocity}
\end{figure*}

 Fig.~\ref{fig:11.Jets Mass-Velocity} shows temperature-density slices of jets in the uniform medium at $t = 68 \, \mathrm{Myr}$, produced by the three codes, with different injection velocities shown in separate rows. For reference, we overlay the contours of the lobes in the case of the fiducial fixed-neighbour-number scheme (Fig.~\ref{fig:10.Jets Injection Velocity}). 

\par In {\sc swift}, the lobe geometry (and thermodynamic properties) are very similar in the two different injection schemes. This is primarily due to the fact that SPH particles can only strictly increase in mass, up to a factor of two, before being split. As a result, {\sc swift} is effective at preserving the mass of the resolution elements in the injection region, leading to minimal variation between the two injection schemes.

\par In the case of {\sc arepo}, we do see some slight deviations from the fiducial model, but we cannot find any systemic differences in lobe geometry, energetics or the development of instabilities and backflows. Thanks to its mass refinement algorithm, {\sc arepo} constrains most resolution elements to within a factor of two of their original mass. However, since mass can flow into or out of cells, and refinement is restrained by an additional geometric regularity criterion (making sure that very distorted cells do not get refined), some variation in cell masses is expected. As a caveat, we had to disable this extra geometric criterion for the lowest velocity ($v_\mathrm{j}=2\times10^4 \, \mathrm{km/s}$) (and highest mass loading) runs with this injection module to allow the code to converge to the target neighbour mass. The observed discrepancies between the two injection schemes are also dependent on the stochasticity of individual runs and sensitive to the generation of random angles in the jet launching module.

\par In contrast, {\sc pluto} shows a consistent and systematic difference; the lobes are longer and cooler than in the fiducial injection scheme. This results from injecting into a (much) higher neighbour mass, which, for a fixed kinetic energy injection, leads to lower jet velocities. Reduced velocities result in weaker shocks, which deposit less thermal energy into the lobes, making them cooler. Moreover, lobes retain more kinetic energy and so preferentially expand along the jet axis rather than laterally. 

\par Interestingly, in the fixed-neighbour-mass injection scheme, {\sc arepo} and {\sc pluto} behave more similarly. This is evident in the bottom row of Fig.~\ref{fig:9.Lobe Length General Plot}, where the evolution of lobe length with time shows the two codes exhibiting convergent behaviour across all injection velocities. The contrast is particularly clear when compared to the middle row, which shows results from the fixed-neighbour-number injection scheme, where {\sc arepo} produces consistently longer lobes than {\sc pluto}. 

\par This test highlights how differences in mass resolution between numerical methods can influence jet evolution, especially when combined with injection schemes where neighbour mass is not held fixed \citep[see also][]{Bourne_2015}. It underscores the need for careful prescription of injection modules when comparing simulations across codes with different resolution strategies.

\subsection{Stratified medium jets}
\label{ssec:stratified medium jets}

\par While most of our simulations were run in a uniform medium, realistic galaxy cluster atmospheres exhibit density gradients, with density decreasing from the cluster centre. Including such a setup is particularly important for a code comparison project, as each code will achieve different initial mass and/or spatial resolution. Our assumed density profile is an isothermal beta profile, as described in Section~\ref{ssec:initial conditions}. For this setup, we run our simulations for $44 \, \mathrm{Myr}$ to produce similar-length jets, code by code, as in the uniform-density medium. 

\par The initial conditions for {\sc swift} and {\sc arepo} are generated by sampling resolution elements using inverse transform sampling \citep{Devroye_1986} inside a sphere, and then retaining those that reside inside our box for a total of $N = 2.9 \times 10^7$ ($M_\mathrm{gas} = 10^5 \, \mathrm{M}_\odot$) elements. We test that the initial gas distribution remains stable, in equilibrium with the external gravitational potential, for at least four times the duration of the jet episode. Additionally, we let the system relax adiabatically for $\sim 200 \, \mathrm{Myr}$, before launching the jet to smooth out any initial over/under densities. In the case of {\sc pluto}, its static grid  allows the analytic density profile to be directly applied in the initial configuration. 

\par It is important to note that these two distinct setups---sampling equal-mass elements for {\sc swift} and {\sc arepo}, and applying an analytic density model on a grid for {\sc pluto}---affect effective resolutions in different ways. {\sc swift} and {\sc arepo} will start the simulation with more resolution elements, of a fixed mass, in the densest regions, i.e., the centre of the domain, with progressively fewer elements as the density falls off with radius. The initial spatial resolution in these codes will therefore decrease with radius. In contrast, {\sc pluto} uses the same spatial resolution in all of the domain, but its mass mass resolution varies to follow the density profile. This makes it harder to achieve a fair comparison; we find that {\sc pluto} consistently exhibits lower effective resolution in this setup if we use the same number of resolution elements. To work around this, we include a {\sc pluto} simulation with the same number of grid cells as initial resolution elements in {\sc swift} and {\sc arepo} (standard resolution: SR) and another one where the grid spatial resolution matches the mean spatial resolution inside the density profile core ($r<10 \, \mathrm{kpc}$) in the {\sc swift} and {\sc arepo} initial conditions (high resolution: HR). We achieve the latter by setting the cell volumes in {\sc pluto} to be the same as the mean cell volume in the core of the relaxed {\sc arepo} initial conditions.

\par
\begin{figure*}
    \centering
    \begin{subfigure}{\textwidth}
        \centering
        \includegraphics[width=\textwidth]{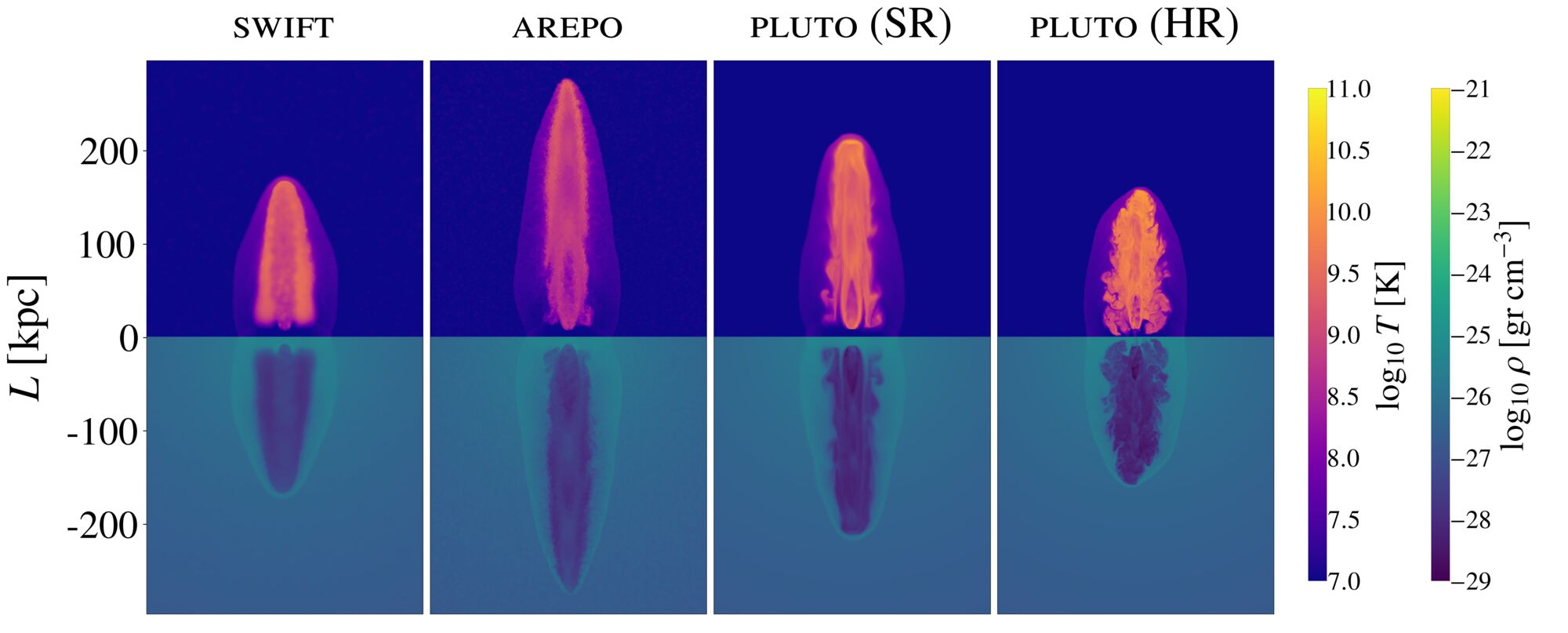}
    \end{subfigure}
     \makebox[0pt][l]{\hspace{-10cm} 
    \begin{subfigure}{\textwidth}
        \centering
        \includegraphics[width=\textwidth]{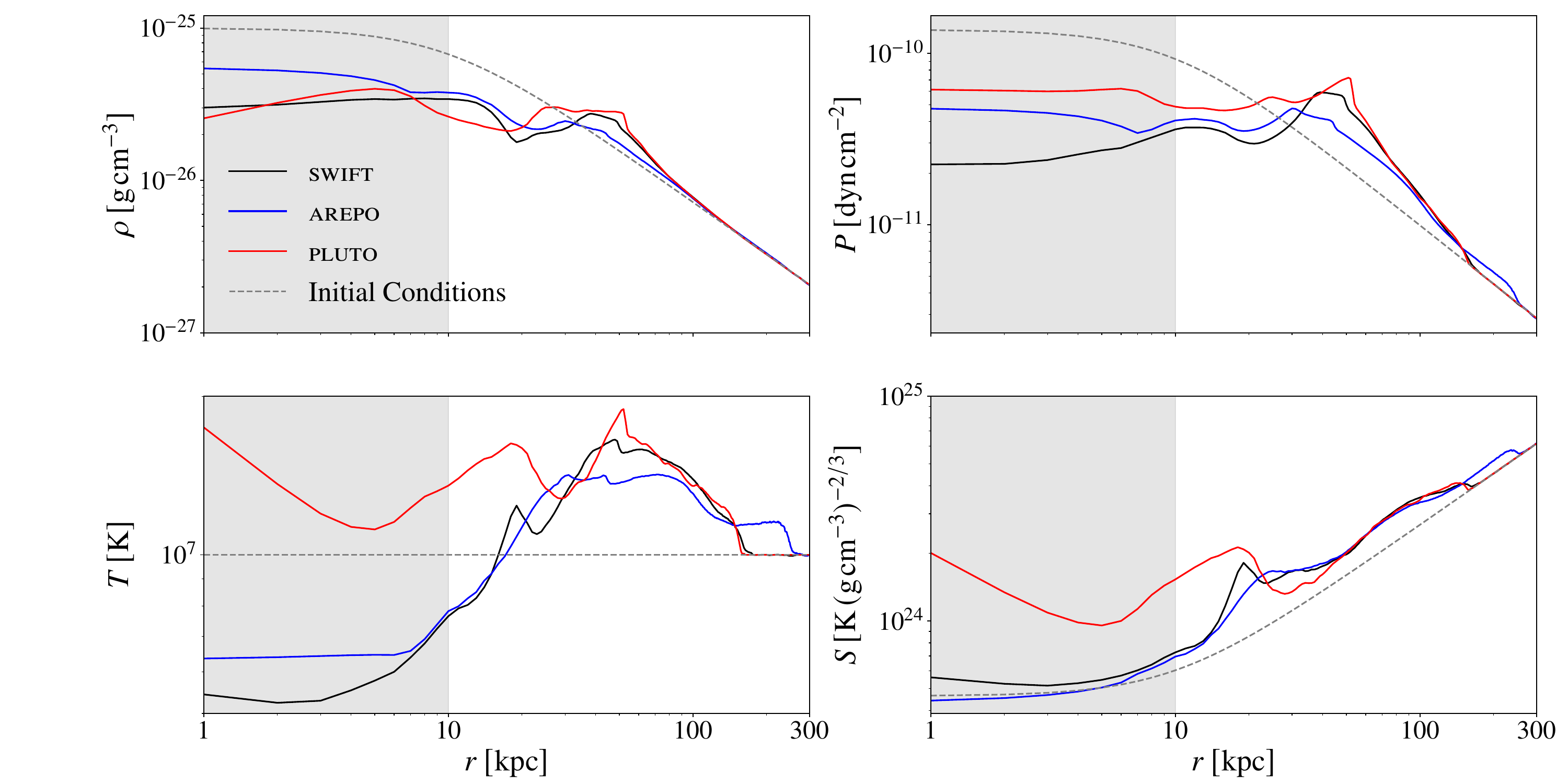}
    \end{subfigure}}
    \caption{\textbf{Top:} Temperature-density slices of {\sc swift}, {\sc arepo} and {\sc pluto} jets in a stratified medium after $44 \, \mathrm{Myr}$. We include two resolution cases for {\sc pluto} for a fairer comparison; SR denotes standard resolution and HR high resolution as described in the main text. All jet lobes appear longer and thinner compared to the uniform medium lobes and the analytic self-similar solution for the stratified medium. \textbf{Bottom:} Ambient medium profiles for density (top left), pressure (top right), temperature (bottom left) and entropy (bottom left). The dashed line represents the initial conditions, black the {\sc swift} simulation, blue the {\sc arepo} and red the HR {\sc pluto}. All jets increase gas temperature and entropy and cause a rise in density and pressure followed by a drop in the inner regions.}
    \label{fig:12.Jets Stratified Medium}
\end{figure*}

 In the top panel of Fig.~\ref{fig:12.Jets Stratified Medium}, we show temperature-density slices of jets simulated with the three codes in the stratified medium at $t = 44 \, \mathrm{Myr}$. We note that our stratified medium reaches lower densities than the uniform medium beyond $\sim 32 \mathrm{kpc}$. Hence, all of the lobes appear longer and thinner than in the uniform medium, remain clearly separated and do not develop the strong backflows we observed in the previous sections. We find a more significant departure from the analytic self-similar solution (especially for {\sc arepo}), with the lobes increasing their aspect ratio throughout the run. Moreover, we note less lobe deceleration but more consistent lobe volumes and masses across the codes. Additionally, we notice a larger portion of the lobe energy being in kinetic form in all the codes, with {\sc arepo} specifically, retaining more overall energy in the lobes. These trends reflect the jets propagating into progressively lower-density regions, encountering less external ram pressure at the jet head and advancing more freely.

\par Comparing the three codes, we observe very similar trends to those in the uniform-medium runs: {\sc swift} inflates short and wide lobes; {\sc arepo} produces long and thin lobes; and the SR of {\sc pluto} produces intermediate-length lobes. Additionally, the HR resolution lobes are shorter, with more pronounced KH instabilities, following the trend we observed in the resolution study in the uniform medium in Section~\ref{ssec:resolution}.

\par In the bottom panel of Fig.~\ref{fig:12.Jets Stratified Medium} we show the ambient medium profiles for density (top left), pressure (top right), temperature (bottom left), and entropy (bottom right). The density and temperature profiles represent volume and mass weighted averages, respectively, for different radial bins, while the pressure and entropy curves are derived afterwards as $P = (\gamma-1)\, u\, \rho$, $S = T\, \rho^{-2/3}$, where $S$ is a measure of entropy that is monotonically related to the thermodynamic entropy. We make sure to exclude the lobes when calculating the profiles to isolate the impact of the jets on the ambient medium. The dashed line represents the initial conditions, black the {\sc swift} simulation, blue the {\sc arepo} and red the HR {\sc pluto} runs. The SR {\sc pluto} run presents only minor differences from the HR {\sc pluto} run. 

\par All simulations exhibit clear deviations from hydrostatic equilibrium within the inner $100$--$200$ $\, \mathrm{kpc}$, as the jets displace and heat the ambient medium. However, care must be taken when interpreting the innermost $10 \, \mathrm{kpc}$, indicated by the shaded region, where direct injections of feedback may influence the results. Moreover, since the jets expand along a single axis, the effects seen in the spherically averaged profiles are diluted. This will be especially relevant for {\sc arepo} jets which are more elongated and thus fill a smaller solid angle. 

\par Despite differences in morphology, all jets have a broadly similar impact on the ambient medium: they heat the gas in the inner regions and displace it to larger radii. This results in elevated mean gas temperature and entropy along with a rise in density and pressure, followed by a drop as the jets propagate outward. {\sc swift} and {\sc pluto} jets have a more pronounced effect on gas density and pressure at intermediate radii  ($r \sim 50 \, \mathrm{kpc}$). while {\sc pluto} jets, in particular, produce the strongest temperature and entropy spikes at small radii ($r < 20 , \mathrm{kpc}$). In contrast, the longer {\sc arepo} jets extend their influence to larger radii, leaving noticeable imprints well beyond the core. This suggests that shorter jets with wider and hotter lobes are more efficient at heating and uplifting gas at small radii, potentially more directly influencing gas with the shortest cooling times.

\subsection{Remnants}
\label{ssec:remnants} 

\par As AGN jets propagate through the ambient medium, they inflate hot, low-density lobes that expand outward, displacing the dense central regions of the stratified atmosphere. This process generates shocks and sound waves and drives bulk motions and turbulence into the surrounding gas, potentially decreasing or disrupting the large-scale accretion flow onto the central black hole powering the jets \citep{Bourne_2023}. Once the jets shut off, the lobes no longer receive a fresh supply of momentum and energy and begin to rise buoyantly through the stratified atmosphere (Stuart et al., in prep.). Candidates for such AGN remnants have been identified in observations \citep[e.g.][]{Parma_2007, Murgia_2011, Godfrey_2017, Mahatma_2018, Dutta_2023}, based on the balance of accretion and feedback \citep{Jurlin_2020, Shabala_2020}, and have been studied in simulations under the influence of cluster weather \citep{Heinz_2006, Morsony_2010, Bourne_2019, Bourne_2021} 

\par To study the properties of jet remnants, we use a larger box with a volume of $900\times900\times1200 \, \mathrm{kpc^3}$. We use the stratified medium to allow for the buoyancy of the hot remnants. We set up our initial conditions as described in the previous section, this time with $N = 1.8 \times 10^7$ resolution elements ($M_\mathrm{gas} = 10^6 \, \mathrm{M}_\odot$). We turn on the jets for $44 \, \mathrm{Myr}$, and subsequently allow the remnants to evolve during the quiescent stage for another $152 \, \mathrm{Myr}$ (for a total of $196 \, \mathrm{Myr}$). 

\par
\begin{figure*}
    \centering
    \begin{subfigure}{\textwidth}
        \centering
        \includegraphics[width=0.9\textwidth]{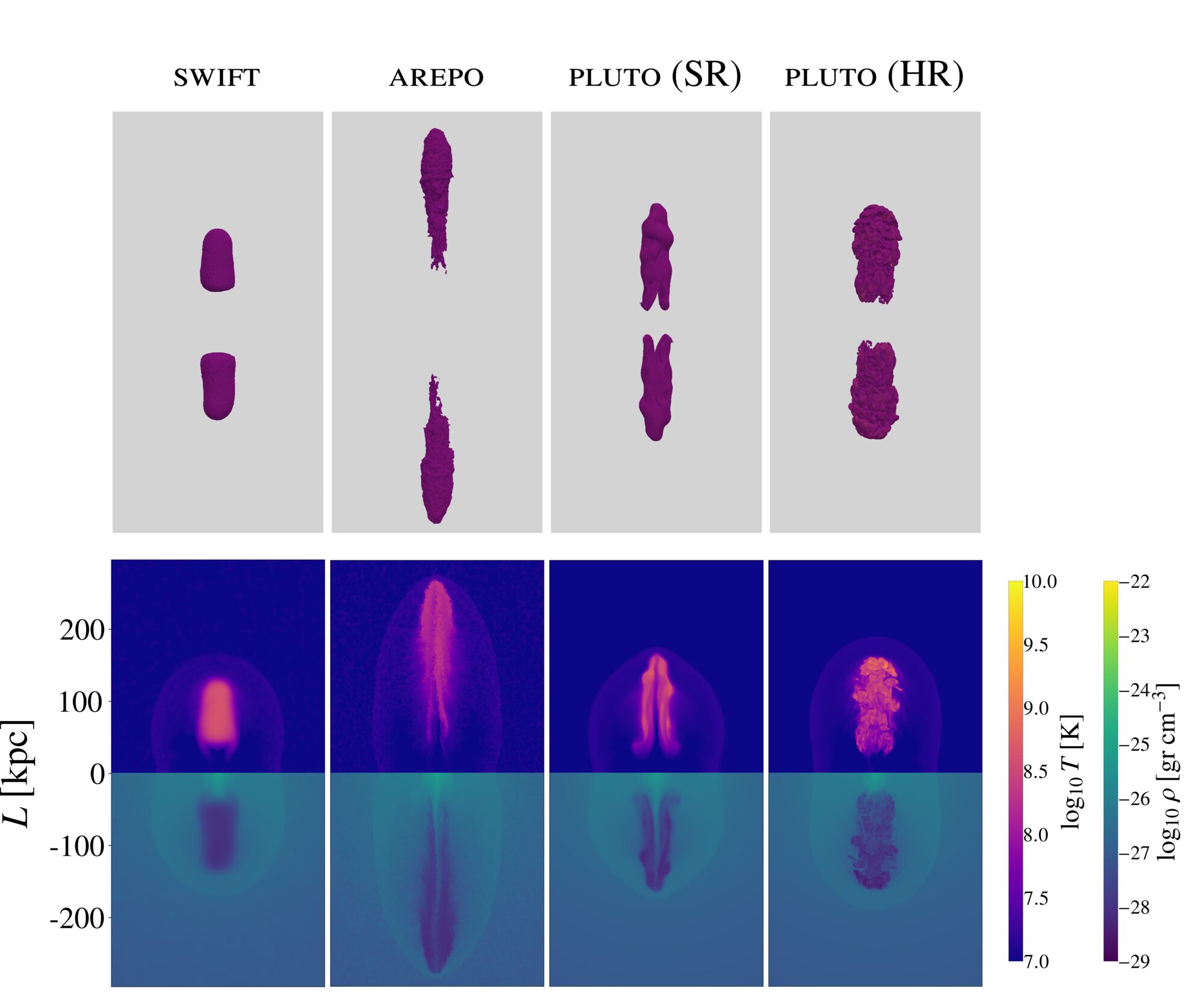}
    \end{subfigure}
     \makebox[0pt][l]{\hspace{-10cm} 
    \begin{subfigure}{\textwidth}
        \centering
        \includegraphics[width=0.9\textwidth]{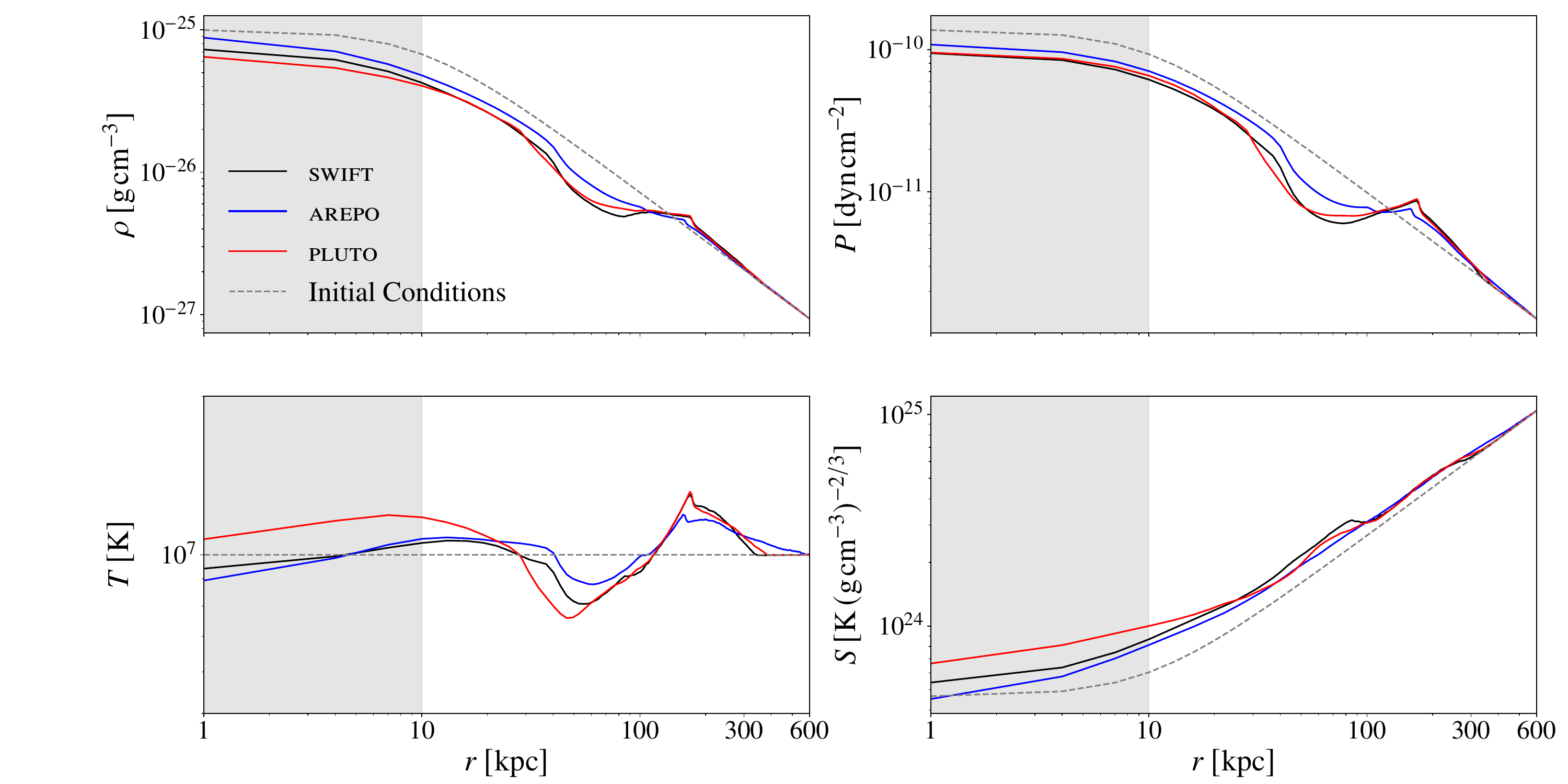}
    \end{subfigure}}
    \caption{\textbf{Top:} 3D Volume renderings of remnant lobes simulated with the three codes in the stratified medium, after $196 \, \mathrm{Myr}$ (when jets turn off after $44 \, \mathrm{Myr}$). \textbf{Middle:} Temperature-density slices of jet remnants and their ambient medium. {\sc swift} produces cylindrical bubbles while {\sc arepo} long filamentary remnants. {\sc pluto} SR presents a thin hollow remnant with internal mixing, while its HR run produces a more coherent structure with instabilities and trailing gas. \textbf{Bottom:} Ambient medium profiles after $196 \, \mathrm{Myr}$. The profiles have relaxed closer to their new equilibrium in the absence of cooling. All jets and remnants have produced similar effects at this stage.}
    \label{fig:13.Jets Remnants}
\end{figure*}

 In the top panel of Fig.~\ref{fig:13.Jets Remnants}, we show 3D volume renderings of the lobes of our jet remnants, while in the middle panel we present temperature-density slices of the remnants and the stratified medium, at $t = 196 \, \mathrm{Myr}$. The physical evolution of these remnants depends on the state of the jets at the end of the active jet phase, as well as their interaction with the surrounding medium and gravitational field. 

\par {\sc swift} jets evolve into stubby, smooth, and cylindrical bubbles with limited small-scale structure. In contrast, {\sc arepo} jets produce long, thin, and filamentary remnants that exhibit more internal mixing. The SR {\sc pluto} simulation produces intermediate-length remnants where much of their interior material has already mixed into the ambient medium. This mixing, also apparent, although to a lesser degree, in the {\sc arepo} remnants, is the result of the development and growth of a large-scale Rayleigh-Taylor (RT) instability. This instability develops because the gravitational field accelerates denser ambient gas into the lighter lobe material. Since the lobe is embedded in the ambient gas, the RT instability manifests as circular overturning motions that start to mix the lobes from the base and progress inward/upward. These motions can be coupled to the growth of KH modes that further enhance mixing \citep{Reynolds_2005, Husko_2023b}. By contrast, the HR {\sc pluto} remnants remain more coherent, taking on a wider cylindrical shape, though signs of mixing are also apparent, with some trailing gas and prominent KH instabilities.

\par
\begin{figure*}
    \centering
    \includegraphics[width=0.7\textheight]{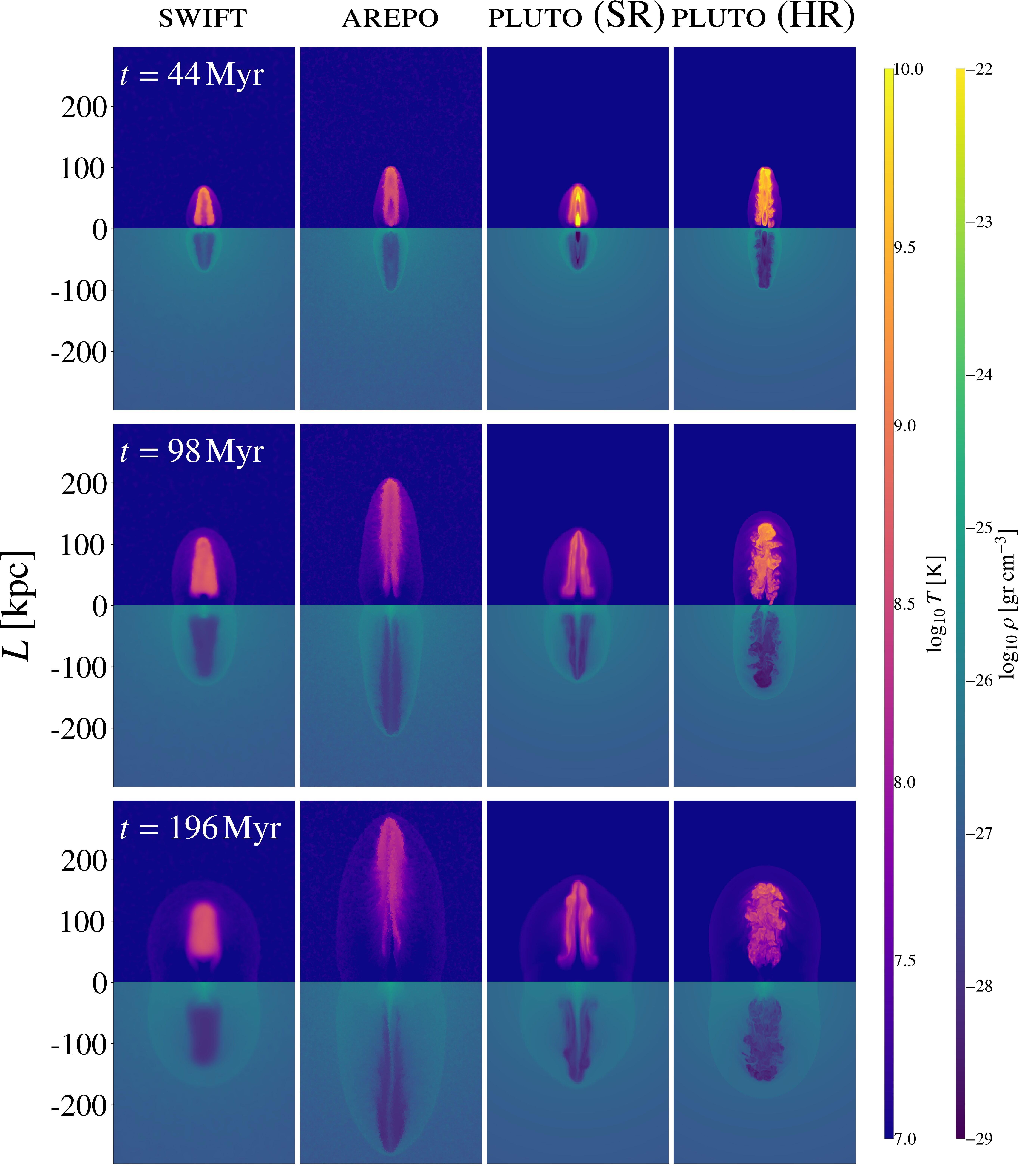}
    \caption{Temperature-density slices of the time evolution of the jets and their remnants simulated with the three codes in the stratified medium. Different rows showcase the different jets and remnants at progressively later times, with the jets turning off at $t = 44 \, \mathrm{Myr}$.}
    \label{fig:14.Jets Remnants Evolution}
\end{figure*}

 In Fig.~\ref{fig:14.Jets Remnants Evolution} we display temperature-density slices for the three jets and remnants, for progressively later times in each row. The first row corresponds to the time of jet turn-off and is a lower resolution version of the top panel of Fig.~\ref{fig:12.Jets Stratified Medium}. In general, the properties of jets and their lobes at turn-off appear to play an important role in shaping the differences between remnants of different codes. The relatively smooth and short {\sc swift} lobes of the active phase evolve into stubby, cylindrical remnants, while the long, thin, and cool {\sc arepo} lobes develop into elongated, cool remnants. The two {\sc pluto} runs start with very hot lobes that eventually become cooler via mixing and adiabatic expansion as the jets turn off and the lobes start expanding and rising in the stratified medium. The KH instabilities visible in the HR remnant can also be seen in the lobe during the active jet phase. We find that {\sc arepo} jets produce highly elongated remnants even though their active lobes seem to be as long as the HR {\sc pluto} lobes, which instead develop more cylindrical remnants. This difference likely arises from the thinness of the {\sc arepo} lobes and their higher kinetic energy content. By contrast, the HR {\sc pluto} lobes convert a larger fraction of their energy in thermal form, as evidenced by the higher lobe temperatures, which leads to a more isotropic expansion.

\par Rising buoyantly through the stratified medium, these hot, under-dense remnants are slowed by drag forces and shaped by the presence of the wake and drift \citep{Pope_2010, Husko_2023b}, ultimately attaining their terminal velocity after approximately $75$--$125 \, \mathrm{Myr}$. Across all codes, the remnants gradually reach remarkably similar terminal velocities and lobe masses. Over the course of the remnant phase, most of the energy in the lobes is transferred to the ambient medium. By the end of the run, less than $10\%$ of the injected energy remains as thermal and kinetic energy in the lobes, while around $70$--$80\%$ ends up in the medium, primarily as thermal energy. Moreover, a significant fraction of the injected energy, about $20\%$, contributes to the gravitational potential energy of the displaced gas. These results are broadly consistent between {\sc swift}, {\sc arepo} and both resolutions of {\sc pluto} and are not particularly influenced by our definition of the lobe. 

\par In the bottom panel of Fig.~\ref{fig:13.Jets Remnants} we plot the ambient medium profiles for density, temperature, pressure and entropy after $196 \, \mathrm{Myr}$. As expected, once the jets shut off, the profiles begin to move closer to their new equilibrium under the adiabatic conditions of our simulations. Even without cooling, some residual effects of jet activity still remain. Both density and pressure remain slightly suppressed at small to intermediate radii, while the entropy profile shows a clear enhancement at intermediate radii, consistent with irreversible heating caused by earlier jet-driven shocks. Notably, all three codes predict broadly similar profiles, with only minor variations.

\section{Discussion}
\label{sec:discussion}

\subsection{Comparison with previous jet simulations}
\label{ssec:comparison with other jet simulations}
\par Our code comparison project builds upon a wealth of previous high-resolution AGN jet simulations \citep[e.g.][]{Hardcastle_2013, English_2016, Weinberger_2017, Bourne_2017, Yates_2018, Husko_2023, Husko_2023b, Weinberger_2023}. Comparing our new jet model (see Section~\ref{ssec:jet model}) with select implementations from the literature can prove useful in understanding how differences in jet launching and injection schemes influence jet propagation, lobe inflation, and the deposition of feedback energy in the ambient medium. This can also shed light on how individual AGN feedback schemes can be effectively designed and honed for specific codes.

\subsubsection{{\sc swift} jets}
\label{sssec:swift jets}
\par \citet{Husko_2023} developed a novel jet injection technique within the {\sc swift} code that employs the SPH method. The jets were implemented as a series of discrete injection events, realised by shooting hydrodynamically decoupled SPH particles and recoupling them on scales of a few kpc, once they cleared the central jet-launching region. The key difference from our model lies in the treatment of decoupled particles: rather than using virtual particles and distributing their mass, momentum, and energy to existing neighbouring particles (as done in our model), their scheme directly hydrodynamically recouples SPH particles to represent injection events. While both approaches can effectively model jet launching and lobe inflation, their method introduces additional SPH particles into the lobes, while ours increases the masses of existing particles (up to twice the initial value), thereby slightly reducing the effective resolution in the lobes. Nonetheless, direct comparison of the methods is difficult, as the different injection techniques lead to distinct couplings between the jet parameters; for instance, the (effective) jet injection velocity in their model is set as an independent parameter, whereas in ours it depends on neighbour number/mass and initial velocities as well.

\subsubsection{{\sc arepo} jets}
\label{sssec:arepo jets}
\par The flexible refinement techniques in {\sc arepo} have enabled the development of various high-resolution jet models. In \cite{Bourne_2017}, jets were assumed to be pre-collimated and were launched by continuously injecting mass, momentum, and energy into existing gas cells within a small, highly refined cylinder along the jet direction. The authors employed multiple refinement techniques to better resolve both the injection region and the jet lobes by using jet tracer- and gradient-dependent refinement criteria. This allowed for more computationally efficient simulations in which lobe structure and instabilities were captured at higher resolution. Such refinement criteria were demonstrated to be critical to model continuously injected, light jets, whereas pure Lagrangian refinement led to poorly resolved and/or stunted jets.

\par An alternative approach was presented in \cite{Weinberger_2017}, where the authors set up (magneto) hydrodynamical simulations of jets and remnants in {\sc arepo}. In this case, the jets were injected by setting the exact thermodynamic state (density and internal energy) and gas velocity within two regions, close to the centre of the domain and along the jet direction. Similar to \citet{Bourne_2017}, the jet was assumed to be pre-collimated. They also included jet tracer- and volume-dependent refinement to override the standard mass refinement to achieve better resolution in the lobes, without which the jets and lobes were poorly resolved.

\par The above studies highlight the importance of refinement in moving-mesh simulations, especially in simulating fast, light jets that inflate low-density lobes. These have been supplemented by a range of other works that consider jets in realistic cluster environments \citep{Bourne_2019, Bourne_2021}, the impact of cosmic rays and/or magnetic fields \citep{Ehlert_2018}, and self-regulated feedback in clusters \citep{Ehlert_2022} or on smaller scales assuming spin-driven jets \citep{Talbot_2021, Talbot_2022, Talbot_2024}. Specialised refinement is crucial in jet models that include a continuous injection at low powers and/or light jets, where insufficient resolution could otherwise lead to excessive numerical mixing. We expect our discrete feedback scheme to mitigate this issue, even with mass refinement alone, though the resolution of the lobes will still be affected. Nonetheless, low-resolution simulations remain valuable as refinement techniques can be very computationally expensive over long timescales.

\subsubsection{{\sc pluto} jets}
\label{sssec:pluto jets}
\par Most simulations of AGN feedback with {\sc pluto} \citep[see e.g.][]{Hardcastle_2013, Hardcastle_2014, English_2016, English_2019, Yates-Jones_2021, Yates-Jones_2023} inject jets continuously by applying inflow boundary conditions and, in some cases, overwriting the hydrodynamic values (density, pressure, velocity) of the grid cells that define the injection region. This approach leads to more stable and reproducible results (than launching virtual particles at random angles), but, at least in the latter case, it has the drawback of not self-consistently evolving the central region. {\sc pluto} jet simulations almost always employ the special relativistic module (RHD) in the code to capture relativistic effects, and sometimes even the magnetohydrodynamics module (RMHD) alongside it, to self-consistently evolve magnetic fields. As mentioned earlier, magnetic fields and relativistic effects were not included in our simulations, in order to facilitate a fair and clear hydrodynamics comparison of codes.

\par It is worth mentioning that some studies of astrophysical jets with {\sc pluto} \citep[e.g.][]{Mattia_2023} use the HLL Riemann solver instead of the more standard HLLC (or HLLD). The HLL solver is more diffusive but also more stable, which can prove advantageous in simulations of jets. Our main suite of {\sc pluto} simulations was run with the more accurate HLLC (while switching to the HLL in the presence of strong shocks), but we additionally tested the HLL solver. In general, most of the trends (see Section~\ref{sssec:general behaviour}) and code differences remained unchanged, although the jets and lobes tended to be slightly longer and exhibited weaker KH instabilities. 


\subsubsection{General behaviour}
\label{sssec:general behaviour}

Across all simulation codes and different studies, we observe several consistent trends in AGN jet behaviour:

\begin{itemize}

\item \textbf{Self-similar evolution:} Certain jet parameters can lead to large-scale jet and lobe evolution that tends to self-similarity. In our uniform-medium runs, higher injection velocities led to more consistent self-similar behaviour in all the codes. In the stratified medium, the jets diverged further from the analytic predictions, inflating more elongated lobes with growing aspect ratios. This is in line with previous high-resolution simulations \citep{Hardcastle_2013, English_2016, Husko_2023}.

\item \textbf{Jet velocity:} Assuming equal powers, fast-light jets tend to be shorter and inflate wider, hotter lobes, as more of their kinetic energy can be thermalised through shocks and adiabatic expansion ($P\,dV$) work in the ambient medium. In contrast, slow-heavy jets propagate further and produce thinner, cooler lobes, due to their higher inertia. This trend was evident in our results and has been noted in previous studies \citep{English_2016, Weinberger_2017, Husko_2023, Huško_2023}.

\item \textbf{Lobe energetics:} Jet lobes typically retain around $40$--$50\%$ of the injected energy \citep{Hardcastle_2013, English_2016, Weinberger_2017, Bourne_2017, Bourne_2019, Bourne_2021}---although some studies find slightly lower fractions \citep{Husko_2023} and others report significantly lower values when including relativistic effects \citep{Perucho_2017}. In our simulations, most cases were consistent with around $40$--$45\%$ energy retained in the lobes by the end of the run. The partition of this energy (thermal or kinetic) depends on the jet model, parameters, and environment. In general, fast-light jets convert more of their energy into thermal energy than slow-heavy jets, due to stronger shocks and increased $P\,dV$ work. 

\item \textbf{Jet length and resolution:} Jet length is a complex function of resolution. Lower resolution causes artificial SPH smoothing in particle codes and numerical mixing and diffusion in grid codes. In general, this results in shorter jets \citep{Weinberger_2017, Bourne_2017, Weinberger_2023, Husko_2023}. This effect was noted in our own simulations, especially for the {\sc arepo} jets. Moreover, it was particularly noticeable for all of the codes at low injection velocities ($v_\mathrm{j} < 2 \times 10^4 \, \mathrm{km/s}$). In resolution tests not shown here, such jets evolved ballistically---maintaining nearly constant bulk lobe speed---and became significantly longer as resolution increased. A similar trend was observed in \citet{Husko_2023}. However, at very high resolutions, Kelvin–Helmholtz instabilities may become better resolved, potentially reducing jet length again \citep{Yates_2018, Husko_2023, Weinberger_2023}. We saw this in the {\sc pluto} jets in both the uniform and the stratified medium. Still, interpreting these instabilities requires care \citep{Bourne_2017}. Overall, jet length likely increases with resolution up to a point, then may decrease once again, as physical instabilities are better resolved.

\item \textbf{Instabilities in remnant bubbles:} The hot, low-density remnant bubbles left behind by jets can develop RT and KH instabilities as they rise through denser ambient gas. These instabilities lead to increased mixing and potential dissolution of the bubbles. A large-scale RT instability was observed in our {\sc arepo} and even more so in our SR {\sc pluto} remnants, while KH instabilities were seen in the HR run. Their effects depend strongly on resolution, initial conditions, and relevant timescales, and may even be suppressed in more realistic setups \citep{Reynolds_2005, Weinberger_2017, English_2019, Husko_2023b}.

\end{itemize}

\subsubsection{Numerical effects}
\label{sssec:impact of numerics}

\par There is now growing evidence that different numerical solvers or codes can lead to differences in jet propagation, bubble inflation and regulation of cooling in clusters. In \citet{Martizzi_2018}, the authors compared the HLLC and HLLE Riemann solvers of the grid code {\sc athena}, concluding that the former could better resolve the cold, over-dense gas filaments that formed as a result of the interaction of jet propagation and cooling. Similarly, \citet{Husko_2023b} showed that remnant bubbles simulated with minimal SPH methods differed in appearance and temperature from those simulated with modern SPH methods that include artificial viscosity limiters and artificial conduction. Additionally, \citet{Ogiya_2018} found that AGN inflated bubbles are unstable to Kelvin-Helmholtz instabilities when modelled with grid, meshless codes or modern SPH methods, but that a minimal SPH model could result in bubbles that do not mix efficiently and remain well-formed, due to artificial surface tension. Finally, in \citet{Weinberger_2023}, the authors compared two codes ({\sc arepo} and {\sc gizmo}) in modelling AGN jet feedback. They used a different jet scheme and cooling function for each code and found that {\sc arepo} jets, compared to {\sc gizmo} jets, were more efficient at delaying cooling flows in clusters (though this is also model dependent). These studies illustrate how differences in numerical hydrodynamics propagate into the deposition of mass, momentum and energy in astrophysical systems, potentially influencing the interplay of feedback processes, cooling flows and accretion. Therefore, careful comparison of different AGN models and different codes as applied in such systems can help elucidate the actual microphysical processes at play and discover robust ways to model them.

\subsection{Areas for further development}
\label{ssec:shortcomings/future work}

\par In this work, we have focused on idealised, purely hydrodynamical AGN jet simulations in simplified environments. While these clean setups allow for careful comparison of the hydrodynamics of the different jets, we necessarily sacrificed complexity in different aspects of modern jet modelling \citep[see e.g.,][for a recent review]{Bourne_2023}. 

\par As emphasised earlier, our study compares three different codes using the same jet model, rather than implementing subgrid feedback schemes tailored to each code. In recent years, hydrodynamical codes are increasingly being used in tandem with feedback models specific to their numerical framework, especially in more complex, high-resolution simulations, like zoom-in simulations of individual clusters. In future work, we aim to extend our comparison project by including feedback models tuned to each code and adopting setups that are more reflective of realistic astrophysical environments.

\par In galaxy formation simulations, AGN feedback is usually coupled to the supermassive black hole accretion rate to determine its power. Incorporating this coupling in our comparison framework would offer insights into how differences in jet propagation and deposition of momentum and energy at small radii influence the self-regulated cycle of black hole accretion and jet launching. Another feature of cosmological simulations is their use of additional physics modules, such as radiative cooling, star formation and magnetic fields, which can significantly impact jet evolution and lobe morphology. Including such modules would allow for more realistic modelling of AGN feedback and could highlight more subtle differences between codes. Crucially, cosmological environments include complex, evolving structures, characterised by inhomogeneities and turbulence, which are fundamentally different from idealised setups, impact jet evolution \citep[e.g.][]{Heinz_2006, Morsony_2010, Bourne_2019, Bourne_2021, Yates-Jones_2023} and must be taken into account for a self-consistent galaxy formation perspective.

\par We should also note that our comparisons were focused on three widely used hydrodynamical codes, spanning the spectrum from fully Lagrangian ({\sc swift}), to moving-mesh ({\sc arepo}), to Eulerian grid-based ({\sc pluto}) methods. However, these represent a subset of a wealth of modern codes that are frequently used in astrophysical simulations \citep[e.g.][]{Teyssier_2002, Springel_2005b, Menon_2015, Springel_2021}. Notably, we have not included the meshless finite-volume/finite-mass method ({\sc gizmo}) developed by \cite{Hopkins_2015}. Given the competitiveness of the meshless methods in standard test problems and the development of the particle spawning jet model \citep{Su_2021}, it would be valuable to include in comparisons of AGN jet models in the future.

\par Finally, we have not considered comparisons of the speed and efficiency of the different codes, as we were focusing on the physical properties and evolution of jets and the lobes that they inflate. We did not attempt to optimise our setups on these aspects and thus could have introduced unnecessary biases or inefficiencies that do not reflect the true performance of the methods. We reserve such an analysis for future work. 

\section{Conclusions}
\label{sec:conclusions}

\par We performed simulations of idealised AGN jets using a novel launching module with three widely used astrophysical hydrodynamical codes, each representing one of the three main classes of computational fluid codes: {\sc swift} (SPH), {\sc arepo} (moving-mesh) and {\sc pluto} (static grid). 

\begin{itemize}
    \item We simulated jets in a uniform medium, using identical parameters and initial conditions. All jets inflated lobes of hot gas that followed a self-similar scaling. {\sc swift} generated the shortest jets, with wide and hot lobes. {\sc arepo} produced the longest jets that formed the thinnest and coolest lobes. {\sc pluto} jets and lobes displayed intermediate characteristics.
    \item We tested three different jet injection velocities. Fast-light jets were more efficient at shock heating and adiabatic expansion and produced shorter, hotter lobes with stronger backflows. In contrast, slow-heavy jets had more inertia and higher momentum flux and thus produced longer, cooler lobes.
    \item In a stratified medium with an external gravitational field, all jets diverged from the self-similar scaling and inflated longer and thinner lobes due to lower external ram pressure. The relative differences between codes closely mirrored those in the uniform case. Despite variations, all jets and lobes had a comparable impact on the ambient medium. 
    \item After the jets were switched off, {\sc swift} produced smooth cylindrical bubbles, {\sc arepo} formed long, thin filaments, and {\sc pluto} jets evolved into intermediate-length remnants with fluid instabilities. In the absence of cooling, ambient gas profiles relaxed close to a similar new equilibrium across all codes.
     \item Code differences depended on how feedback coupled to the computational domain and the effective mass and spatial resolution of each method. 
     \begin{itemize}
          \item The constant spatial resolution in {\sc pluto} enabled very well-resolved flows within the lobes, showing clear signs of fluid instabilities.
          \item The adaptive spatial resolution in {\sc swift} and {\sc arepo} yielded lower resolution in the low density lobes (but higher near the bow shock). The lack of mixing and the kernel-based smoothing in {\sc swift} reduced the effective spatial resolution in the lobes more so than the moving-mesh in {\sc arepo}. Thus, {\sc swift} lobes remained smooth while {\sc arepo} lobes displayed some fluid instabilities. 
          \item Our comparison of equal numbers of particles and cells disadvantaged {\sc swift}, since the kernel based smoothing in SPH inherently results in lower spatial resolution compared to other codes. 
          \item In our fiducial fixed-neighbour-number injection scheme, {\sc pluto} (with adaptive mass resolution) achieved higher effective jet velocities than {\sc arepo} (with constant mass resolution). Using a fixed-neighbour-mass scheme reduced differences between these two codes.
     \end{itemize}
\end{itemize}

\par All in all, our results indicate that, with identical injection models and initial conditions, high-resolution simulations of AGN jets and remnants can still exhibit notable differences depending on the choice of hydrodynamical code.

\par While such differences are important for high-resolution, detailed studies of jet and lobe evolution, they may be less important in the context of cosmological simulations of structure formation, where the main focus is on capturing the impact of jet feedback on resolvable scales. Owing to resolution limitations, these simulations have traditionally focused on reproducing the global statistics of galaxy populations, rather than resolving the small-scale physics of feedback. In this regime, simplified AGN feedback models continue to be widespread, and the principal uncertainties stem from model selection, parameter tuning, and the interactions with prescriptions in other modules, while the specific treatment of hydrodynamics is likely less important.

\par However, modern advances in numerical modelling are enabling a gradual shift in this paradigm. Recent developments in injection methods and refinement strategies across different codes have brought to light novel ways for simulating jets across a variety of scales and environments. While resolution constraints in structure formation simulations still require simplified AGN feedback models, these are now increasingly being supplemented with advanced numerical methods that provide a stronger link between physical processes and numerical implementation \citep[see e.g.][]{Guo_2025}. It is therefore prudent to test the consistency of subgrid models across different hydrodynamical solvers; this is best achieved with high-resolution simulations, as employed in this work, which can serve as a guide to inform the prescriptions in simplified models. This approach not only helps qualify the robustness of AGN feedback models but also facilitates the development and calibration of future jet subgrid prescriptions. These prescriptions should deliver consistent results across different hydrodynamical codes and potentially approach the fidelity of high-resolution jet simulations. 

\section*{Acknowledgements}
We thank the anonymous referee for their careful reading of the paper and their thoughtful comments. M.A.B. acknowledges support from a UKRI Stephen Hawking Fellowship (EP/X04257X/1) as well as the Science and Technology Facilities Council (STFC), grant code ST/W000997/1. CP acknowledges the support of the ARC Centre of Excellence for All Sky Astrophysics in 3 Dimensions (ASTRO 3D), through project number CE170100013. This research was undertaken with the assistance of resources from the National Computational Infrastructure (NCI Australia), an NCRIS enabled capability supported by the Australian Government. This work was supported by resources provided by the Pawsey Supercomputing Research Centre’s Setonix Supercomputer (\url{https://doi.org/10.48569/18sb-8s43}), with funding from the Australian Government and the Government of Western Australia.

\textit{Software citations}: We thank the \textsc{swift}, \textsc{arepo} and \textsc{pluto} development teams for making these codes public. This research also relies heavily on the \textsc{python} \citep{vanRossum.1995} open source community, in particular, \textsc{numpy} \citep{Harris2020}, \textsc{matplotlib} \citep{Hunter2007}, \textsc{scipy} \citep{Virtanen2020} , \textsc{swiftsimio} \citep{Borrow_2020} and \textsc{pyvista} \citep{Sullivan_2019}. 

\section*{Data Availability}
The data used in producing this article will be shared on reasonable request to the corresponding author.



\bibliographystyle{mnras}
\bibliography{Bibliography} 

@ARTICLE{Silk_1998,
       author = {{Silk}, Joseph and {Rees}, Martin J.},
        title = "{Quasars and galaxy formation}",
      journal = {\aap},
         year = 1998,
        month = mar,
       volume = {331},
        pages = {L1-L4},
          doi = {10.48550/arXiv.astro-ph/9801013},
archivePrefix = {arXiv},
       eprint = {astro-ph/9801013},
 primaryClass = {astro-ph},
       adsurl = {https://ui.adsabs.harvard.edu/abs/1998A&A...331L...1S}
}

@book{Devroye_1986,
  title={Non-Uniform Random Variate Generation},
  author={Devroye, Luc},
  year={1986},
  publisher={Springer},
  address={New York},
  isbn={978-0-387-96245-0}
}

@Article{Eckert_2021,
AUTHOR = {Eckert, Dominique and Gaspari, Massimo and Gastaldello, Fabio and Le Brun, Amandine M. C. and O’Sullivan, Ewan},
TITLE = {Feedback from Active Galactic Nuclei in Galaxy Groups},
JOURNAL = {Universe},
VOLUME = {7},
YEAR = {2021},
NUMBER = {5},
ARTICLE-NUMBER = {142},
URL = {https://www.mdpi.com/2218-1997/7/5/142},
ISSN = {2218-1997},
DOI = {10.3390/universe7050142}
}

@ARTICLE{Arepo,
       author = {{Springel}, Volker},
        title = "{E pur si muove: Galilean-invariant cosmological hydrodynamical simulations on a moving mesh}",
      journal = {\mnras},
         year = 2010,
        month = jan,
       volume = {401},
       number = {2},
        pages = {791-851},
          doi = {10.1111/j.1365-2966.2009.15715.x},
archivePrefix = {arXiv},
       eprint = {0901.4107},
 primaryClass = {astro-ph.CO},
       adsurl = {https://ui.adsabs.harvard.edu/abs/2010MNRAS.401..791S}
}

@ARTICLE{Arepo2,
       author = {{Pakmor}, R{\"u}diger and {Springel}, Volker and {Bauer}, Andreas and {Mocz}, Philip and {Munoz}, Diego J. and {Ohlmann}, Sebastian T. and {Schaal}, Kevin and {Zhu}, Chenchong},
        title = "{Improving the convergence properties of the moving-mesh code AREPO}",
      journal = {\mnras},
         year = 2016,
        month = jan,
       volume = {455},
       number = {1},
        pages = {1134-1143},
          doi = {10.1093/mnras/stv2380},
archivePrefix = {arXiv},
       eprint = {1503.00562},
 primaryClass = {astro-ph.GA},
       adsurl = {https://ui.adsabs.harvard.edu/abs/2016MNRAS.455.1134P}
}

@ARTICLE{ArepoCode,
       author = {{Weinberger}, Rainer and {Springel}, Volker and {Pakmor}, R{\"u}diger},
        title = "{The AREPO Public Code Release}",
      journal = {\apjs},
         year = 2020,
        month = jun,
       volume = {248},
       number = {2},
          eid = {32},
        pages = {32},
          doi = {10.3847/1538-4365/ab908c},
archivePrefix = {arXiv},
       eprint = {1909.04667},
 primaryClass = {astro-ph.IM},
       adsurl = {https://ui.adsabs.harvard.edu/abs/2020ApJS..248...32W}
}

@ARTICLE{Talbot_2021,
       author = {{Talbot}, Rosie Y. and {Bourne}, Martin A. and {Sijacki}, Debora},
        title = "{Blandford-Znajek jets in galaxy formation simulations: method and implementation}",
      journal = {\mnras},
         year = 2021,
        month = jul,
       volume = {504},
       number = {3},
        pages = {3619-3650},
          doi = {10.1093/mnras/stab804},
archivePrefix = {arXiv},
       eprint = {2011.10580},
 primaryClass = {astro-ph.GA},
       adsurl = {https://ui.adsabs.harvard.edu/abs/2021MNRAS.504.3619T}
}

@ARTICLE{Talbot_2022,
       author = {{Talbot}, Rosie Y. and {Sijacki}, Debora and {Bourne}, Martin A.},
        title = "{Blandford-Znajek jets in galaxy formation simulations: exploring the diversity of outflows produced by spin-driven AGN jets in Seyfert galaxies}",
      journal = {\mnras},
         year = 2022,
        month = aug,
       volume = {514},
       number = {3},
        pages = {4535-4559},
          doi = {10.1093/mnras/stac1566},
archivePrefix = {arXiv},
       eprint = {2111.01801},
 primaryClass = {astro-ph.GA},
       adsurl = {https://ui.adsabs.harvard.edu/abs/2022MNRAS.514.4535T}
}

@ARTICLE{Talbot_2024,
       author = {{Talbot}, Rosie Y. and {Sijacki}, Debora and {Bourne}, Martin A.},
        title = "{Simulations of spin-driven AGN jets in gas-rich galaxy mergers}",
      journal = {\mnras},
         year = 2024,
        month = mar,
       volume = {528},
       number = {3},
        pages = {5432-5451},
          doi = {10.1093/mnras/stae392},
archivePrefix = {arXiv},
       eprint = {2306.07316},
 primaryClass = {astro-ph.GA},
       adsurl = {https://ui.adsabs.harvard.edu/abs/2024MNRAS.528.5432T}
}

@ARTICLE{Sijacki_2015,
       author = {{Sijacki}, Debora and {Vogelsberger}, Mark and {Genel}, Shy and {Springel}, Volker and {Torrey}, Paul and {Snyder}, Gregory F. and {Nelson}, Dylan and {Hernquist}, Lars},
        title = "{The Illustris simulation: the evolving population of black holes across cosmic time}",
      journal = {\mnras},
         year = 2015,
        month = sep,
       volume = {452},
       number = {1},
        pages = {575-596},
          doi = {10.1093/mnras/stv1340},
archivePrefix = {arXiv},
       eprint = {1408.6842},
 primaryClass = {astro-ph.GA},
       adsurl = {https://ui.adsabs.harvard.edu/abs/2015MNRAS.452..575S}
}

@ARTICLE{Booth_2010,
       author = {{Booth}, C.~M. and {Schaye}, Joop},
        title = "{Dark matter haloes determine the masses of supermassive black holes}",
      journal = {\mnras},
         year = 2010,
        month = jun,
       volume = {405},
       number = {1},
        pages = {L1-L5},
          doi = {10.1111/j.1745-3933.2010.00832.x},
archivePrefix = {arXiv},
       eprint = {0911.0935},
 primaryClass = {astro-ph.CO},
       adsurl = {https://ui.adsabs.harvard.edu/abs/2010MNRAS.405L...1B}
}

@ARTICLE{Hardcastle_2020,
       author = {{Hardcastle}, M.~J. and {Croston}, J.~H.},
        title = "{Radio galaxies and feedback from AGN jets}",
      journal = {\nar},
         year = 2020,
        month = jun,
       volume = {88},
          eid = {101539},
        pages = {101539},
          doi = {10.1016/j.newar.2020.101539},
archivePrefix = {arXiv},
       eprint = {2003.06137},
 primaryClass = {astro-ph.HE},
       adsurl = {https://ui.adsabs.harvard.edu/abs/2020NewAR..8801539H}
}

@ARTICLE{King_2015,
       author = {{King}, Andrew and {Pounds}, Ken},
        title = "{Powerful Outflows and Feedback from Active Galactic Nuclei}",
      journal = {\araa},
         year = 2015,
        month = aug,
       volume = {53},
        pages = {115-154},
          doi = {10.1146/annurev-astro-082214-122316},
archivePrefix = {arXiv},
       eprint = {1503.05206},
 primaryClass = {astro-ph.GA},
       adsurl = {https://ui.adsabs.harvard.edu/abs/2015ARA&A..53..115K}
}

@ARTICLE{Costa_2020,
       author = {{Costa}, Tiago and {Pakmor}, R{\"u}diger and {Springel}, Volker},
        title = "{Powering galactic superwinds with small-scale AGN winds}",
      journal = {\mnras},
         year = 2020,
        month = oct,
       volume = {497},
       number = {4},
        pages = {5229-5255},
          doi = {10.1093/mnras/staa2321},
archivePrefix = {arXiv},
       eprint = {2006.05997},
 primaryClass = {astro-ph.GA},
       adsurl = {https://ui.adsabs.harvard.edu/abs/2020MNRAS.497.5229C}
}

@ARTICLE{Costa_2018,
       author = {{Costa}, Tiago and {Rosdahl}, Joakim and {Sijacki}, Debora and {Haehnelt}, Martin G.},
        title = "{Quenching star formation with quasar outflows launched by trapped IR radiation}",
      journal = {\mnras},
         year = 2018,
        month = sep,
       volume = {479},
       number = {2},
        pages = {2079-2111},
          doi = {10.1093/mnras/sty1514},
archivePrefix = {arXiv},
       eprint = {1709.08638},
 primaryClass = {astro-ph.GA},
       adsurl = {https://ui.adsabs.harvard.edu/abs/2018MNRAS.479.2079C}
}

@article{Raouf_2017,
    author = {Raouf, Mojtaba and Shabala, Stanislav S. and Croton, Darren J. and Khosroshahi, Habib G. and Bernyk, Maksym},
    title = {The many lives of active galactic nuclei–II: The formation and evolution of radio jets and their impact on galaxy evolution},
    journal = {Monthly Notices of the Royal Astronomical Society},
    volume = {471},
    number = {1},
    pages = {658-670},
    year = {2017},
    month = {06},
    issn = {0035-8711},
    doi = {10.1093/mnras/stx1598},
    url = {https://doi.org/10.1093/mnras/stx1598},
    eprint = {https://academic.oup.com/mnras/article-pdf/471/1/658/19376155/stx1598.pdf},
}

@article{Fanidakis_2010,
    author = {Fanidakis, N. and Baugh, C. M. and Benson, A. J. and Bower, R. G. and Cole, S. and Done, C. and Frenk, C. S.},
    title = {Grand unification of AGN activity in the ΛCDM cosmology},
    journal = {Monthly Notices of the Royal Astronomical Society},
    volume = {410},
    number = {1},
    pages = {53-74},
    year = {2010},
    month = {12},
    issn = {0035-8711},
    doi = {10.1111/j.1365-2966.2010.17427.x},
    url = {https://doi.org/10.1111/j.1365-2966.2010.17427.x},
    eprint = {https://academic.oup.com/mnras/article-pdf/410/1/53/18439155/mnras0410-0053.pdf},
}

@article{Thomas_2021,
    author = {Thomas, Nicole and Davé, Romeel and Jarvis, Matt J and Anglés-Alcázar, Daniel},
    title = {The radio galaxy population in the simba simulations},
    journal = {Monthly Notices of the Royal Astronomical Society},
    volume = {503},
    number = {3},
    pages = {3492-3509},
    year = {2021},
    month = {03},
    issn = {0035-8711},
    doi = {10.1093/mnras/stab654},
    url = {https://doi.org/10.1093/mnras/stab654},
    eprint = {https://academic.oup.com/mnras/article-pdf/503/3/3492/36843201/stab654.pdf},
}

@ARTICLE{Bieri_2017,
       author = {{Bieri}, Rebekka and {Dubois}, Yohan and {Rosdahl}, Joakim and {Wagner}, Alexander and {Silk}, Joseph and {Mamon}, Gary A.},
        title = "{Outflows driven by quasars in high-redshift galaxies with radiation hydrodynamics}",
      journal = {\mnras},
         year = 2017,
        month = jan,
       volume = {464},
       number = {2},
        pages = {1854-1873},
          doi = {10.1093/mnras/stw2380},
archivePrefix = {arXiv},
       eprint = {1606.06281},
 primaryClass = {astro-ph.GA},
       adsurl = {https://ui.adsabs.harvard.edu/abs/2017MNRAS.464.1854B}
}

@ARTICLE{Ishibashi_2015,
       author = {{Ishibashi}, W. and {Fabian}, A.~C.},
        title = "{AGN feedback: galactic-scale outflows driven by radiation pressure on dust}",
      journal = {\mnras},
         year = 2015,
        month = jul,
       volume = {451},
       number = {1},
        pages = {93-102},
          doi = {10.1093/mnras/stv944},
archivePrefix = {arXiv},
       eprint = {1504.07393},
 primaryClass = {astro-ph.GA},
       adsurl = {https://ui.adsabs.harvard.edu/abs/2015MNRAS.451...93I}
}

@ARTICLE{Read_2012,
       author = {{Read}, J.~I. and {Hayfield}, T.},
        title = "{SPHS: smoothed particle hydrodynamics with a higher order dissipation switch}",
      journal = {\mnras},
         year = 2012,
        month = jun,
       volume = {422},
       number = {4},
        pages = {3037-3055},
          doi = {10.1111/j.1365-2966.2012.20819.x},
archivePrefix = {arXiv},
       eprint = {1111.6985},
 primaryClass = {astro-ph.CO},
       adsurl = {https://ui.adsabs.harvard.edu/abs/2012MNRAS.422.3037R}
}

@ARTICLE{Mukherjee_2025,
       author = {{Mukherjee}, Dipanjan},
        title = "{Jet-Feedback on kpc scales: a review}",
      journal = {arXiv e-prints},
         year = 2025,
        month = jun,
          eid = {arXiv:2506.03888},
        pages = {arXiv:2506.03888},
          doi = {10.48550/arXiv.2506.03888},
archivePrefix = {arXiv},
       eprint = {2506.03888},
 primaryClass = {astro-ph.GA},
       adsurl = {https://ui.adsabs.harvard.edu/abs/2025arXiv250603888M}
}

@ARTICLE{PLUTO_AMR,
       author = {{Mignone}, A. and {Zanni}, C. and {Tzeferacos}, P. and {van Straalen}, B. and {Colella}, P. and {Bodo}, G.},
        title = "{The PLUTO Code for Adaptive Mesh Computations in Astrophysical Fluid Dynamics}",
      journal = {\apjs},
         year = 2012,
        month = jan,
       volume = {198},
       number = {1},
          eid = {7},
        pages = {7},
          doi = {10.1088/0067-0049/198/1/7},
archivePrefix = {arXiv},
       eprint = {1110.0740},
 primaryClass = {astro-ph.HE},
       adsurl = {https://ui.adsabs.harvard.edu/abs/2012ApJS..198....7M}
}

@ARTICLE{PLUTO,
       author = {{Mignone}, A. and {Bodo}, G. and {Massaglia}, S. and {Matsakos}, T. and {Tesileanu}, O. and {Zanni}, C. and {Ferrari}, A.},
        title = "{PLUTO: A Numerical Code for Computational Astrophysics}",
      journal = {\apjs},
         year = 2007,
        month = may,
       volume = {170},
       number = {1},
        pages = {228-242},
          doi = {10.1086/513316},
archivePrefix = {arXiv},
       eprint = {astro-ph/0701854},
 primaryClass = {astro-ph},
       adsurl = {https://ui.adsabs.harvard.edu/abs/2007ApJS..170..228M}
}

@ARTICLE{SWIFT,
       author = {{Schaller}, Matthieu and {Borrow}, Josh and {Draper}, Peter W. and {Ivkovic}, Mladen and {McAlpine}, Stuart and {Vandenbroucke}, Bert and {Bah{\'e}}, Yannick and {Chaikin}, Evgenii and {Chalk}, Aidan B.~G. and {Chan}, Tsang Keung and {Correa}, Camila and {van Daalen}, Marcel and {Elbers}, Willem and {Gonnet}, Pedro and {Hausammann}, Lo{\"\i}c and {Helly}, John and {Hu{\v{s}}ko}, Filip and {Kegerreis}, Jacob A. and {Nobels}, Folkert S.~J. and {Ploeckinger}, Sylvia and {Revaz}, Yves and {Roper}, William J. and {Ruiz-Bonilla}, Sergio and {Sandnes}, Thomas D. and {Uyttenhove}, Yolan and {Willis}, James S. and {Xiang}, Zhen},
        title = "{SWIFT: A modern highly-parallel gravity and smoothed particle hydrodynamics solver for astrophysical and cosmological applications}",
      journal = {\mnras},
         year = 2024,
        month = mar,
          doi = {10.1093/mnras/stae922},
archivePrefix = {arXiv},
       eprint = {2305.13380},
 primaryClass = {astro-ph.IM},
       adsurl = {https://ui.adsabs.harvard.edu/abs/2024MNRAS.tmp..925S}
}

@article{Hopkins_2015,
    author = {Hopkins, Philip F.},
    title = {A new class of accurate, mesh-free hydrodynamic simulation methods},
    journal = {Monthly Notices of the Royal Astronomical Society},
    volume = {450},
    number = {1},
    pages = {53-110},
    year = {2015},
    month = {04},
    issn = {0035-8711},
    doi = {10.1093/mnras/stv195},
    url = {https://doi.org/10.1093/mnras/stv195},
    eprint = {https://academic.oup.com/mnras/article-pdf/450/1/53/18507840/stv195.pdf},
}

@article{Teyssier_2002,
	author = {{Teyssier, R.}},
	title = {Cosmological hydrodynamics with adaptive mesh refinement  - A new high resolution code called RAMSES },
	DOI= "10.1051/0004-6361:20011817",
	url= "https://doi.org/10.1051/0004-6361:20011817",
	journal = {A&A},
	year = 2002,
	volume = 385,
	number = 1,
	pages = "337-364",
}

@article{Di_Matteo_2005,
   title={Energy input from quasars regulates the growth and activity of black holes and their host galaxies},
   volume={433},
   ISSN={1476-4687},
   url={http://dx.doi.org/10.1038/nature03335},
   DOI={10.1038/nature03335},
   number={7026},
   journal={Nature},
   publisher={Springer Science and Business Media LLC},
   author={Di Matteo, Tiziana and Springel, Volker and Hernquist, Lars},
   year={2005},
   month=feb, pages={604–607} }

@ARTICLE{Di_Matteo_2008,
       author = {{Di Matteo}, Tiziana and {Colberg}, J{\"o}rg and {Springel}, Volker and {Hernquist}, Lars and {Sijacki}, Debora},
        title = "{Direct Cosmological Simulations of the Growth of Black Holes and Galaxies}",
      journal = {\apj},
         year = 2008,
        month = mar,
       volume = {676},
       number = {1},
        pages = {33-53},
          doi = {10.1086/524921},
archivePrefix = {arXiv},
       eprint = {0705.2269},
 primaryClass = {astro-ph},
       adsurl = {https://ui.adsabs.harvard.edu/abs/2008ApJ...676...33D}
}

@article{Springel_2005b,
    author = {Springel, Volker},
    title = {The cosmological simulation code gadget-2},
    journal = {Monthly Notices of the Royal Astronomical Society},
    volume = {364},
    number = {4},
    pages = {1105-1134},
    year = {2005},
    month = {12},
    issn = {0035-8711},
    doi = {10.1111/j.1365-2966.2005.09655.x},
    url = {https://doi.org/10.1111/j.1365-2966.2005.09655.x},
    eprint = {https://academic.oup.com/mnras/article-pdf/364/4/1105/18657201/364-4-1105.pdf},
}

@article{Springel_2021,
    author = {Springel, Volker and Pakmor, Rüdiger and Zier, Oliver and Reinecke, Martin},
    title = {Simulating cosmic structure formation with the gadget-4 code},
    journal = {Monthly Notices of the Royal Astronomical Society},
    volume = {506},
    number = {2},
    pages = {2871-2949},
    year = {2021},
    month = {07},
    issn = {0035-8711},
    doi = {10.1093/mnras/stab1855},
    url = {https://doi.org/10.1093/mnras/stab1855},
    eprint = {https://academic.oup.com/mnras/article-pdf/506/2/2871/39271725/stab1855.pdf},
}

@ARTICLE{Menon_2015,
       author = {{Menon}, Harshitha and {Wesolowski}, Lukasz and {Zheng}, Gengbin and {Jetley}, Pritish and {Kale}, Laxmikant and {Quinn}, Thomas and {Governato}, Fabio},
        title = "{Adaptive techniques for clustered N-body cosmological simulations}",
      journal = {Computational Astrophysics and Cosmology},
         year = 2015,
        month = mar,
       volume = {2},
          eid = {1},
        pages = {1},
          doi = {10.1186/s40668-015-0007-9},
archivePrefix = {arXiv},
       eprint = {1409.1929},
 primaryClass = {astro-ph.IM},
       adsurl = {https://ui.adsabs.harvard.edu/abs/2015ComAC...2....1M}
}

@article{Guo_2025,
doi = {10.3847/1538-4357/add1da},
url = {https://dx.doi.org/10.3847/1538-4357/add1da},
year = {2025},
month = {jul},
publisher = {The American Astronomical Society},
volume = {987},
number = {2},
pages = {202},
author = {Guo, Minghao and Stone, James M. and Quataert, Eliot and Springel, Volker},
title = {Cyclic Zoom: Multiscale GRMHD Modeling of Black Hole Accretion and Feedback},
journal = {The Astrophysical Journal}
}

@article{Laing_2013,
    author = {Laing, R. A. and Bridle, A. H.},
    title = {Systematic properties of decelerating relativistic jets in low-luminosity radio galaxies},
    journal = {Monthly Notices of the Royal Astronomical Society},
    volume = {437},
    number = {4},
    pages = {3405-3441},
    year = {2013},
    month = {12},
    issn = {0035-8711},
    doi = {10.1093/mnras/stt2138},
    url = {https://doi.org/10.1093/mnras/stt2138},
    eprint = {https://academic.oup.com/mnras/article-pdf/437/4/3405/13763049/stt2138.pdf},
}

@article{Volonteri_2016,
    author = {Volonteri, M. and Dubois, Y. and Pichon, C. and Devriendt, J.},
    title = {The cosmic evolution of massive black holes in the Horizon-AGN simulation},
    journal = {Monthly Notices of the Royal Astronomical Society},
    volume = {460},
    number = {3},
    pages = {2979-2996},
    year = {2016},
    month = {05},
    issn = {0035-8711},
    doi = {10.1093/mnras/stw1123},
    url = {https://doi.org/10.1093/mnras/stw1123},
    eprint = {https://academic.oup.com/mnras/article-pdf/460/3/2979/8129829/stw1123.pdf},
}

@ARTICLE{Ehlert_2018,
       author = {{Ehlert}, K. and {Weinberger}, R. and {Pfrommer}, C. and {Pakmor}, R. and {Springel}, V.},
        title = "{Simulations of the dynamics of magnetized jets and cosmic rays in galaxy clusters}",
      journal = {\mnras},
         year = 2018,
        month = dec,
       volume = {481},
       number = {3},
        pages = {2878-2900},
          doi = {10.1093/mnras/sty2397},
archivePrefix = {arXiv},
       eprint = {1806.05679},
 primaryClass = {astro-ph.CO},
       adsurl = {https://ui.adsabs.harvard.edu/abs/2018MNRAS.481.2878E}
}

@article{Ehlert_2022,
    author = {Ehlert, K and Weinberger, R and Pfrommer, C and Pakmor, R and Springel, V},
    title = {Self-regulated AGN feedback of light jets in cool-core galaxy clusters},
    journal = {Monthly Notices of the Royal Astronomical Society},
    volume = {518},
    number = {3},
    pages = {4622-4645},
    year = {2022},
    month = {10},
    issn = {0035-8711},
    doi = {10.1093/mnras/stac2860},
    url = {https://doi.org/10.1093/mnras/stac2860},
    eprint = {https://academic.oup.com/mnras/article-pdf/518/3/4622/47750003/stac2860.pdf},
}

@article{Bourne_2017,
    author = {Bourne, Martin A. and Sijacki, Debora},
    title = "{AGN jet feedback on a moving mesh: cocoon inflation, gas flows and turbulence}",
    journal = {Monthly Notices of the Royal Astronomical Society},
    volume = {472},
    number = {4},
    pages = {4707-4735},
    year = {2017},
    month = {09},
    issn = {0035-8711},
    doi = {10.1093/mnras/stx2269},
    url = {https://doi.org/10.1093/mnras/stx2269},
    eprint = {https://academic.oup.com/mnras/article-pdf/472/4/4707/49203881/mnras\_472\_4\_4707.pdf},
}

@article{Husko_2023,
    author = {Huško, Filip and Lacey, Cedric G},
    title = "{Active galactic nuclei jets simulated with smoothed particle hydrodynamics}",
    journal = {Monthly Notices of the Royal Astronomical Society},
    volume = {520},
    number = {4},
    pages = {5090-5109},
    year = {2023},
    month = {02},
    issn = {0035-8711},
    doi = {10.1093/mnras/stad450},
    url = {https://doi.org/10.1093/mnras/stad450},
    eprint = {https://academic.oup.com/mnras/article-pdf/520/4/5090/49319289/stad450.pdf},
}

@article{Husko_2023b,
    author = {Huško, Filip and Lacey, Cedric G},
    title = "{The complex interplay of AGN jet-inflated bubbles and the intracluster medium}",
    journal = {Monthly Notices of the Royal Astronomical Society},
    volume = {521},
    number = {3},
    pages = {4375-4394},
    year = {2023},
    month = {03},
    issn = {0035-8711},
    doi = {10.1093/mnras/stad793},
    url = {https://doi.org/10.1093/mnras/stad793},
    eprint = {https://academic.oup.com/mnras/article-pdf/521/3/4375/49688988/stad793.pdf},
}

@article{Yates-Jones_2021,
    author = {Yates-Jones, Patrick M and Shabala, Stanislav S and Krause, Martin G H},
    title = "{Dynamics of relativistic radio jets in asymmetric environments}",
    journal = {Monthly Notices of the Royal Astronomical Society},
    volume = {508},
    number = {4},
    pages = {5239-5250},
    year = {2021},
    month = {10},
    issn = {0035-8711},
    doi = {10.1093/mnras/stab2917},
    url = {https://doi.org/10.1093/mnras/stab2917},
    eprint = {https://academic.oup.com/mnras/article-pdf/508/4/5239/40883442/stab2917.pdf},
}

@article{English_2016,
    author = {English, W. and Hardcastle, M. J. and Krause, M. G. H.},
    title = "{Numerical modelling of the lobes of radio galaxies in cluster environments – III. Powerful relativistic and non-relativistic jets}",
    journal = {Monthly Notices of the Royal Astronomical Society},
    volume = {461},
    number = {2},
    pages = {2025-2043},
    year = {2016},
    month = {06},
    issn = {0035-8711},
    doi = {10.1093/mnras/stw1407},
    url = {https://doi.org/10.1093/mnras/stw1407},
    eprint = {https://academic.oup.com/mnras/article-pdf/461/2/2025/8107638/stw1407.pdf},
}

@article{English_2019,
    author = {English, W and Hardcastle, M J and Krause, M G H},
    title = {Numerical modelling of the lobes of radio galaxies in cluster environments – IV. Remnant radio galaxies},
    journal = {Monthly Notices of the Royal Astronomical Society},
    volume = {490},
    number = {4},
    pages = {5807-5819},
    year = {2019},
    month = {10},
    issn = {0035-8711},
    doi = {10.1093/mnras/stz2978},
    url = {https://doi.org/10.1093/mnras/stz2978},
    eprint = {https://academic.oup.com/mnras/article-pdf/490/4/5807/30820733/stz2978.pdf},
}

@article{Hardcastle_2013,
    author = {Hardcastle, M. J. and Krause, M. G. H.},
    title = {Numerical modelling of the lobes of radio galaxies in cluster environments},
    journal = {Monthly Notices of the Royal Astronomical Society},
    volume = {430},
    number = {1},
    pages = {174-196},
    year = {2013},
    month = {01},
    issn = {0035-8711},
    doi = {10.1093/mnras/sts564},
    url = {https://doi.org/10.1093/mnras/sts564},
    eprint = {https://academic.oup.com/mnras/article-pdf/430/1/174/3071578/sts564.pdf},
}

@article{Hardcastle_2014,
    author = {Hardcastle, M. J. and Krause, M. G. H.},
    title = {Numerical modelling of the lobes of radio galaxies in cluster environments – II. Magnetic field configuration and observability},
    journal = {Monthly Notices of the Royal Astronomical Society},
    volume = {443},
    number = {2},
    pages = {1482-1499},
    year = {2014},
    month = {07},
    issn = {0035-8711},
    doi = {10.1093/mnras/stu1229},
    url = {https://doi.org/10.1093/mnras/stu1229},
    eprint = {https://academic.oup.com/mnras/article-pdf/443/2/1482/3669066/stu1229.pdf},
}

@Article{Turner_2023a,
AUTHOR = {Turner, Ross J. and Shabala, Stanislav S.},
TITLE = {Dynamics of Powerful Radio Galaxies},
JOURNAL = {Galaxies},
VOLUME = {11},
YEAR = {2023},
NUMBER = {4},
ARTICLE-NUMBER = {87},
URL = {https://www.mdpi.com/2075-4434/11/4/87},
ISSN = {2075-4434},
DOI = {10.3390/galaxies11040087}
}

@article{Vernaleo_2006,
doi = {10.1086/504029},
url = {https://dx.doi.org/10.1086/504029},
year = {2006},
month = {jul},
publisher = {},
volume = {645},
number = {1},
pages = {83},
author = {John C. Vernaleo and Christopher S. Reynolds},
title = {AGN Feedback and Cooling Flows: Problems with Simple Hydrodynamic Models},
journal = {The Astrophysical Journal}
}

@article{Reynolds_2005,
    author = {Reynolds, Christopher S. and McKernan, Barry and Fabian, Andrew C. and Stone, James M. and Vernaleo, John C.},
    title = {Buoyant radio lobes in a viscous intracluster medium},
    journal = {Monthly Notices of the Royal Astronomical Society},
    volume = {357},
    number = {1},
    pages = {242-250},
    year = {2005},
    month = {02},
    issn = {0035-8711},
    doi = {10.1111/j.1365-2966.2005.08643.x},
    url = {https://doi.org/10.1111/j.1365-2966.2005.08643.x},
    eprint = {https://academic.oup.com/mnras/article-pdf/357/1/242/3481921/357-1-242.pdf},
}

@article{Yates_2018,
    author = {Yates, Patrick M and Shabala, Stanislav S and Krause, Martin G H},
    title = {Observability of intermittent radio sources in galaxy groups and clusters},
    journal = {Monthly Notices of the Royal Astronomical Society},
    volume = {480},
    number = {4},
    pages = {5286-5306},
    year = {2018},
    month = {08},
    issn = {0035-8711},
    doi = {10.1093/mnras/sty2191},
    url = {https://doi.org/10.1093/mnras/sty2191},
    eprint = {https://academic.oup.com/mnras/article-pdf/480/4/5286/25636611/sty2191.pdf},
}

@article{Shabala_2020,
    author = {Shabala, Stanislav S and Jurlin, Nika and Morganti, Raffaella and Brienza, Marisa and Hardcastle, Martin J and Godfrey, Leith E H and Krause, Martin G H and Turner, Ross J},
    title = {The duty cycle of radio galaxies revealed by LOFAR: remnant and restarted radio source populations in the Lockman Hole},
    journal = {Monthly Notices of the Royal Astronomical Society},
    volume = {496},
    number = {2},
    pages = {1706-1717},
    year = {2020},
    month = {04},
    issn = {0035-8711},
    doi = {10.1093/mnras/staa1172},
    url = {https://doi.org/10.1093/mnras/staa1172},
    eprint = {https://academic.oup.com/mnras/article-pdf/496/2/1706/33483618/staa1172.pdf},
}

@article{Jurlin_2020,
   title={The life cycle of radio galaxies in the LOFAR Lockman Hole field},
   volume={638},
   ISSN={1432-0746},
   url={http://dx.doi.org/10.1051/0004-6361/201936955},
   DOI={10.1051/0004-6361/201936955},
   journal={Astronomy &amp; Astrophysics},
   publisher={EDP Sciences},
   author={Jurlin, N. and Morganti, R. and Brienza, M. and Mandal, S. and Maddox, N. and Duncan, K. J. and Shabala, S. S. and Hardcastle, M. J. and Prandoni, I. and Röttgering, H. J. A. and Mahatma, V. and Best, P. N. and Mingo, B. and Sabater, J. and Shimwell, T. W. and Tasse, C.},
   year={2020},
   month=jun, pages={A34} }

@article{Falle_1991,
    author = {Falle, S. A. E. G.},
    title = "{Self-similar jets}",
    journal = {Monthly Notices of the Royal Astronomical Society},
    volume = {250},
    number = {3},
    pages = {581-596},
    year = {1991},
    month = {06},
    issn = {0035-8711},
    doi = {10.1093/mnras/250.3.581},
    url = {https://doi.org/10.1093/mnras/250.3.581},
    eprint = {https://academic.oup.com/mnras/article-pdf/250/3/581/3189879/mnras250-0581.pdf},
}

@article{Kaiser_1997,
    author = {Kaiser, Christian R. and Alexander, Paul},
    title = "{A self-similar model for extragalactic radio sources}",
    journal = {Monthly Notices of the Royal Astronomical Society},
    volume = {286},
    number = {1},
    pages = {215-222},
    year = {1997},
    month = {03},
    issn = {0035-8711},
    doi = {10.1093/mnras/286.1.215},
    url = {https://doi.org/10.1093/mnras/286.1.215},
    eprint = {https://academic.oup.com/mnras/article-pdf/286/1/215/5553393/286-1-215.pdf},
}

@article{Komissarov_1998,
    author = {Komissarov, S. S. and Falle, S. A. E. G.},
    title = "{The large-scale structure of FR-II radio sources}",
    journal = {Monthly Notices of the Royal Astronomical Society},
    volume = {297},
    number = {4},
    pages = {1087-1108},
    year = {1998},
    month = {07},
    issn = {0035-8711},
    doi = {10.1046/j.1365-8711.1998.01547.x},
    url = {https://doi.org/10.1046/j.1365-8711.1998.01547.x},
    eprint = {https://academic.oup.com/mnras/article-pdf/297/4/1087/3464761/297-4-1087.pdf},
}

@article{Kaiser_2007,
    author = {Kaiser, Christian R. and Best, Philip N.},
    title = "{Luminosity function, sizes and FR dichotomy of radio‐loud AGN}",
    journal = {Monthly Notices of the Royal Astronomical Society},
    volume = {381},
    number = {4},
    pages = {1548-1560},
    year = {2007},
    month = {10},
    issn = {0035-8711},
    doi = {10.1111/j.1365-2966.2007.12350.x},
    url = {https://doi.org/10.1111/j.1365-2966.2007.12350.x},
    eprint = {https://academic.oup.com/mnras/article-pdf/381/4/1548/2936969/mnras0381-1548.pdf},
}

@article{Cattaneo_2007,
    author = {Cattaneo, A. and Teyssier, R.},
    title = "{AGN self-regulation in cooling flow clusters}",
    journal = {Monthly Notices of the Royal Astronomical Society},
    volume = {376},
    number = {4},
    pages = {1547-1556},
    year = {2007},
    month = {04},
    issn = {0035-8711},
    doi = {10.1111/j.1365-2966.2007.11512.x},
    url = {https://doi.org/10.1111/j.1365-2966.2007.11512.x},
    eprint = {https://academic.oup.com/mnras/article-pdf/376/4/1547/18665261/mnras0376-1547.pdf},
}

@article{Li_2014,
doi = {10.1088/0004-637X/789/1/54},
url = {https://dx.doi.org/10.1088/0004-637X/789/1/54},
year = {2014},
month = {jun},
publisher = {The American Astronomical Society},
volume = {789},
number = {1},
pages = {54},
author = {Yuan Li and Greg L. Bryan},
title = {MODELING ACTIVE GALACTIC NUCLEUS FEEDBACK IN COOL-CORE CLUSTERS: THE BALANCE BETWEEN HEATING AND COOLING},
journal = {The Astrophysical Journal}
}

@ARTICLE{KarenYang_2016,
       author = {{Yang}, H. -Y. Karen and {Reynolds}, Christopher S.},
        title = "{Interplay Among Cooling, AGN Feedback, and Anisotropic Conduction in the Cool Cores of Galaxy Clusters}",
      journal = {\apj},
         year = 2016,
        month = feb,
       volume = {818},
       number = {2},
          eid = {181},
        pages = {181},
          doi = {10.3847/0004-637X/818/2/181},
archivePrefix = {arXiv},
       eprint = {1512.05796},
 primaryClass = {astro-ph.GA},
       adsurl = {https://ui.adsabs.harvard.edu/abs/2016ApJ...818..181Y}
}

@article{Agertz_2007,
    author = {Agertz, Oscar and Moore, Ben and Stadel, Joachim and Potter, Doug and Miniati, Francesco and Read, Justin and Mayer, Lucio and Gawryszczak, Artur and Kravtsov, Andrey and Nordlund, Åke and Pearce, Frazer and Quilis, Vicent and Rudd, Douglas and Springel, Volker and Stone, James and Tasker, Elizabeth and Teyssier, Romain and Wadsley, James and Walder, Rolf},
    title = "{Fundamental differences between SPH and grid methods}",
    journal = {Monthly Notices of the Royal Astronomical Society},
    volume = {380},
    number = {3},
    pages = {963-978},
    year = {2007},
    month = {08},
    issn = {0035-8711},
    doi = {10.1111/j.1365-2966.2007.12183.x},
    url = {https://doi.org/10.1111/j.1365-2966.2007.12183.x},
    eprint = {https://academic.oup.com/mnras/article-pdf/380/3/963/2796387/mnras0380-0963.pdf},
}

@article{Sijacki_2012,
    author = {Sijacki, Debora and Vogelsberger, Mark and Kereš, Dušan and Springel, Volker and Hernquist, Lars},
    title = "{Moving mesh cosmology: the hydrodynamics of galaxy formation}",
    journal = {Monthly Notices of the Royal Astronomical Society},
    volume = {424},
    number = {4},
    pages = {2999-3027},
    year = {2012},
    month = {08},
    issn = {0035-8711},
    doi = {10.1111/j.1365-2966.2012.21466.x},
    url = {https://doi.org/10.1111/j.1365-2966.2012.21466.x},
    eprint = {https://academic.oup.com/mnras/article-pdf/424/4/2999/18713553/424-4-2999.pdf},
}

@ARTICLE{Frenk_1999,
       author = {{Frenk}, C.~S. and {White}, S.~D.~M. and {Bode}, P. and {Bond}, J.~R. and {Bryan}, G.~L. and {Cen}, R. and {Couchman}, H.~M.~P. and {Evrard}, A.~E. and {Gnedin}, N. and {Jenkins}, A. and {Khokhlov}, A.~M. and {Klypin}, A. and {Navarro}, J.~F. and {Norman}, M.~L. and {Ostriker}, J.~P. and {Owen}, J.~M. and {Pearce}, F.~R. and {Pen}, U. -L. and {Steinmetz}, M. and {Thomas}, P.~A. and {Villumsen}, J.~V. and {Wadsley}, J.~W. and {Warren}, M.~S. and {Xu}, G. and {Yepes}, G.},
        title = "{The Santa Barbara Cluster Comparison Project: A Comparison of Cosmological Hydrodynamics Solutions}",
      journal = {\apj},
         year = 1999,
        month = nov,
       volume = {525},
       number = {2},
        pages = {554-582},
          doi = {10.1086/307908},
archivePrefix = {arXiv},
       eprint = {astro-ph/9906160},
 primaryClass = {astro-ph},
       adsurl = {https://ui.adsabs.harvard.edu/abs/1999ApJ...525..554F}
}

@article{Sembolini_2016,
    author = {Sembolini, Federico and Yepes, Gustavo and Pearce, Frazer R. and Knebe, Alexander and Kay, Scott T. and Power, Chris and Cui, Weiguang and Beck, Alexander M. and Borgani, Stefano and Dalla Vecchia, Claudio and Davé, Romeel and Elahi, Pascal Jahan and February, Sean and Huang, Shuiyao and Hobbs, Alex and Katz, Neal and Lau, Erwin and McCarthy, Ian G. and Murante, Guiseppe and Nagai, Daisuke and Nelson, Kaylea and Newton, Richard D. A. and Perret, Valentin and Puchwein, Ewald and Read, Justin I. and Saro, Alexandro and Schaye, Joop and Teyssier, Romain and Thacker, Robert J.},
    title = "{nIFTy galaxy cluster simulations – I. Dark matter and non-radiative models}",
    journal = {Monthly Notices of the Royal Astronomical Society},
    volume = {457},
    number = {4},
    pages = {4063-4080},
    year = {2016},
    month = {02},
    issn = {0035-8711},
    doi = {10.1093/mnras/stw250},
    url = {https://doi.org/10.1093/mnras/stw250},
    eprint = {https://academic.oup.com/mnras/article-pdf/457/4/4063/18515441/stw250.pdf},
}

@article{Voit_2005,
    author = {Voit, G. Mark and Kay, Scott T. and Bryan, Greg L.},
    title = "{The baseline intracluster entropy profile from gravitational structure formation}",
    journal = {Monthly Notices of the Royal Astronomical Society},
    volume = {364},
    number = {3},
    pages = {909-916},
    year = {2005},
    month = {12},
    issn = {0035-8711},
    doi = {10.1111/j.1365-2966.2005.09621.x},
    url = {https://doi.org/10.1111/j.1365-2966.2005.09621.x},
    eprint = {https://academic.oup.com/mnras/article-pdf/364/3/909/4014787/364-3-909.pdf},
}

@article{Wadsley_2008,
    author = {Wadsley, J. W. and Veeravalli, G. and Couchman, H. M. P.},
    title = "{On the treatment of entropy mixing in numerical cosmology}",
    journal = {Monthly Notices of the Royal Astronomical Society},
    volume = {387},
    number = {1},
    pages = {427-438},
    year = {2008},
    month = {05},
    issn = {0035-8711},
    doi = {10.1111/j.1365-2966.2008.13260.x},
    url = {https://doi.org/10.1111/j.1365-2966.2008.13260.x},
    eprint = {https://academic.oup.com/mnras/article-pdf/387/1/427/3214920/mnras0387-0427.pdf},
}

@article{Power_2014,
    author = {Power, C. and Read, J. I. and Hobbs, A.},
    title = "{The formation of entropy cores in non-radiative galaxy cluster simulations: smoothed particle hydrodynamics versus adaptive mesh refinement}",
    journal = {Monthly Notices of the Royal Astronomical Society},
    volume = {440},
    number = {4},
    pages = {3243-3256},
    year = {2014},
    month = {04},
    issn = {0035-8711},
    doi = {10.1093/mnras/stu418},
    url = {https://doi.org/10.1093/mnras/stu418},
    eprint = {https://academic.oup.com/mnras/article-pdf/440/4/3243/3840389/stu418.pdf},
}

@article{Böehringer_1993,
    author = {Böehringer, H. and Voges, W. and Fabian, A. C. and Edge, A. C. and Neumann, D. M.},
    title = {A ROSAT HRI study of the interaction of the X-ray-emitting gas and radio lobes of NGC 1275},
    journal = {Monthly Notices of the Royal Astronomical Society},
    volume = {264},
    number = {1},
    pages = {L25-L28},
    year = {1993},
    month = {09},
    issn = {0035-8711},
    doi = {10.1093/mnras/264.1.L25},
    url = {https://doi.org/10.1093/mnras/264.1.L25},
    eprint = {https://academic.oup.com/mnras/article-pdf/264/1/L25/3735623/mnras264-0L25.pdf},
}

@article{Fabian_2000,
    author = {Fabian, A. C. and Sanders, J. S. and Ettori, S. and Taylor, G. B. and Allen, S. W. and Crawford, C. S. and Iwasawa, K. and Johnstone, R. M. and Ogle, P. M.},
    title = {Chandra imaging of the complex X-ray core of the Perseus cluster},
    journal = {Monthly Notices of the Royal Astronomical Society},
    volume = {318},
    number = {4},
    pages = {L65-L68},
    year = {2000},
    month = {11},
    issn = {0035-8711},
    doi = {10.1046/j.1365-8711.2000.03904.x},
    url = {https://doi.org/10.1046/j.1365-8711.2000.03904.x},
    eprint = {https://academic.oup.com/mnras/article-pdf/318/4/L65/2830936/318-4-L65.pdf},
}

@article{Scheuer_1974,
    author = {Scheuer, P. A. G.},
    title = {Models of Extragalactic Radio Sources with a Continuous Energy Supply from a Central Object},
    journal = {Monthly Notices of the Royal Astronomical Society},
    volume = {166},
    number = {3},
    pages = {513-528},
    year = {1974},
    month = {03},
    issn = {0035-8711},
    doi = {10.1093/mnras/166.3.513},
    url = {https://doi.org/10.1093/mnras/166.3.513},
    eprint = {https://academic.oup.com/mnras/article-pdf/166/3/513/8079635/mnras166-0513.pdf},
}

@article{McNamara_2007,
   author = "McNamara, B.R. and Nulsen, P.E.J.",
   title = "Heating Hot Atmospheres with Active Galactic Nuclei", 
   journal= "Annual Review of Astronomy and Astrophysics",
   year = "2007",
   volume = "45",
   number = "Volume 45, 2007",
   pages = "117-175",
   doi = "https://doi.org/10.1146/annurev.astro.45.051806.110625",
   url = "https://www.annualreviews.org/content/journals/10.1146/annurev.astro.45.051806.110625",
   publisher = "Annual Reviews",
   issn = "1545-4282",
   type = "Journal Article",
  }

@article{Husko_2022,
    author = {Huško, Filip and Lacey, Cedric G and Schaye, Joop and Schaller, Matthieu and Nobels, Folkert S J},
    title = {Spin-driven jet feedback in idealized simulations of galaxy groups and clusters},
    journal = {Monthly Notices of the Royal Astronomical Society},
    volume = {516},
    number = {3},
    pages = {3750-3772},
    year = {2022},
    month = {08},
    issn = {0035-8711},
    doi = {10.1093/mnras/stac2278},
    url = {https://doi.org/10.1093/mnras/stac2278},
    eprint = {https://academic.oup.com/mnras/article-pdf/516/3/3750/46359742/stac2278.pdf},
}

@article{Oei_2024,
   title={Black hole jets on the scale of the cosmic web},
   volume={633},
   ISSN={1476-4687},
   url={http://dx.doi.org/10.1038/s41586-024-07879-y},
   DOI={10.1038/s41586-024-07879-y},
   number={8030},
   journal={Nature},
   publisher={Springer Science and Business Media LLC},
   author={Oei, Martijn S. S. L. and Hardcastle, Martin J. and Timmerman, Roland and Gast, Aivin R. D. J. G. I. B. and Botteon, Andrea and Rodriguez, Antonio C. and Stern, Daniel and Calistro Rivera, Gabriela and van Weeren, Reinout J. and Röttgering, Huub J. A. and Intema, Huib T. and de Gasperin, Francesco and Djorgovski, S. G.},
   year={2024},
   month=sep, pages={537–541} }

@article{Dave_2019,
    author = {Davé, Romeel and Anglés-Alcázar, Daniel and Narayanan, Desika and Li, Qi and Rafieferantsoa, Mika H and Appleby, Sarah},
    title = {simba: Cosmological simulations with black hole growth and feedback},
    journal = {Monthly Notices of the Royal Astronomical Society},
    volume = {486},
    number = {2},
    pages = {2827-2849},
    year = {2019},
    month = {04},
    issn = {0035-8711},
    doi = {10.1093/mnras/stz937},
    url = {https://doi.org/10.1093/mnras/stz937},
    eprint = {https://academic.oup.com/mnras/article-pdf/486/2/2827/28524018/stz937.pdf},
}

@article{Cole_2000,
    author = {Cole, Shaun and Lacey, Cedric G. and Baugh, Carlton M. and Frenk, Carlos S.},
    title = {Hierarchical galaxy formation},
    journal = {Monthly Notices of the Royal Astronomical Society},
    volume = {319},
    number = {1},
    pages = {168-204},
    year = {2000},
    month = {11},
    issn = {0035-8711},
    doi = {10.1046/j.1365-8711.2000.03879.x},
    url = {https://doi.org/10.1046/j.1365-8711.2000.03879.x},
    eprint = {https://academic.oup.com/mnras/article-pdf/319/1/168/3734609/319-1-168.pdf},
}

@article{Henden_2018,
    author = {Henden, Nicholas A and Puchwein, Ewald and Shen, Sijing and Sijacki, Debora},
    title = {The FABLE simulations: a feedback model for galaxies, groups, and clusters},
    journal = {Monthly Notices of the Royal Astronomical Society},
    volume = {479},
    number = {4},
    pages = {5385-5412},
    year = {2018},
    month = {07},
    issn = {0035-8711},
    doi = {10.1093/mnras/sty1780},
    url = {https://doi.org/10.1093/mnras/sty1780},
    eprint = {https://academic.oup.com/mnras/article-pdf/479/4/5385/25218425/sty1780.pdf},
}

@article{McCarthy_2016,
    author = {McCarthy, Ian G. and Schaye, Joop and Bird, Simeon and Le Brun, Amandine M. C.},
    title = {The bahamas project: calibrated hydrodynamical simulations for large-scale structure cosmology},
    journal = {Monthly Notices of the Royal Astronomical Society},
    volume = {465},
    number = {3},
    pages = {2936-2965},
    year = {2016},
    month = {10},
    issn = {0035-8711},
    doi = {10.1093/mnras/stw2792},
    url = {https://doi.org/10.1093/mnras/stw2792},
    eprint = {https://academic.oup.com/mnras/article-pdf/465/3/2936/8485530/stw2792.pdf},
}

@article{Dubois_2013,
    author = {Dubois, Yohan and Gavazzi, Raphaël and Peirani, Sébastien and Silk, Joseph},
    title = {AGN-driven quenching of star formation: morphological and dynamical implications for early-type galaxies},
    journal = {Monthly Notices of the Royal Astronomical Society},
    volume = {433},
    number = {4},
    pages = {3297-3313},
    year = {2013},
    month = {06},
    issn = {0035-8711},
    doi = {10.1093/mnras/stt997},
    url = {https://doi.org/10.1093/mnras/stt997},
    eprint = {https://academic.oup.com/mnras/article-pdf/433/4/3297/4933412/stt997.pdf},
}

@article{Donnari_2020,
    author = {Donnari, Martina and Pillepich, Annalisa and Joshi, Gandhali D and Nelson, Dylan and Genel, Shy and Marinacci, Federico and Rodriguez-Gomez, Vicente and Pakmor, Rüdiger and Torrey, Paul and Vogelsberger, Mark and Hernquist, Lars},
    title = {Quenched fractions in the IllustrisTNG simulations: the roles of AGN feedback, environment, and pre-processing},
    journal = {Monthly Notices of the Royal Astronomical Society},
    volume = {500},
    number = {3},
    pages = {4004-4024},
    year = {2020},
    month = {10},
    issn = {0035-8711},
    doi = {10.1093/mnras/staa3006},
    url = {https://doi.org/10.1093/mnras/staa3006},
    eprint = {https://academic.oup.com/mnras/article-pdf/500/3/4004/34695286/staa3006.pdf},
}

@article{Goubert_2024,
    author = {Goubert, Paul H and Bluck, Asa F L and Piotrowska, Joanna M and Maiolino, Roberto},
    title = {The role of environment and AGN feedback in quenching local galaxies: comparing cosmological hydrodynamical simulations to the SDSS},
    journal = {Monthly Notices of the Royal Astronomical Society},
    volume = {528},
    number = {3},
    pages = {4891-4921},
    year = {2024},
    month = {01},
    issn = {0035-8711},
    doi = {10.1093/mnras/stae269},
    url = {https://doi.org/10.1093/mnras/stae269},
    eprint = {https://academic.oup.com/mnras/article-pdf/528/3/4891/56670506/stae269.pdf},
}

@article{Byrne_2024,
doi = {10.3847/1538-4357/ad67ca},
url = {https://dx.doi.org/10.3847/1538-4357/ad67ca},
year = {2024},
month = {sep},
publisher = {The American Astronomical Society},
volume = {973},
number = {2},
pages = {149},
author = {Byrne, Lindsey and Faucher-Giguère, Claude-André and Wellons, Sarah and Hopkins, Philip F. and Anglés-Alcázar, Daniel and Sultan, Imran and Wijers, Nastasha and Moreno, Jorge and Ponnada, Sam},
title = {Effects of Multichannel Active Galactic Nuclei Feedback in FIRE Cosmological Simulations of Massive Galaxies},
journal = {The Astrophysical Journal}
}

@ARTICLE{Kugel_2023,
       author = {{Kugel}, Roi and {Schaye}, Joop and {Schaller}, Matthieu and {Helly}, John C. and {Braspenning}, Joey and {Elbers}, Willem and {Frenk}, Carlos S. and {McCarthy}, Ian G. and {Kwan}, Juliana and {Salcido}, Jaime and {van Daalen}, Marcel P. and {Vandenbroucke}, Bert and {Bah{\'e}}, Yannick M. and {Borrow}, Josh and {Chaikin}, Evgenii and {Hu{\v{s}}ko}, Filip and {Jenkins}, Adrian and {Lacey}, Cedric G. and {Nobels}, Folkert S.~J. and {Vernon}, Ian},
        title = "{FLAMINGO: calibrating large cosmological hydrodynamical simulations with machine learning}",
      journal = {\mnras},
         year = 2023,
        month = dec,
       volume = {526},
       number = {4},
        pages = {6103-6127},
          doi = {10.1093/mnras/stad2540},
archivePrefix = {arXiv},
       eprint = {2306.05492},
 primaryClass = {astro-ph.CO},
       adsurl = {https://ui.adsabs.harvard.edu/abs/2023MNRAS.526.6103K}
}

@ARTICLE{Bigwood_2025,
       author = {{Bigwood}, Leah and {Bourne}, Martin A. and {Irsic}, Vid and {Amon}, Alexandra and {Sijacki}, Debora},
        title = "{The case for large-scale AGN feedback in galaxy formation simulations: insights from XFABLE}",
      journal = {arXiv e-prints},
         year = 2025,
        month = jan,
          eid = {arXiv:2501.16983},
        pages = {arXiv:2501.16983},
          doi = {10.48550/arXiv.2501.16983},
archivePrefix = {arXiv},
       eprint = {2501.16983},
 primaryClass = {astro-ph.CO},
       adsurl = {https://ui.adsabs.harvard.edu/abs/2025arXiv250116983B}
}

@article{Weinberger_2016,
    author = {Weinberger, Rainer and Springel, Volker and Hernquist, Lars and Pillepich, Annalisa and Marinacci, Federico and Pakmor, Rüdiger and Nelson, Dylan and Genel, Shy and Vogelsberger, Mark and Naiman, Jill and Torrey, Paul},
    title = {Simulating galaxy formation with black hole driven thermal and kinetic feedback},
    journal = {Monthly Notices of the Royal Astronomical Society},
    volume = {465},
    number = {3},
    pages = {3291-3308},
    year = {2016},
    month = {11},
    issn = {0035-8711},
    doi = {10.1093/mnras/stw2944},
    url = {https://doi.org/10.1093/mnras/stw2944},
    eprint = {https://academic.oup.com/mnras/article-pdf/465/3/3291/8518280/stw2944.pdf},
}

@article{Dubois_2012,
    author = {Dubois, Yohan and Devriendt, Julien and Slyz, Adrianne and Teyssier, Romain},
    title = {Self-regulated growth of supermassive black holes by a dual jet–heating active galactic nucleus feedback mechanism: methods, tests and implications for cosmological simulations},
    journal = {Monthly Notices of the Royal Astronomical Society},
    volume = {420},
    number = {3},
    pages = {2662-2683},
    year = {2012},
    month = {02},
    issn = {0035-8711},
    doi = {10.1111/j.1365-2966.2011.20236.x},
    url = {https://doi.org/10.1111/j.1365-2966.2011.20236.x},
    eprint = {https://academic.oup.com/mnras/article-pdf/420/3/2662/3028346/mnras0420-2662.pdf},
}

@article{Sijacki_2007,
    author = {Sijacki, Debora and Springel, Volker and Di Matteo, Tiziana and Hernquist, Lars},
    title = {A unified model for AGN feedback in cosmological simulations of structure formation},
    journal = {Monthly Notices of the Royal Astronomical Society},
    volume = {380},
    number = {3},
    pages = {877-900},
    year = {2007},
    month = {08},
    issn = {0035-8711},
    doi = {10.1111/j.1365-2966.2007.12153.x},
    url = {https://doi.org/10.1111/j.1365-2966.2007.12153.x},
    eprint = {https://academic.oup.com/mnras/article-pdf/380/3/877/2792738/mnras0380-0877.pdf},
}

@article{Gaspari_2020,
       author = {{Gaspari}, Massimo and {Tombesi}, Francesco and {Cappi}, Massimo},
        title = "{Linking macro-, meso- and microscales in multiphase AGN feeding and feedback}",
      journal = {Nature Astronomy},
         year = 2020,
        month = jan,
       volume = {4},
        pages = {10-13},
          doi = {10.1038/s41550-019-0970-1},
archivePrefix = {arXiv},
       eprint = {2001.04985},
 primaryClass = {astro-ph.GA},
       adsurl = {https://ui.adsabs.harvard.edu/abs/2020NatAs...4...10G}
}

@article{Bourne_2015,
    author = {Bourne, Martin A. and Zubovas, Kastytis and Nayakshin, Sergei},
    title = {The resolution bias: low-resolution feedback simulations are better at destroying galaxies},
    journal = {Monthly Notices of the Royal Astronomical Society},
    volume = {453},
    number = {2},
    pages = {1829-1842},
    year = {2015},
    month = {08},
    issn = {0035-8711},
    doi = {10.1093/mnras/stv1730},
    url = {https://doi.org/10.1093/mnras/stv1730},
    eprint = {https://academic.oup.com/mnras/article-pdf/453/2/1829/3950115/stv1730.pdf},
}

@ARTICLE{Bourne_2019,
       author = {{Bourne}, Martin A. and {Sijacki}, Debora and {Puchwein}, Ewald},
        title = "{AGN jet feedback on a moving mesh: lobe energetics and X-ray properties in a realistic cluster environment}",
      journal = {\mnras},
         year = 2019,
        month = nov,
       volume = {490},
       number = {1},
        pages = {343-349},
          doi = {10.1093/mnras/stz2604},
archivePrefix = {arXiv},
       eprint = {1901.11030},
 primaryClass = {astro-ph.GA},
       adsurl = {https://ui.adsabs.harvard.edu/abs/2019MNRAS.490..343B}
}

@ARTICLE{Bourne_2021,
       author = {{Bourne}, Martin A. and {Sijacki}, Debora},
        title = "{AGN jet feedback on a moving mesh: gentle cluster heating by weak shocks and lobe disruption}",
      journal = {\mnras},
         year = 2021,
        month = sep,
       volume = {506},
       number = {1},
        pages = {488-513},
          doi = {10.1093/mnras/stab1662},
archivePrefix = {arXiv},
       eprint = {2008.12784},
 primaryClass = {astro-ph.HE},
       adsurl = {https://ui.adsabs.harvard.edu/abs/2021MNRAS.506..488B}
}

@article{Bourne_2023,
AUTHOR = {Bourne, Martin A. and Yang, Hsiang-Yi Karen},
TITLE = {Recent Progress in Modeling the Macro- and Micro-Physics of Radio Jet Feedback in Galaxy Clusters},
JOURNAL = {Galaxies},
VOLUME = {11},
YEAR = {2023},
NUMBER = {3},
ARTICLE-NUMBER = {73},
URL = {https://www.mdpi.com/2075-4434/11/3/73},
ISSN = {2075-4434},
DOI = {10.3390/galaxies11030073}
}

@article{Weinberger_2017,
    author = {Weinberger, Rainer and Ehlert, Kristian and Pfrommer, Christoph and Pakmor, Rüdiger and Springel, Volker},
    title = {Simulating the interaction of jets with the intracluster medium},
    journal = {Monthly Notices of the Royal Astronomical Society},
    volume = {470},
    number = {4},
    pages = {4530-4546},
    year = {2017},
    month = {06},
    issn = {0035-8711},
    doi = {10.1093/mnras/stx1409},
    url = {https://doi.org/10.1093/mnras/stx1409},
    eprint = {https://academic.oup.com/mnras/article-pdf/470/4/4530/19273866/stx1409.pdf},
}

@article{Vogelsberger_2020,
  title={Cosmological simulations of galaxy formation},
  author={Vogelsberger, Mark and Marinacci, Federico and Torrey, Paul and Puchwein, Ewald},
  journal={Nature Reviews Physics},
  volume={2},
  number={1},
  pages={42--66},
  year={2020},
  publisher={Nature Publishing Group UK London}
}

@article{Martizzi_2018,
    author = {Martizzi, Davide and Quataert, Eliot and Faucher-Giguère, Claude-André and Fielding, Drummond},
    title = {Simulations of jet heating in galaxy clusters: successes and challenges},
    journal = {Monthly Notices of the Royal Astronomical Society},
    volume = {483},
    number = {2},
    pages = {2465-2486},
    year = {2018},
    month = {11},
    issn = {0035-8711},
    doi = {10.1093/mnras/sty3273},
    url = {https://doi.org/10.1093/mnras/sty3273},
    eprint = {https://academic.oup.com/mnras/article-pdf/483/2/2465/27184698/sty3273.pdf},
}

@article{Sullivan_2019, doi = {10.21105/joss.01450}, url = {https://doi.org/10.21105/joss.01450}, year = {2019}, publisher = {The Open Journal}, volume = {4}, number = {37}, pages = {1450}, author = {Sullivan, C. Bane and Kaszynski, Alexander A.}, title = {PyVista: 3D plotting and mesh analysis through a streamlined interface for the Visualization Toolkit (VTK)}, journal = {Journal of Open Source Software} }

@article{Rupke_2011,
doi = {10.1088/2041-8205/729/2/L27},
url = {https://doi.org/10.1088/2041-8205/729/2/L27},
year = {2011},
month = {feb},
publisher = {The American Astronomical Society},
volume = {729},
number = {2},
pages = {L27},
author = {Rupke, David S. N. and Veilleux, Sylvain},
title = {INTEGRAL FIELD SPECTROSCOPY OF MASSIVE, KILOPARSEC-SCALE OUTFLOWS IN THE INFRARED-LUMINOUS QSO Mrk 231},
journal = {The Astrophysical Journal Letters}
}

@ARTICLE{Hardee_2007,
       author = {{Hardee}, Philip E.},
        title = "{Stability Properties of Strongly Magnetized Spine-Sheath Relativistic Jets}",
      journal = {\apj},
         year = 2007,
        month = jul,
       volume = {664},
       number = {1},
        pages = {26-46},
          doi = {10.1086/518409},
archivePrefix = {arXiv},
       eprint = {0704.1621},
 primaryClass = {astro-ph},
       adsurl = {https://ui.adsabs.harvard.edu/abs/2007ApJ...664...26H}
}

@article{Spilker_2025,
doi = {10.3847/1538-4357/adb750},
url = {https://doi.org/10.3847/1538-4357/adb750},
year = {2025},
month = {mar},
publisher = {The American Astronomical Society},
volume = {982},
number = {2},
pages = {72},
author = {Spilker, Justin S. and Champagne, Jaclyn B. and Fan, Xiaohui and Fujimoto, Seiji and van der Werf, Paul P. and Yang, Jinyi and Yue, Minghao},
title = {Direct Evidence for Active Galactic Nuclei Feedback from Fast Molecular Outflows in Reionization-era Quasars},
journal = {The Astrophysical Journal}
}

@article{Yates-Jones_2023, 
title={CosmoDRAGoN simulations—I. Dynamics and observable signatures of radio jets in cosmological environments}, volume={40}, DOI={10.1017/pasa.2023.10}, journal={Publications of the Astronomical Society of Australia}, author={Yates-Jones, Patrick M. and Shabala, Stanislav S. and Power, Chris and Krause, Martin G. H. and Hardcastle, Martin J. and Mohd Noh Velastín, Elena A. N. and Stewart, Georgia S. C.}, year={2023}, pages={e014}}

@misc{Ogiya_2018,
      title={Physical and numerical stability and instability of AGN bubbles in a hot intracluster medium}, 
      author={Go Ogiya and Pawel Biernacki and Oliver Hahn and Romain Teyssier},
      year={2018},
      eprint={1802.02177},
      archivePrefix={arXiv},
      primaryClass={astro-ph.GA},
      url={https://arxiv.org/abs/1802.02177}, 
}

@article{Bower_2006,
    author = {Bower, R. G. and Benson, A. J. and Malbon, R. and Helly, J. C. and Frenk, C. S. and Baugh, C. M. and Cole, S. and Lacey, C. G.},
    title = {Breaking the hierarchy of galaxy formation},
    journal = {Monthly Notices of the Royal Astronomical Society},
    volume = {370},
    number = {2},
    pages = {645-655},
    year = {2006},
    month = {06},
    issn = {0035-8711},
    doi = {10.1111/j.1365-2966.2006.10519.x},
    url = {https://doi.org/10.1111/j.1365-2966.2006.10519.x},
    eprint = {https://academic.oup.com/mnras/article-pdf/370/2/645/2898993/mnras0370-0645.pdf},
}

@article{Forman_2005,
doi = {10.1086/429746},
url = {https://doi.org/10.1086/429746},
year = {2005},
month = {dec},
publisher = {},
volume = {635},
number = {2},
pages = {894},
author = {Forman, W. and Nulsen, P. and Heinz, S. and Owen, F. and Eilek, J. and Vikhlinin, A. and Markevitch, M. and Kraft, R. and Churazov, E. and Jones, C.},
title = {Reflections of Active Galactic Nucleus Outbursts in the Gaseous Atmosphere of M87},
journal = {The Astrophysical Journal}
}

@article{Forman_2007,
doi = {10.1086/519480},
url = {https://doi.org/10.1086/519480},
year = {2007},
month = {aug},
publisher = {},
volume = {665},
number = {2},
pages = {1057},
author = {Forman, W. and Jones, C. and Churazov, E. and Markevitch, M. and Nulsen, P. and Vikhlinin, A. and Begelman, M. and Böhringer, H. and Eilek, J. and Heinz, S. and Kraft, R. and Owen, F. and Pahre, M.},
title = {Filaments, Bubbles, and Weak Shocks in the Gaseous Atmosphere of M87},
journal = {The Astrophysical Journal}
}

@article{Beck_2015,
    author = {Beck, A. M. and Murante, G. and Arth, A. and Remus, R.-S. and Teklu, A. F. and Donnert, J. M. F. and Planelles, S. and Beck, M. C. and Förster, P. and Imgrund, M. and Dolag, K. and Borgani, S.},
    title = {An improved SPH scheme for cosmological simulations},
    journal = {Monthly Notices of the Royal Astronomical Society},
    volume = {455},
    number = {2},
    pages = {2110-2130},
    year = {2015},
    month = {11},
    issn = {0035-8711},
    doi = {10.1093/mnras/stv2443},
    url = {https://doi.org/10.1093/mnras/stv2443},
    eprint = {https://academic.oup.com/mnras/article-pdf/455/2/2110/18514729/stv2443.pdf},
}

@article{Su_2021,
    author = {Su, Kung-Yi and Hopkins, Philip F and Bryan, Greg L and Somerville, Rachel S and Hayward, Christopher C and Anglés-Alcázar, Daniel and Faucher-Giguère, Claude-André and Wellons, Sarah and Stern, Jonathan and Terrazas, Bryan A and Chan, T K and Orr, Matthew E and Hummels, Cameron and Feldmann, Robert and Kereš, Dušan},
    title = {Which AGN jets quench star formation in massive galaxies?},
    journal = {Monthly Notices of the Royal Astronomical Society},
    volume = {507},
    number = {1},
    pages = {175-204},
    year = {2021},
    month = {07},
    issn = {0035-8711},
    doi = {10.1093/mnras/stab2021},
    url = {https://doi.org/10.1093/mnras/stab2021},
    eprint = {https://academic.oup.com/mnras/article-pdf/507/1/175/39770376/stab2021.pdf},
}

@article{Weinberger_2023,
    author = {Weinberger, Rainer and Su, Kung-Yi and Ehlert, Kristian and Pfrommer, Christoph and Hernquist, Lars and Bryan, Greg L and Springel, Volker and Li, Yuan and Burkhart, Blakesley and Choi, Ena and Faucher-Giguère, Claude-André},
    title = {Active galactic nucleus jet feedback in hydrostatic haloes},
    journal = {Monthly Notices of the Royal Astronomical Society},
    volume = {523},
    number = {1},
    pages = {1104-1125},
    year = {2023},
    month = {05},
    issn = {0035-8711},
    doi = {10.1093/mnras/stad1396},
    url = {https://doi.org/10.1093/mnras/stad1396},
    eprint = {https://academic.oup.com/mnras/article-pdf/523/1/1104/50488688/stad1396.pdf},
}

@article{Dutta_2023,
doi = {10.3847/1538-4357/acaf01},
url = {https://dx.doi.org/10.3847/1538-4357/acaf01},
year = {2023},
month = {feb},
publisher = {The American Astronomical Society},
volume = {944},
number = {2},
pages = {176},
author = {Dutta, Sushant and Singh, Veeresh and Chandra, C. H. Ishwara and Wadadekar, Yogesh and Kayal, Abhijit and Heywood, Ian},
title = {Search and Characterization of Remnant Radio Galaxies in the XMM-LSS Deep Field},
journal = {The Astrophysical Journal}
}

@article{Perucho_2017,
    author = {Perucho, Manel and Martí, José-María and Quilis, Vicent and Borja-Lloret, Marina},
    title = {Radio mode feedback: Does relativity matter?},
    journal = {Monthly Notices of the Royal Astronomical Society: Letters},
    volume = {471},
    number = {1},
    pages = {L120-L124},
    year = {2017},
    month = {07},
    issn = {1745-3925},
    doi = {10.1093/mnrasl/slx115},
    url = {https://doi.org/10.1093/mnrasl/slx115},
    eprint = {https://academic.oup.com/mnrasl/article-pdf/471/1/L120/56953643/mnrasl\_471\_1\_l120.pdf},
}

@article{Mattia_2023,
	author = {Mattia, G. and Del Zanna, L. and Bugli, M. and Pavan, A. and Ciolfi, R. and Bodo, G. and Mignone, A.},
	title = {Resistive relativistic MHD simulations of astrophysical jets⋆},
	DOI= "10.1051/0004-6361/202347126",
	url= "https://doi.org/10.1051/0004-6361/202347126",
	journal = {A&A},
	year = 2023,
	volume = 679,
	pages = "A49",
}

@article{Godfrey_2017,
    author = {Godfrey, L. E. H. and Morganti, R. and Brienza, M.},
    title = {On the population of remnant Fanaroff–Riley type II radio galaxies and implications for radio source dynamics},
    journal = {Monthly Notices of the Royal Astronomical Society},
    volume = {471},
    number = {1},
    pages = {891-907},
    year = {2017},
    month = {06},
    issn = {0035-8711},
    doi = {10.1093/mnras/stx1538},
    url = {https://doi.org/10.1093/mnras/stx1538},
    eprint = {https://academic.oup.com/mnras/article-pdf/471/1/891/19386263/stx1538.pdf},
}

@article{Mahatma_2018,
    author = {Mahatma, V H and Hardcastle, M J and Williams, W L and Brienza, M and Brüggen, M and Croston, J H and Gurkan, G and Harwood, J J and Kunert-Bajraszewska, M and Morganti, R and Röttgering, H J A and Shimwell, T W and Tasse, C},
    title = {Remnant radio-loud AGN in the Herschel-ATLAS field},
    journal = {Monthly Notices of the Royal Astronomical Society},
    volume = {475},
    number = {4},
    pages = {4557-4578},
    year = {2018},
    month = {01},
    issn = {0035-8711},
    doi = {10.1093/mnras/sty025},
    url = {https://doi.org/10.1093/mnras/sty025},
    eprint = {https://academic.oup.com/mnras/article-pdf/475/4/4557/23995949/sty025.pdf},
}

@ARTICLE{Begelman_1989,
       author = {{Begelman}, Mitchell C. and {Cioffi}, Denis F.},
        title = "{Overpressured Cocoons in Extragalactic Radio Sources}",
      journal = {\apjl},
         year = 1989,
        month = oct,
       volume = {345},
        pages = {L21},
          doi = {10.1086/185542},
       adsurl = {https://ui.adsabs.harvard.edu/abs/1989ApJ...345L..21B}
}

@ARTICLE{Hardcastle_2018,
       author = {{Hardcastle}, M.~J.},
        title = "{A simulation-based analytic model of radio galaxies}",
      journal = {\mnras},
         year = 2018,
        month = apr,
       volume = {475},
       number = {2},
        pages = {2768-2786},
          doi = {10.1093/mnras/stx3358},
archivePrefix = {arXiv},
       eprint = {1801.00667},
 primaryClass = {astro-ph.HE},
       adsurl = {https://ui.adsabs.harvard.edu/abs/2018MNRAS.475.2768H}
}

@ARTICLE{Turner_2023b,
       author = {{Turner}, Ross J. and {Yates-Jones}, Patrick M. and {Shabala}, Stanislav S. and {Quici}, Benjamin and {Stewart}, Georgia S.~C.},
        title = "{RAiSE: simulation-based analytical model of AGN jets and lobes}",
      journal = {\mnras},
         year = 2023,
        month = jan,
       volume = {518},
       number = {1},
        pages = {945-964},
          doi = {10.1093/mnras/stac2998},
archivePrefix = {arXiv},
       eprint = {2206.09573},
 primaryClass = {astro-ph.HE},
       adsurl = {https://ui.adsabs.harvard.edu/abs/2023MNRAS.518..945T}
}

@article{Murgia_2011,
	author = {Murgia, M. and Parma, P. and Mack, K.-H. and de Ruiter, H. R. and Fanti, R. and Govoni, F. and Tarchi, A. and Giacintucci, S. and Markevitch, M.},
	title = {Dying radio galaxies in clusters},
	DOI= "10.1051/0004-6361/201015302",
	url= "https://doi.org/10.1051/0004-6361/201015302",
	journal = {A&A},
	year = 2011,
	volume = 526,
	pages = "A148",
	month = "",
}

@article{Parma_2007,
	author = {{Parma, P.} and {Murgia, M.} and {de Ruiter, H. R.} and {Fanti, R.} and {Mack, K.-H.} and {Govoni, F.}},
	title = {In search of dying radio sources in the local universe*},
	DOI= "10.1051/0004-6361:20077592",
	url= "https://doi.org/10.1051/0004-6361:20077592",
	journal = {A&A},
	year = 2007,
	volume = 470,
	number = 3,
	pages = "875-888",
}

@article{Pope_2010,
    author = {Pope, Edward C. D. and Babul, Arif and Pavlovski, Georgi and Bower, Richard G. and Dotter, Aaron},
    title = {Mass transport by buoyant bubbles in galaxy clusters},
    journal = {Monthly Notices of the Royal Astronomical Society},
    volume = {406},
    number = {3},
    pages = {2023-2037},
    year = {2010},
    month = {08},
    issn = {0035-8711},
    doi = {10.1111/j.1365-2966.2010.16816.x},
    url = {https://doi.org/10.1111/j.1365-2966.2010.16816.x},
    eprint = {https://academic.oup.com/mnras/article-pdf/406/3/2023/3015534/mnras0406-2023.pdf},
}

@article{SPHENIX,
    author = {Borrow, Josh and Schaller, Matthieu and Bower, Richard G and Schaye, Joop},
    title = {Sphenix: smoothed particle hydrodynamics for the next generation of galaxy formation simulations},
    journal = {Monthly Notices of the Royal Astronomical Society},
    volume = {511},
    number = {2},
    pages = {2367-2389},
    year = {2021},
    month = {11},
    issn = {0035-8711},
    doi = {10.1093/mnras/stab3166},
    url = {https://doi.org/10.1093/mnras/stab3166},
    eprint = {https://academic.oup.com/mnras/article-pdf/511/2/2367/42504037/stab3166.pdf},
}

@article{Huško_2023,
    author = {Huško, Filip and Lacey, Cedric G and Schaye, Joop and Nobels, Folkert S J and Schaller, Matthieu},
    title = {Winds versus jets: a comparison between black hole feedback modes in simulations of idealized galaxy groups and clusters},
    journal = {Monthly Notices of the Royal Astronomical Society},
    volume = {527},
    number = {3},
    pages = {5988-6020},
    year = {2023},
    month = {11},
    issn = {0035-8711},
    doi = {10.1093/mnras/stad3548},
    url = {https://doi.org/10.1093/mnras/stad3548},
    eprint = {https://academic.oup.com/mnras/article-pdf/527/3/5988/54022033/stad3548.pdf},
}

@article{Schaye_2014,
    author = {Schaye, Joop and Crain, Robert A. and Bower, Richard G. and Furlong, Michelle and Schaller, Matthieu and Theuns, Tom and Dalla Vecchia, Claudio and Frenk, Carlos S. and McCarthy, I. G. and Helly, John C. and Jenkins, Adrian and Rosas-Guevara, Y. M. and White, Simon D. M. and Baes, Maarten and Booth, C. M. and Camps, Peter and Navarro, Julio F. and Qu, Yan and Rahmati, Alireza and Sawala, Till and Thomas, Peter A. and Trayford, James},
    title = {The EAGLE project: simulating the evolution and assembly of galaxies and their environments},
    journal = {Monthly Notices of the Royal Astronomical Society},
    volume = {446},
    number = {1},
    pages = {521-554},
    year = {2014},
    month = {11},
    issn = {0035-8711},
    doi = {10.1093/mnras/stu2058},
    url = {https://doi.org/10.1093/mnras/stu2058},
    eprint = {https://academic.oup.com/mnras/article-pdf/446/1/521/4139718/stu2058.pdf},
}

@article{Chauhan_2020,
    author = {Chauhan, Garima and Lagos, Claudia del P and Stevens, Adam R H and Obreschkow, Danail and Power, Chris and Meyer, Martin},
    title = {The physical drivers of the atomic hydrogen–halo mass relation},
    journal = {Monthly Notices of the Royal Astronomical Society},
    volume = {498},
    number = {1},
    pages = {44-67},
    year = {2020},
    month = {08},
    issn = {0035-8711},
    doi = {10.1093/mnras/staa2251},
    url = {https://doi.org/10.1093/mnras/staa2251},
    eprint = {https://academic.oup.com/mnras/article-pdf/498/1/44/33703577/staa2251.pdf},
}

@ARTICLE{Croton_2006,
       author = {{Croton}, Darren J. and {Springel}, Volker and {White}, Simon D.~M. and {De Lucia}, G. and {Frenk}, C.~S. and {Gao}, L. and {Jenkins}, A. and {Kauffmann}, G. and {Navarro}, J.~F. and {Yoshida}, N.},
        title = "{The many lives of active galactic nuclei: cooling flows, black holes and the luminosities and colours of galaxies}",
      journal = {\mnras},
         year = 2006,
        month = jan,
       volume = {365},
       number = {1},
        pages = {11-28},
          doi = {10.1111/j.1365-2966.2005.09675.x},
archivePrefix = {arXiv},
       eprint = {astro-ph/0508046},
 primaryClass = {astro-ph},
       adsurl = {https://ui.adsabs.harvard.edu/abs/2006MNRAS.365...11C}
}

@ARTICLE{Durier_2012,
       author = {{Durier}, Fabrice and {Dalla Vecchia}, Claudio},
        title = "{Implementation of feedback in smoothed particle hydrodynamics: towards concordance of methods}",
      journal = {\mnras},
         year = 2012,
        month = jan,
       volume = {419},
       number = {1},
        pages = {465-478},
          doi = {10.1111/j.1365-2966.2011.19712.x},
archivePrefix = {arXiv},
       eprint = {1105.3729},
 primaryClass = {astro-ph.CO},
       adsurl = {https://ui.adsabs.harvard.edu/abs/2012MNRAS.419..465D}
}

@ARTICLE{Saitoh_2009,
       author = {{Saitoh}, Takayuki R. and {Makino}, Junichiro},
        title = "{A Necessary Condition for Individual Time Steps in SPH Simulations}",
      journal = {\apjl},
         year = 2009,
        month = jun,
       volume = {697},
       number = {2},
        pages = {L99-L102},
          doi = {10.1088/0004-637X/697/2/L99},
archivePrefix = {arXiv},
       eprint = {0808.0773},
 primaryClass = {astro-ph},
       adsurl = {https://ui.adsabs.harvard.edu/abs/2009ApJ...697L..99S}
}

@article{Pounds_2003,
    author = {Pounds, K. A. and Reeves, J. N. and King, A. R. and Page, K. L. and O'Brien, P. T. and Turner, M. J. L.},
    title = {A high-velocity ionized outflow and XUV photosphere in the narrow emission line quasar PG1211+143},
    journal = {Monthly Notices of the Royal Astronomical Society},
    volume = {345},
    number = {3},
    pages = {705-713},
    year = {2003},
    month = {11},
    issn = {0035-8711},
    doi = {10.1046/j.1365-8711.2003.07006.x},
    url = {https://doi.org/10.1046/j.1365-8711.2003.07006.x},
    eprint = {https://academic.oup.com/mnras/article-pdf/345/3/705/2945820/345-3-705.pdf},
}

@article{King_2003,
    author = {King, A. R. and Pounds, K. A.},
    title = {Black hole winds},
    journal = {Monthly Notices of the Royal Astronomical Society},
    volume = {345},
    number = {2},
    pages = {657-659},
    year = {2003},
    month = {10},
    issn = {0035-8711},
    doi = {10.1046/j.1365-8711.2003.06980.x},
    url = {https://doi.org/10.1046/j.1365-8711.2003.06980.x},
    eprint = {https://academic.oup.com/mnras/article-pdf/345/2/657/4887870/345-2-657.pdf},
}

@article{Arrigoni_2018,
    author = {Arrigoni Battaia, Fabrizio and Hennawi, Joseph F and Prochaska, J Xavier and Oñorbe, Jose and Farina, Emanuele P and Cantalupo, Sebastiano and Lusso, Elisabeta},
    title = {QSO MUSEUM I: a sample of 61 extended Ly α-emission nebulae surrounding z ∼ 3 quasars},
    journal = {Monthly Notices of the Royal Astronomical Society},
    volume = {482},
    number = {3},
    pages = {3162-3205},
    year = {2018},
    month = {10},
    issn = {0035-8711},
    doi = {10.1093/mnras/sty2827},
    url = {https://doi.org/10.1093/mnras/sty2827},
    eprint = {https://academic.oup.com/mnras/article-pdf/482/3/3162/26653804/sty2827.pdf},
}

@article{Sazonov_2005,
    author = {Sazonov, S. Yu. and Ostriker, J. P. and Ciotti, L. and Sunyaev, R. A.},
    title = {Radiative feedback from quasars and the growth of massive black holes in stellar spheroids},
    journal = {Monthly Notices of the Royal Astronomical Society},
    volume = {358},
    number = {1},
    pages = {168-180},
    year = {2005},
    month = {03},
    issn = {0035-8711},
    doi = {10.1111/j.1365-2966.2005.08763.x},
    url = {https://doi.org/10.1111/j.1365-2966.2005.08763.x},
    eprint = {https://academic.oup.com/mnras/article-pdf/358/1/168/3478164/358-1-168.pdf},
}

@article{Rosswog_2007,
    author = {Rosswog, Stephan and Price, Daniel},
    title = {magma: a three-dimensional, Lagrangian magnetohydrodynamics code for merger applications},
    journal = {Monthly Notices of the Royal Astronomical Society},
    volume = {379},
    number = {3},
    pages = {915-931},
    year = {2007},
    month = {07},
    issn = {0035-8711},
    doi = {10.1111/j.1365-2966.2007.11984.x},
    url = {https://doi.org/10.1111/j.1365-2966.2007.11984.x},
    eprint = {https://academic.oup.com/mnras/article-pdf/379/3/915/3498552/mnras0379-0915.pdf},
}

@article{Sandnes_2025,
title = {REMIX SPH – improving mixing in smoothed particle hydrodynamics simulations using a generalised, material-independent approach},
journal = {Journal of Computational Physics},
volume = {532},
pages = {113907},
year = {2025},
issn = {0021-9991},
doi = {https://doi.org/10.1016/j.jcp.2025.113907},
url = {https://www.sciencedirect.com/science/article/pii/S0021999125001901},
author = {T.D. Sandnes and V.R. Eke and J.A. Kegerreis and R.J. Massey and S. Ruiz-Bonilla and M. Schaller and L.F.A. Teodoro}
}

@article{Braspenning_2023,
    author = {Braspenning, Joey and Schaye, Joop and Borrow, Josh and Schaller, Matthieu},
    title = {Sensitivity of non-radiative cloud–wind interactions to the hydrodynamic solver},
    journal = {Monthly Notices of the Royal Astronomical Society},
    volume = {523},
    number = {1},
    pages = {1280-1295},
    year = {2023},
    month = {05},
    issn = {0035-8711},
    doi = {10.1093/mnras/stad1243},
    url = {https://doi.org/10.1093/mnras/stad1243},
    eprint = {https://academic.oup.com/mnras/article-pdf/523/1/1280/50489531/stad1243.pdf},
}

@article{Borrow_2020, doi = {10.21105/joss.02430}, url = {https://doi.org/10.21105/joss.02430}, year = {2020}, publisher = {The Open Journal}, volume = {5}, number = {52}, pages = {2430}, author = {Josh Borrow and Alexei Borrisov}, title = {swiftsimio: A Python library for reading SWIFT data}, journal = {Journal of Open Source Software} }

@article{Borrow_2021,
    author = {Borrow, Josh and Schaller, Matthieu and Bower, Richard G and Schaye, Joop},
    title = {Sphenix: smoothed particle hydrodynamics for the next generation of galaxy formation simulations},
    journal = {Monthly Notices of the Royal Astronomical Society},
    volume = {511},
    number = {2},
    pages = {2367-2389},
    year = {2021},
    month = {11},
    issn = {0035-8711},
    doi = {10.1093/mnras/stab3166},
    url = {https://doi.org/10.1093/mnras/stab3166},
    eprint = {https://academic.oup.com/mnras/article-pdf/511/2/2367/42504037/stab3166.pdf},
}

@article{Saitoh_2013,
doi = {10.1088/0004-637X/768/1/44},
url = {https://dx.doi.org/10.1088/0004-637X/768/1/44},
year = {2013},
month = {apr},
publisher = {The American Astronomical Society},
volume = {768},
number = {1},
pages = {44},
author = {Saitoh, Takayuki R. and Makino, Junichiro},
title = {A DENSITY-INDEPENDENT FORMULATION OF SMOOTHED PARTICLE HYDRODYNAMICS},
journal = {The Astrophysical Journal}
}

@article{Price_2018, 
title={Phantom: A Smoothed Particle Hydrodynamics and Magnetohydrodynamics Code for Astrophysics}, volume={35}, DOI={10.1017/pasa.2018.25}, journal={Publications of the Astronomical Society of Australia}, author={Price, Daniel J. and Wurster, James and Tricco, Terrence S. and Nixon, Chris and Toupin, Stéven and Pettitt, Alex and Chan, Conrad and Mentiplay, Daniel and Laibe, Guillaume and Glover, Simon and et al.}, year={2018}, pages={e031}}

@article{Wadsley_2017,
    author = {Wadsley, James W. and Keller, Benjamin W. and Quinn, Thomas R.},
    title = {Gasoline2: a modern smoothed particle hydrodynamics code},
    journal = {Monthly Notices of the Royal Astronomical Society},
    volume = {471},
    number = {2},
    pages = {2357-2369},
    year = {2017},
    month = {06},
    issn = {0035-8711},
    doi = {10.1093/mnras/stx1643},
    url = {https://doi.org/10.1093/mnras/stx1643},
    eprint = {https://academic.oup.com/mnras/article-pdf/471/2/2357/19491158/stx1643.pdf},
}

@article{Heinz_2006,
    author = {Heinz, S. and Brüggen, M. and Young, A. and Levesque, E.},
    title = {The answer is blowing in the wind: simulating the interaction of jets with dynamic cluster atmospheres},
    journal = {Monthly Notices of the Royal Astronomical Society: Letters},
    volume = {373},
    number = {1},
    pages = {L65-L69},
    year = {2006},
    month = {11},
    issn = {1745-3925},
    doi = {10.1111/j.1745-3933.2006.00243.x},
    url = {https://doi.org/10.1111/j.1745-3933.2006.00243.x},
    eprint = {https://academic.oup.com/mnrasl/article-pdf/373/1/L65/54688636/mnrasl\_373\_1\_l65.pdf},
}

@article{Morsony_2010,
    author = {Morsony, Brian J. and Heinz, Sebastian and Brüggen, Marcus and Ruszkowski, Mateusz},
    title = {Swimming against the current: simulations of central AGN evolution in dynamic galaxy clusters},
    journal = {Monthly Notices of the Royal Astronomical Society},
    volume = {407},
    number = {2},
    pages = {1277-1289},
    year = {2010},
    month = {09},
    issn = {0035-8711},
    doi = {10.1111/j.1365-2966.2010.17059.x},
    url = {https://doi.org/10.1111/j.1365-2966.2010.17059.x},
    eprint = {https://academic.oup.com/mnras/article-pdf/407/2/1277/3912893/mnras0407-1277.pdf},
}

@article{Fabian_2003,
    author = {Fabian, A. C. and Sanders, J. S. and Allen, S. W. and Crawford, C. S. and Iwasawa, K. and Johnstone, R. M. and Schmidt, R. W. and Taylor, G. B.},
    title = "{A deep Chandra observation of the Perseus cluster: shocks and ripples}",
    journal = {MNRAS},
    volume = {344},
    number = {3},
    pages = {L43-L47},
    year = {2003},
    month = {09},
    issn = {0035-8711},
    doi = {10.1046/j.1365-8711.2003.06902.x},
    url = {https://doi.org/10.1046/j.1365-8711.2003.06902.x},
    eprint = {https://academic.oup.com/mnras/article-pdf/344/3/L43/3315205/344-3-L43.pdf},
}

@article{Fabian_2006,
    author = {Fabian, A. C. and Sanders, J. S. and Taylor, G. B. and Allen, S. W. and Crawford, C. S. and Johnstone, R. M. and Iwasawa, K.},
    title = "{A very deep Chandra observation of the Perseus cluster: shocks, ripples and conduction}",
    journal = {MNRAS},
    volume = {366},
    number = {2},
    pages = {417-428},
    year = {2006},
    month = {02},
    issn = {0035-8711},
    doi = {10.1111/j.1365-2966.2005.09896.x},
    url = {https://doi.org/10.1111/j.1365-2966.2005.09896.x},
    eprint = {https://academic.oup.com/mnras/article-pdf/366/2/417/4047326/366-2-417.pdf},
}

@ARTICLE{Hitomi_2016,
       author = {{Hitomi Collaboration} and {Aharonian}, Felix and {Akamatsu}, Hiroki and {Akimoto}, Fumie and {Allen}, Steven W. and {Anabuki}, Naohisa and {Angelini}, Lorella and {Arnaud}, Keith and {Audard}, Marc and {Awaki}, Hisamitsu and {Axelsson}, Magnus and {Bamba}, Aya and {Bautz}, Marshall and {Blandford}, Roger and {Brenneman}, Laura and {Brown}, Gregory V. and {Bulbul}, Esra and {Cackett}, Edward and {Chernyakova}, Maria and {Chiao}, Meng and {Coppi}, Paolo and {Costantini}, Elisa and {de Plaa}, Jelle and {den Herder}, Jan-Willem and {Done}, Chris and {Dotani}, Tadayasu and {Ebisawa}, Ken and {Eckart}, Megan and {Enoto}, Teruaki and {Ezoe}, Yuichiro and {Fabian}, Andrew C. and {Ferrigno}, Carlo and {Foster}, Adam and {Fujimoto}, Ryuichi and {Fukazawa}, Yasushi and {Furuzawa}, Akihiro and {Galeazzi}, Massimiliano and {Gallo}, Luigi and {Gandhi}, Poshak and {Giustini}, Margherita and {Goldwurm}, Andrea and {Gu}, Liyi and {Guainazzi}, Matteo and {Haba}, Yoshito and {Hagino}, Kouichi and {Hamaguchi}, Kenji and {Harrus}, Ilana and {Hatsukade}, Isamu and {Hayashi}, Katsuhiro and {Hayashi}, Takayuki and {Hayashida}, Kiyoshi and {Hiraga}, Junko and {Hornschemeier}, Ann and {Hoshino}, Akio and {Hughes}, John and {Iizuka}, Ryo and {Inoue}, Hajime and {Inoue}, Yoshiyuki and {Ishibashi}, Kazunori and {Ishida}, Manabu and {Ishikawa}, Kumi and {Ishisaki}, Yoshitaka and {Itoh}, Masayuki and {Iyomoto}, Naoko and {Kaastra}, Jelle and {Kallman}, Timothy and {Kamae}, Tuneyoshi and {Kara}, Erin and {Kataoka}, Jun and {Katsuda}, Satoru and {Katsuta}, Junichiro and {Kawaharada}, Madoka and {Kawai}, Nobuyuki and {Kelley}, Richard and {Khangulyan}, Dmitry and {Kilbourne}, Caroline and {King}, Ashley and {Kitaguchi}, Takao and {Kitamoto}, Shunji and {Kitayama}, Tetsu and {Kohmura}, Takayoshi and {Kokubun}, Motohide and {Koyama}, Shu and {Koyama}, Katsuji and {Kretschmar}, Peter and {Krimm}, Hans and {Kubota}, Aya and {Kunieda}, Hideyo and {Laurent}, Philippe and {Lebrun}, Fran{\c{c}}ois and {Lee}, Shiu-Hang and {Leutenegger}, Maurice and {Limousin}, Olivier and {Loewenstein}, Michael and {Long}, Knox S. and {Lumb}, David and {Madejski}, Grzegorz and {Maeda}, Yoshitomo and {Maier}, Daniel and {Makishima}, Kazuo and {Markevitch}, Maxim and {Matsumoto}, Hironori and {Matsushita}, Kyoko and {McCammon}, Dan and {McNamara}, Brian and {Mehdipour}, Missagh and {Miller}, Eric and {Miller}, Jon and {Mineshige}, Shin and {Mitsuda}, Kazuhisa and {Mitsuishi}, Ikuyuki and {Miyazawa}, Takuya and {Mizuno}, Tsunefumi and {Mori}, Hideyuki and {Mori}, Koji and {Moseley}, Harvey and {Mukai}, Koji and {Murakami}, Hiroshi and {Murakami}, Toshio and {Mushotzky}, Richard and {Nagino}, Ryo and {Nakagawa}, Takao and {Nakajima}, Hiroshi and {Nakamori}, Takeshi and {Nakano}, Toshio and {Nakashima}, Shinya and {Nakazawa}, Kazuhiro and {Nobukawa}, Masayoshi and {Noda}, Hirofumi and {Nomachi}, Masaharu and {O'Dell}, Steve and {Odaka}, Hirokazu and {Ohashi}, Takaya and {Ohno}, Masanori and {Okajima}, Takashi and {Ota}, Naomi and {Ozaki}, Masanobu and {Paerels}, Frits and {Paltani}, Stephane and {Parmar}, Arvind and {Petre}, Robert and {Pinto}, Ciro and {Pohl}, Martin and {Porter}, F. Scott and {Pottschmidt}, Katja and {Ramsey}, Brian and {Reynolds}, Christopher and {Russell}, Helen and {Safi-Harb}, Samar and {Saito}, Shinya and {Sakai}, Kazuhiro and {Sameshima}, Hiroaki and {Sato}, Goro and {Sato}, Kosuke and {Sato}, Rie and {Sawada}, Makoto and {Schartel}, Norbert and {Serlemitsos}, Peter and {Seta}, Hiromi and {Shidatsu}, Megumi and {Simionescu}, Aurora and {Smith}, Randall and {Soong}, Yang and {Stawarz}, Lukasz and {Sugawara}, Yasuharu and {Sugita}, Satoshi and {Szymkowiak}, Andrew and {Tajima}, Hiroyasu and {Takahashi}, Hiromitsu and {Takahashi}, Tadayuki and {Takeda}, Shin'Ichiro and {Takei}, Yoh and {Tamagawa}, Toru and {Tamura}, Keisuke and {Tamura}, Takayuki and {Tanaka}, Takaaki and {Tanaka}, Yasuo and {Tanaka}, Yasuyuki and {Tashiro}, Makoto and {Tawara}, Yuzuru and {Terada}, Yukikatsu and {Terashima}, Yuichi and {Tombesi}, Francesco and {Tomida}, Hiroshi and {Tsuboi}, Yohko and {Tsujimoto}, Masahiro and {Tsunemi}, Hiroshi and {Tsuru}, Takeshi and {Uchida}, Hiroyuki and {Uchiyama}, Hideki and {Uchiyama}, Yasunobu and {Ueda}, Shutaro and {Ueda}, Yoshihiro and {Ueno}, Shiro and {Uno}, Shin'Ichiro and {Urry}, Meg and {Ursino}, Eugenio and {de Vries}, Cor and {Watanabe}, Shin and {Werner}, Norbert},
        title = "{The quiescent intracluster medium in the core of the Perseus cluster}",
      journal = {\nat},
         year = 2016,
        month = jul,
       volume = {535},
       number = {7610},
        pages = {117-121},
          doi = {10.1038/nature18627},
archivePrefix = {arXiv},
       eprint = {1607.04487},
 primaryClass = {astro-ph.GA},
       adsurl = {https://ui.adsabs.harvard.edu/abs/2016Natur.535..117H}
}

@article{van_Weeren_2024,
	author = {{van Weeren, R. J.} and {Timmerman, R.} and {Vaidya, V.} and {Gendron-Marsolais, M.-L.} and {Botteon, A.} and {Roberts, I. D.} and {Hlavacek-Larrondo, J.} and {Bonafede, A.} and {Brüggen, M.} and {Brunetti, G.} and {Cassano, R.} and {Cuciti, V.} and {Edge, A. C.} and {Gastaldello, F.} and {Groeneveld, C.} and {Shimwell, T. W.}},
	title = {LOFAR high-band antenna observations of the Perseus cluster - The discovery of a giant radio halo},
	DOI= "10.1051/0004-6361/202451618",
	url= "https://doi.org/10.1051/0004-6361/202451618",
	journal = {A&A},
	year = 2024,
	volume = 692,
	pages = "A12",
}

@article{Fabian_2012,
   author = "Fabian, A.C.",
   title = "Observational Evidence of Active Galactic Nuclei Feedback", 
   journal= "Annual Review of Astronomy and Astrophysics",
   year = "2012",
   volume = "50",
   number = "Volume 50, 2012",
   pages = "455-489",
   doi = "https://doi.org/10.1146/annurev-astro-081811-125521",
   url = "https://www.annualreviews.org/content/journals/10.1146/annurev-astro-081811-125521",
   publisher = "Annual Reviews",
   issn = "1545-4282",
   type = "Journal Article",
  }

@Inbook{Hlavacek-Larrondo_2022,
author="Hlavacek-Larrondo, Julie
and Li, Yuan
and Churazov, Eugene",
editor="Bambi, Cosimo
and Santangelo, Andrea",
title="AGN Feedback in Groups and Clusters of Galaxies",
bookTitle="Handbook of X-ray and Gamma-ray Astrophysics",
year="2022",
publisher="Springer Nature Singapore",
address="Singapore",
pages="1--66",
isbn="978-981-16-4544-0",
doi="10.1007/978-981-16-4544-0_122-1",
url="https://doi.org/10.1007/978-981-16-4544-0_122-1"
}

@article{Genel_2017,
    author = {Genel, Shy and Nelson, Dylan and Pillepich, Annalisa and Springel, Volker and Pakmor, Rüdiger and Weinberger, Rainer and Hernquist, Lars and Naiman, Jill and Vogelsberger, Mark and Marinacci, Federico and Torrey, Paul},
    title = {The size evolution of star-forming and quenched galaxies in the IllustrisTNG simulation},
    journal = {Monthly Notices of the Royal Astronomical Society},
    volume = {474},
    number = {3},
    pages = {3976-3996},
    year = {2017},
    month = {11},
    issn = {0035-8711},
    doi = {10.1093/mnras/stx3078},
    url = {https://doi.org/10.1093/mnras/stx3078},
    eprint = {https://academic.oup.com/mnras/article-pdf/474/3/3976/23002809/stx3078.pdf},
}

@article{Arjona_2025,
	author = {{Arjona-Gálvez, Elena} and {Cardona-Barrero, Salvador} and {Grand, Robert J. J.} and {Di Cintio, Arianna} and {Dalla Vecchia, Claudio} and {Benavides, Jose A.} and {Macciò, Andrea V.} and {Libeskind, Noam} and {Knebe, Alexander}},
	title = {A physically motivated galaxy size definition across different state-of-the-art hydrodynamical simulations},
	DOI= "10.1051/0004-6361/202554290",
	url= "https://doi.org/10.1051/0004-6361/202554290",
	journal = {A&A},
	year = 2025,
	volume = 699,
	pages = "A301",
}

@article{Furlong_2016,
    author = {Furlong, M. and Bower, R. G. and Crain, R. A. and Schaye, J. and Theuns, T. and Trayford, J. W. and Qu, Y. and Schaller, M. and Berthet, M. and Helly, J. C.},
    title = {Size evolution of normal and compact galaxies in the EAGLE simulation},
    journal = {Monthly Notices of the Royal Astronomical Society},
    volume = {465},
    number = {1},
    pages = {722-738},
    year = {2016},
    month = {10},
    issn = {0035-8711},
    doi = {10.1093/mnras/stw2740},
    url = {https://doi.org/10.1093/mnras/stw2740},
    eprint = {https://academic.oup.com/mnras/article-pdf/465/1/722/8593488/stw2740.pdf},
}

@article{Benson_2003,
doi = {10.1086/379160},
url = {https://dx.doi.org/10.1086/379160},
year = {2003},
month = {dec},
publisher = {},
volume = {599},
number = {1},
pages = {38},
author = {Benson, A. J. and Bower, R. G. and Frenk, C. S. and Lacey, C. G. and Baugh, C. M. and Cole, S.},
title = {What Shapes the Luminosity Function of Galaxies?},
journal = {The Astrophysical Journal}
}

@article{Genel_2019,
doi = {10.3847/1538-4357/aaf4bb},
url = {https://dx.doi.org/10.3847/1538-4357/aaf4bb},
year = {2019},
month = {jan},
publisher = {The American Astronomical Society},
volume = {871},
number = {1},
pages = {21},
author = {Genel, Shy and Bryan, Greg L. and Springel, Volker and Hernquist, Lars and Nelson, Dylan and Pillepich, Annalisa and Weinberger, Rainer and Pakmor, Rüdiger and Marinacci, Federico and Vogelsberger, Mark},
title = {A Quantification of the Butterfly Effect in Cosmological Simulations and Implications for Galaxy Scaling Relations},
journal = {The Astrophysical Journal}
}

@article{Dave_2020,
    author = {Davé, Romeel and Crain, Robert A and Stevens, Adam R H and Narayanan, Desika and Saintonge, Amelie and Catinella, Barbara and Cortese, Luca},
    title = {Galaxy cold gas contents in modern cosmological hydrodynamic simulations},
    journal = {Monthly Notices of the Royal Astronomical Society},
    volume = {497},
    number = {1},
    pages = {146-166},
    year = {2020},
    month = {07},
    issn = {0035-8711},
    doi = {10.1093/mnras/staa1894},
    url = {https://doi.org/10.1093/mnras/staa1894},
    eprint = {https://academic.oup.com/mnras/article-pdf/497/1/146/33519951/staa1894.pdf},
}

@article{Wright_2024,
    author = {Wright, Ruby J and Somerville, Rachel S and Lagos, Claudia del P and Schaller, Matthieu and Davé, Romeel and Anglés-Alcázar, Daniel and Genel, Shy},
    title = {The baryon cycle in modern cosmological hydrodynamical simulations},
    journal = {Monthly Notices of the Royal Astronomical Society},
    volume = {532},
    number = {3},
    pages = {3417-3440},
    year = {2024},
    month = {07},
    issn = {0035-8711},
    doi = {10.1093/mnras/stae1688},
    url = {https://doi.org/10.1093/mnras/stae1688},
    eprint = {https://academic.oup.com/mnras/article-pdf/532/3/3417/58652438/stae1688.pdf},
}

@article{Pillepich_2017,
    author = {Pillepich, Annalisa and Springel, Volker and Nelson, Dylan and Genel, Shy and Naiman, Jill and Pakmor, Rüdiger and Hernquist, Lars and Torrey, Paul and Vogelsberger, Mark and Weinberger, Rainer and Marinacci, Federico},
    title = {Simulating galaxy formation with the IllustrisTNG model},
    journal = {Monthly Notices of the Royal Astronomical Society},
    volume = {473},
    number = {3},
    pages = {4077-4106},
    year = {2017},
    month = {10},
    issn = {0035-8711},
    doi = {10.1093/mnras/stx2656},
    url = {https://doi.org/10.1093/mnras/stx2656},
    eprint = {https://academic.oup.com/mnras/article-pdf/473/3/4077/21841785/stx2656.pdf},
}

@article{Dubois_2016,
    author = {Dubois, Yohan and Peirani, Sébastien and Pichon, Christophe and Devriendt, Julien and Gavazzi, Raphaël and Welker, Charlotte and Volonteri, Marta},
    title = {The Horizon-AGN simulation: morphological diversity of galaxies promoted by AGN feedback},
    journal = {Monthly Notices of the Royal Astronomical Society},
    volume = {463},
    number = {4},
    pages = {3948-3964},
    year = {2016},
    month = {09},
    issn = {0035-8711},
    doi = {10.1093/mnras/stw2265},
    url = {https://doi.org/10.1093/mnras/stw2265},
    eprint = {https://academic.oup.com/mnras/article-pdf/463/4/3948/18515487/stw2265.pdf},
}

@article{Habouzit_2021,
    author = {Habouzit, Mélanie and Li, Yuan and Somerville, Rachel S and Genel, Shy and Pillepich, Annalisa and Volonteri, Marta and Davé, Romeel and Rosas-Guevara, Yetli and McAlpine, Stuart and Peirani, Sébastien and Hernquist, Lars and Anglés-Alcázar, Daniel and Reines, Amy and Bower, Richard and Dubois, Yohan and Nelson, Dylan and Pichon, Christophe and Vogelsberger, Mark},
    title = {Supermassive black holes in cosmological simulations I: MBH − M⋆ relation and black hole mass function},
    journal = {Monthly Notices of the Royal Astronomical Society},
    volume = {503},
    number = {2},
    pages = {1940-1975},
    year = {2021},
    month = {02},
    issn = {0035-8711},
    doi = {10.1093/mnras/stab496},
    url = {https://doi.org/10.1093/mnras/stab496},
    eprint = {https://academic.oup.com/mnras/article-pdf/503/2/1940/36678260/stab496.pdf},
}

@manual{vanRossum.1995,
 title                = {Python Reference Manual},
 author               = {van Rossum, Guido},
 number               = {R 9525},
 year                 = 1995,
 month                = jan,
 organization            = {Centrum voor Wiskunde en Informatica},
 note                 = {}
}

@Article{Harris2020,
  author    = {Harris, Charles R. and Millman, K. Jarrod and van der Walt, Stéfan J. and Gommers, Ralf and Virtanen, Pauli and Cournapeau, David and Wieser, Eric and Taylor, Julian and Berg, Sebastian and Smith, Nathaniel J. and Kern, Robert and Picus, Matti and Hoyer, Stephan and van Kerkwijk, Marten H. and Brett, Matthew and Haldane, Allan and del Río, Jaime Fernández and Wiebe, Mark and Peterson, Pearu and Gérard-Marchant, Pierre and Sheppard, Kevin and Reddy, Tyler and Weckesser, Warren and Abbasi, Hameer and Gohlke, Christoph and Oliphant, Travis E.},
  journal   = {Nature},
  title     = {Array programming with NumPy},
  year      = {2020},
  issn      = {1476-4687},
  month     = sep,
  number    = {7825},
  pages     = {357--362},
  volume    = {585},
  doi       = {10.1038/s41586-020-2649-2},
  publisher = {Springer Science and Business Media LLC},
}

@ARTICLE{Hunter2007,
  author={Hunter, John D.},
  journal={Computing in Science \& Engineering}, 
  title={Matplotlib: A 2D Graphics Environment}, 
  year={2007},
  volume={9},
  number={3},
  pages={90-95},
  doi={10.1109/MCSE.2007.55}}

@Article{Virtanen2020,
  author    = {Virtanen, Pauli and Gommers, Ralf and Oliphant, Travis E. and Haberland, Matt and Reddy, Tyler and Cournapeau, David and Burovski, Evgeni and Peterson, Pearu and Weckesser, Warren and Bright, Jonathan and van der Walt, Stéfan J. and Brett, Matthew and Wilson, Joshua and Millman, K. Jarrod and Mayorov, Nikolay and Nelson, Andrew R. J. and Jones, Eric and Kern, Robert and Larson, Eric and Carey, C J and Polat, İlhan and Feng, Yu and Moore, Eric W. and VanderPlas, Jake and Laxalde, Denis and Perktold, Josef and Cimrman, Robert and Henriksen, Ian and Quintero, E. A. and Harris, Charles R. and Archibald, Anne M. and Ribeiro, Antônio H. and Pedregosa, Fabian and van Mulbregt, Paul and Vijaykumar, Aditya and Bardelli, Alessandro Pietro and Rothberg, Alex and Hilboll, Andreas and Kloeckner, Andreas and Scopatz, Anthony and Lee, Antony and Rokem, Ariel and Woods, C. Nathan and Fulton, Chad and Masson, Charles and Häggström, Christian and Fitzgerald, Clark and Nicholson, David A. and Hagen, David R. and Pasechnik, Dmitrii V. and Olivetti, Emanuele and Martin, Eric and Wieser, Eric and Silva, Fabrice and Lenders, Felix and Wilhelm, Florian and Young, G. and Price, Gavin A. and Ingold, Gert-Ludwig and Allen, Gregory E. and Lee, Gregory R. and Audren, Hervé and Probst, Irvin and Dietrich, Jörg P. and Silterra, Jacob and Webber, James T and Slavič, Janko and Nothman, Joel and Buchner, Johannes and Kulick, Johannes and Schönberger, Johannes L. and de Miranda Cardoso, José Vinícius and Reimer, Joscha and Harrington, Joseph and Rodríguez, Juan Luis Cano and Nunez-Iglesias, Juan and Kuczynski, Justin and Tritz, Kevin and Thoma, Martin and Newville, Matthew and Kümmerer, Matthias and Bolingbroke, Maximilian and Tartre, Michael and Pak, Mikhail and Smith, Nathaniel J. and Nowaczyk, Nikolai and Shebanov, Nikolay and Pavlyk, Oleksandr and Brodtkorb, Per A. and Lee, Perry and McGibbon, Robert T. and Feldbauer, Roman and Lewis, Sam and Tygier, Sam and Sievert, Scott and Vigna, Sebastiano and Peterson, Stefan and More, Surhud and Pudlik, Tadeusz and Oshima, Takuya and Pingel, Thomas J. and Robitaille, Thomas P. and Spura, Thomas and Jones, Thouis R. and Cera, Tim and Leslie, Tim and Zito, Tiziano and Krauss, Tom and Upadhyay, Utkarsh and Halchenko, Yaroslav O. and Vázquez-Baeza, Yoshiki},
  journal   = {Nature Methods},
  title     = {SciPy 1.0: fundamental algorithms for scientific computing in Python},
  year      = {2020},
  issn      = {1548-7105},
  month     = feb,
  number    = {3},
  pages     = {261--272},
  volume    = {17},
  doi       = {10.1038/s41592-019-0686-2},
  publisher = {Springer Science and Business Media LLC},
}


\appendix

\section{Distribution of velocities}
The coupling between hydrodynamical code and subgrid jet model is an important aspect of jet modelling and should be treated carefully. In our simulations, we used different injection velocities and injection schemes which led to distinct jet behaviours. To examine these differences in detail, we analyse the velocity distributions of individual elements in our simulations and how these distributions vary between runs.

\begin{figure*}
    \centering
    \includegraphics[width=\textwidth]{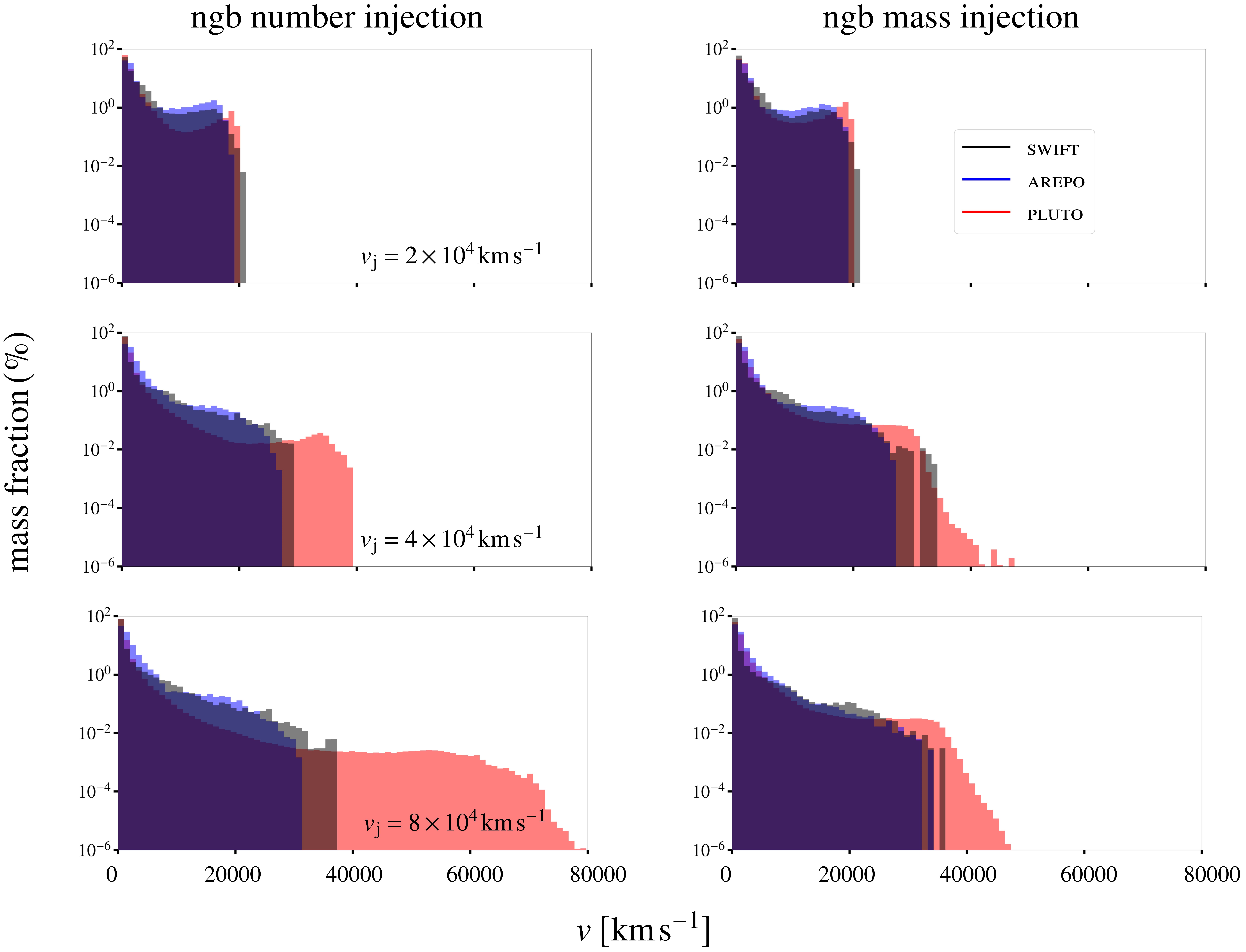}
     \caption{Mass-weighted velocity distributions of resolution elements for the different runs with the three codes. Rows correspond to the different jet injection velocities, while columns correspond to the two injection schemes: fixed neighbour number and fixed neighbour mass. At higher injection velocities, {\sc pluto} cells achieve higher velocities than the other codes but significantly less so for the mass-based injection scheme.}
    \label{fig:Ap.Velocities}
\end{figure*}

In Fig.~\ref{fig:Ap.Velocities}, we plot the mass-weighted velocity distributions of resolution elements for the different runs with the three codes. We include runs with the different jet injection velocities in separate rows and the two injection schemes in the two columns. The left column shows the neighbour-number scheme (see Section~\ref{ssec:jet model})  and the right column shows the neighbour-mass scheme (see Section~\ref{ssec:same neighbour mass injection}). At the lowest injection velocity ($v_\mathrm{j} = 2 \times 10^4 \, \mathrm{km/s}$), the distributions for the three codes show reasonable agreement for both injection schemes. For both the fiducial ($v_\mathrm{j} = 4 \times 10^4 \, \mathrm{km/s}$) and the highest ($v_\mathrm{j} = 8 \times 10^4 \, \mathrm{km/s}$) injection velocities, {\sc pluto} cells achieve higher velocities than the other two codes. This effect is significantly reduced in the case of the mass-based injection scheme. As discussed in the main text, in our fiducial neighbour-number scheme, the neighbour mass in {\sc pluto} will be much lower than in {\sc swift} and {\sc arepo}, as the latter codes employ (roughly) fixed mass resolution. As a result, for fixed kinetic energy injections, {\sc pluto} cells will achieve higher velocities than {\sc swift} particles and {\sc arepo} cells. In the fixed neighbour-mass scheme, however, this effect will naturally be corrected.

\bsp	
\label{lastpage}
\end{document}